\documentclass[a4paper,10pt]{article}
\usepackage[english]{babel}
\usepackage[bookmarksopen,bookmarksdepth=2,breaklinks=true]{hyperref}
\usepackage[ruled,vlined]{algorithm2e} 
\usepackage{booktabs}
\usepackage{mathtools}
\usepackage{amssymb,amsmath}
\usepackage{amsthm}

\usepackage{bbm}
\usepackage{xspace}

\usepackage{framed}
\usepackage{tikz}
\usetikzlibrary{trees}
\usepackage{caption}
\usepackage{subcaption}
\usepackage{fullpage}
\usepackage{amsfonts}
\usepackage[utf8]{inputenc}
\usepackage{authblk}
\usepackage{graphicx}
\graphicspath{{latexFigure/}}
\usepackage{cite}
\usepackage[round,authoryear]{natbib} 

\def\keywords{
{\textit{Keywords}:\,\relax%
}}

\title{Jumping VaR: Order Statistics Volatility Estimator for Jumps Classification and Market Risk Modeling\footnote{The views, thoughts and opinions expressed in this paper are those of the authors in their individual capacity and should not be attributed to Banco BPM S.p.A. or to the authors as representatives or employees of Banco BPM S.p.A.
The authors are grateful to the Elisabetta Benzi, Cecilia Mancini, Roberto Ren\`o for helpful comments and suggestions on the topic.}}

\author[1,2]{Luca Spadafora\thanks{Corresponding Author, \href{mailto:luca.spadafora@gmail.com}{luca.spadafora@gmail.com}}}
\author[1]{Francesca Sivero}
\author[1]{Nicola Picchiotti}
\affil[1]{Internal Model Validation, Banco BPM S.p.A., Piazza F. Meda 4, 20121 Milano, Italy}
\affil[2]{Department of Computer Sciences, University of Verona, Verona, Italy}

\date{}

\begin{document}

\maketitle


\begin{abstract}
\noindent
This paper proposes a new integrated variance estimator based on order statistics within the framework of  jump-diffusion models. Its ability to disentangle the integrated variance from the total process quadratic variation is confirmed by both simulated and empirical tests. 
For practical purposes, we introduce an iterative algorithm\footnote{available on the website \url{https://github.com/sigmaquadro/VolatilityEstimator}}
 to estimate the time-varying volatility  and  the occurred jumps of log-return time series. Such estimates enable 
the definition of a new market risk model for the Value at Risk forecasting. 
We show empirically that this procedure outperforms the standard historical simulation method applying standard back-testing approach.\\
\end{abstract}

\section{Introduction}
The Basel Committee requires that a certain amount of capital is held by financial firms as protection from market risk.  
According to the recent Fundamental Review of Trading Book (FRTB)~\cite{FRTB_1, FRTB_2,FRTB_3}, in case of  internal model approach, the risk measure involved in the computation of the capital requirement moves from Value at Risk (VaR) to  Expected Shortfall (ES). Nevertheless, the Value at Risk remains  the measure required for back-testing purposes. 
In particular, FRTB strengthened the VaR back-testing requirements at portfolio level, where jumping behaviour of time series is not off-setted or smoothed by the aggregation of many instruments. In addition, even if VaR computation is typically required for time horizons that range from $10$ days to one year, in practical applications VaR is usually estimated over a shorter time horizon, e.g. one day, in order to exploit a larger dataset, and then it is projected up to required time horizon, with suitable scaling formulas (see~\cite{Spadafora_1} and the references therein). As a consequence, back-testing analyses are typically performed on short time horizons, where the jumping behaviour can be relevant.
For this reason, the definition of  a market risk model allowing to produce accurate VaR predictions endures as research topic and we chose it as field to which provide our contribution. \\ \\
Noting that the presence of large fluctuations in asset log-returns is responsible for part of the poor back-testing performances of the standard historical simulation VaR method, we propose a new approach that distinguishes the past realizations into ordinary and jump ones and includes this information in the VaR prediction.
More precisely, in order to achieve this goal, as a  first step, we defined a new integrated variance estimator based on order statistics, proposing  an iterative algorithm to estimate the time-varying volatility and the occurred jumps of log-return times series, and, secondly, we introduced a new method for VaR forecasting that models the jumping behaviour of the time series with an ad-hoc approach, specific for the jumping component.
%
\\\\
The outline of the paper is the following. In Section \ref{Literature} we  revise the existing literature related to the topics discussed in the paper.
Section \ref{Vol_Estimators} presents the mathematical framework and the definition of the order statistic volatility estimator, introducing  algorithms for its implementation, 
while Section \ref{NETest} shows  its ability to disentangle the integrated variance from the total process quadratic variation by simulated and empirical tests.
Section \ref{Data_Description} describes a sanity check that we performed on empirical data to verify whether the initial assumptions of our  model held true.
In Section \ref{JumpingVaR_Models} we introduce a new market risk model for VaR estimation based on the outcomes of the  order statistic volatility estimator. Finally, in Section \ref{Results}, we empirically back-test the Jumping VaR model comparing its performances with the ones of other standard and advanced VaR forecasting models.
In Section \ref{Summary_Conclusion} we provide our conclusions.

\section{Literature Review}\label{Literature}
The presence of jumps (large fluctuations) in asset return time series has been confirmed by a wide variety of empirical studies. In many recent works such  asset returns are modeled by jump-diffusion dynamics  and the issue of disentangling the continuous part of the stochastic process from the jump one has been faced. 
For instance, \cite{RePEc:eee:jfinec:v:74:y:2004:i:3:p:487-528} has proven the asymptotic  ability of the MLE to estimate the volatility of log-returns cleaned up from the noise due to the jump component, considering  Poisson jump   diffusion processes (finite jump activity) and  extending the results to  Cauchy jump diffusion processes (specific cases of infinite jump activity). 
\cite{RePEc:oxf:wpaper:2003-w18} have proposed a model-free estimator of the  continuous and the jump part of the  quadratic variation for  stochastic volatility models in case of finite activity jumps, based on the realized power and bipower variations.
This study on the realized multipower variation estimator has also been extended to the case of infinite activity jump processes in \cite{RePEc:eee:spapps:v:116:y:2006:i:5:p:796-806} proving the robustness of such estimator also in case of  infinite activity Lévy processes, with activity  not too high, and providing  asymptotic distributions of the realized multipower variation when there are jumps.
 \cite{RePEc:bla:scjsta:v:36:y:2009:i:2:p:270-296} has presented a non-parametric threshold estimator of the integrated variance for a jump-diffusion model with finite or infinite jump activity, providing also asymptotic results and giving, in case of finite jump activity, a non-parametric estimate of the jump times. We mainly consider  this paper and  \cite{RePEc:eee:econom:v:160:y:2011:i:1:p:77-92} as a starting point for our contribution. Additional details on this estimator can be found in \cite{spadafora2017theoretical}. \\
While several authors have focused their research on jump-diffusion models to describe the asset return dynamics, it is worth noting that \cite{RePEc:ucp:jnlbus:v:75:y:2002:i:2:p:305-332} and \cite{RePEc:eee:jbfina:v:26:y:2002:i:7:p:1297-1316}  have pointed out that pure jump Lévy processes can be used to model asset returns in a more realistic way, substituting the continuous part with infinite small jumps due to the infinite activity jump component of the Lévy process. \\\\
The issue of including asset return jumps in the Value at Risk forecasting has been approached in several papers in the last years. A wide range of these works aims at computing the VaR via parametric methods, supposing jump-diffusion dynamics of the asset log-returns. For instance, \cite{RePEc:spr:finsto:v:5:y:2001:i:2:p:155-180} have suggested computing VaR by applying the Inverse Fourier Transform to the characteristic function of the portfolio log-return process approximated via delta-gamma approach, while \cite{RePEc:fip:fedgfe:2009-40} has implemented the VaR forecasting based on jump-diffusion models, including stochastic volatility and both finite and infinite activity jump components, and on the model specification estimation via Bayesian MCMC (Markov Chain Monte Carlo). Moreover, further related works include \cite{RePEc:taf:quantf:v:4:y:2004:i:2:p:129-139} and \cite{research} proposing analytical VaR computation in jump diffusion model framework. Less research has been performed on the VaR prediction including jumps through the non-parametric method of the  historical simulation. In this regard, \cite{RePEc:fip:fedgfe:2001-17} has proposed  an algorithm to compute VaR based on both the variance-covariance  and on the historical simulation method, modeling the loss process as sum of an ordinary component and a jump one described by a trinomial distribution and also pointing out the need to add the correlation among jump times to better capture the real asset return dynamics.

\section{Volatility Estimator}\label{Vol_Estimators}
\subsection{Introduction}\label{Vol_Estimators_Introduction}
Starting from the non-parametric threshold estimator introduced in  \cite{RePEc:bla:scjsta:v:36:y:2009:i:2:p:270-296}, we propose an estimator of the integrated variance of jump-diffusion stochastic processes, based on order statistics. As a result, estimates of the process time-varying volatility can be obtained by an iterative method and they can be employed for practical purposes, for example for market risk modeling.\\
In this section we provide a simple introduction to the estimator, describing the intuition behind it. As a consequence, we intentionally avoided some accuracy in the approach in favour to a better description of the general idea.  In the following sections, we provide a formal mathematical description of the statistical estimator we want to introduce. \\
In a nutshell, we want to obtain a definition of a jump as a realization that is abnormal and anomalous with respect to the other realizations in our time series. For this reason, we need to consider a reference distribution (in the rest of the paper we consider a Gaussian distribution even if the approach can be easily extended to other distributions) and we need to evaluate if a given realization is compatible with it. A typical approach in this case could be to classify as jump (as anomalous realization) all realizations that lie far from the centre of the distribution. A very naive approach could be for example to consider some multiple of the standard deviation 
of the time series and assume that all the realizations larger (or smaller) than this (negative) threshold are outliers. In our view, this approach is rather rough, as \textit{all} the realizations above (or below) the (negative) threshold are classified as jumps. 
On the contrary, it could happen that many realizations do not lie too far from the centre of the distribution, but they are simply too frequent for the reference distribution. A very simple example could be obtained if one considers a T-Student distribution for the realizations and a Gaussian distribution as a reference. As the T-Student distribution has fat tails, once a sufficiently large (in absolute terms) threshold is defined, many realizations will lie below the threshold, even if their frequency is not compatible with a Gaussian distribution. In other words, for a given threshold, only the really extreme realization will be classified as jump, but the tail of the distribution will remain fatter than the one implied by a Gaussian distribution. In any cases, it can be proved that there exists a smart way to define this threshold in order to obtain estimators that satisfy typical statistical convergence requirements~\cite{RePEc:bla:scjsta:v:36:y:2009:i:2:p:270-296}.
In this paper we want to develop a different approach that is able to go beyond the truncation effect given by the definition of the threshold while maintaining most of the theoretical framework introduced in~\cite{RePEc:bla:scjsta:v:36:y:2009:i:2:p:270-296} that represents our main reference.
Essentially, the main idea of this paper is based on two key points:
\begin{itemize}
\item the mapping of the threshold definition problem into a probability threshold problem
\item the introduction of extreme value theory as a statistical tool to identify jumps.
\end{itemize}
Concerning the first point, we transform the threshold problem in a statistical problem where we want to estimate what is the \textit{probability} of observing a realization so large (or so small). Even if the two problems are equivalent as we just applied a mapping, we think that this step is relevant from a practical point of view because the definition of a probability threshold is typically easier to understand and to interpret than a threshold on a random variable.\\
With respect to the second point, we transform the statistical question from \textit{what is the probability of observing a realization given a reference distribution?} to the question \textit{what is the probability of getting a $k-$th maximum as the one observed if we could sample many times a time series of $n$ realizations?}
By this change in the approach, we are questioning if each ordered realization can be larger (or smaller) than the one observed; in this way, it could happen that the maximum is not large enough to be classified as jump, differently to the $k-$th maximum. In this way, in the above T-Student example, the frequency of largest (smallest) observations can be reduced, avoiding the problem of introducing a truncation of the distribution.\\
In the following sections, we provide a formal description of these simple ideas with both numerical and empirical examples.

\subsection{Framework\label{VE_frame}}
According to the Lévy-It\^{o} Decomposition, a Lévy process is characterized by the superposition of a continuous part, formed by a drift term and a Brownian motion, and  independent Poisson processes, possibly infinitely many, as reported in \cite{tankov2015financial}. Therefore, modeling the stock log-returns as realizations of a Lévy process,  we consider the following set-up.\\\\
Let $\left\{Y_t\right\}_{t \in \mathbbm{R}^+}$  be a stochastic process  with dynamics
\begin{equation}\label{PAeq1}
\left\{\begin{matrix}\begin{split}
&\mathrm{d}Y_t=\sigma_t\mathrm{d}W_t+ \mathrm{d}J_t =\mathrm{d}Y_t^c + \mathrm{d}J_t \quad t >0\\ 
&Y_0=y_0 \in \mathbbm{R}
\end{split}\end{matrix}\right.
\end{equation}
where $W_t$ is a standard Brownian motion, $\sigma_t$ is the volatility of the continuous part of $Y_t$ (named $Y_t^c$) and $J_t$ is a pure jump process.\\
The pure jump process can have either finite activity (FA), i.e. finitely many ``big" jumps can occur in finite time, or infinite activity (IA), i.e. an infinite number of ``compensated small" jumps can occur in finite time interval.\\
Note that we assume the drift term to be negligible and thus we do not include it in the $Y_t$ dynamics in Eq.~\ref{PAeq1}. \\\\ 
The quadratic variation of the process $Y_t$ corresponds to the sum of the quadratic variation of the continuous part,  called  integrated variance (IV),  and the quadratic variation of the jump component (\cite{tankov2015financial}). Formally,
\begin{equation}\label{PAeq2}
\left[Y,Y\right]_t = \left[Y^c,Y^c\right]_t + \sum_{\substack{s\leqslant t \\ \Delta J_s \neq 0}} |\Delta J_s|^2 =\int_0^t\sigma^2_s \mathrm{d}s + \sum_{\substack{s\leqslant t \\ \Delta J_s \neq 0}} |\Delta J_s|^2
\end{equation}
Assuming that the realizations of the $Y_t$ process  are recorded at $n$ discrete equispaced times $\delta t, \dots, n\delta t=T$, our aim is to estimate the integrated variance of the process disentangling it from the quadratic variation of the jump part.\\\\
Naming $Y_{\delta t}, Y_{2\delta t}, \dots, Y_{(n+1)\delta t}$ the $n+1$ random variables obtained sampling $Y_t$ in time,  denote by $\Delta_i Y:= Y_{\left(i+1\right)\delta t}-Y_{i\delta t}$ for $i=1,\dots,n$ the discrete increments of   $Y_t$ and, likewise, $\Delta_i Y^c:= Y_{\left(i+1\right)\delta t}^c-Y_{i\delta t}^c$ as the discrete increments of the continuous part of   $Y_t$. 
The corresponding realizations of such random variables are indicated by lower-case letters, that is $\left\{\Delta_1 y,\dots,\Delta_n y\right\}$, on one hand, and $\left\{\Delta_1 y^c,\dots,\Delta_n y^c\right\}$, on the other hand. This notation will be kept throughout the whole paper.
\subsection{Threshold  Estimator}
The non-parametric threshold estimator proposed in \cite{RePEc:bla:scjsta:v:36:y:2009:i:2:p:270-296} is defined as 
\begin{equation}\label{PAeq3}
\hat{IV}_{Thr} = \sum_{i=1}^n \left(\Delta_i Y\right)^2\mathbbm{1}_{\left\{\left|\Delta_i Y\right|\leqslant \sqrt{\theta \left(\delta t\right)} \right\}}
\end{equation}
where  $\theta \left(\delta t\right)$ is a threshold function that must satisfy $\underset{\delta t \to 0}{\text{lim}}\theta\left(\delta t\right)=0$ and $\underset{\delta t \to 0}{\text{lim}}\frac{\delta t \text{log}\frac{1}{\delta t}}{\theta\left(\delta t\right)}=0$.\\\\
The  Levy's Modulus of Continuity Theorem (\cite{sato1999levy}) is the key result exploited to prove the consistency of this estimator. It states that
\begin{equation}\label{PAeq50}
\underset{\epsilon \to 0}{\text{lim}}\, \underset{\substack{0<t<1\\0<s <\epsilon}}{\text{sup}} \frac{|W_{t+s}-W_{t}|}{\sqrt{2s\text{log}\frac{1}{s}}}=1
\end{equation}
almost surely. Therefore, considering the time grid $0,\delta t, \dots, n\delta t=T$, as enunciated in \cite{RePEc:flo:wpaper:2010-03} and  \cite{RePEc:bla:scjsta:v:36:y:2009:i:2:p:270-296},
\begin{equation}\label{PAeq4}
\mathbb{P}\left(\underset{\delta t \to 0}{\text{lim}}\, \underset{i=1,\dots,n}{\text{sup}} \frac{|W_{\left(i+1\right)\delta t}-W_{i\delta t}|}{\sqrt{2\delta t \text{log}\frac{1}{\delta t}}}\leqslant 1\right)=1
\end{equation}
and, thus, keeping in mind the stochastic integral definition, the absolute value of any path of a stochastic integral with respect to a Brownian motion tends to zero almost surely as the function $\sqrt{2\delta t \text{log}\frac{1}{\delta t}}$. As a result, if $\left(\Delta_i Y\right)^2>\theta \left(\delta t\right)>2\delta t \text{log}\frac{1}{\delta t}$ then,  in the time interval $\delta t$, some jumps in the $Y_t$ process  must be occurred, when $\delta t$ approaches zeros.\\\\
According to \cite{RePEc:bla:scjsta:v:36:y:2009:i:2:p:270-296}, under the assumptions on the drift, the diffusion and the threshold function stated in the mentioned paper, the threshold estimator is \emph{consistent} for both cases of finite and  infinite activity jump part. 
Moreover, considering a process with only finite activity  jump component and under the same set of hypotheses required for the consistency, the threshold estimator has a Normal asymptotic distribution (i.e.~as $\delta t \to 0$, having fixed the grid time horizon $T$) with mean corresponding to the integrated variance and variance proportional 
to the inverse of the number of time steps at which the realizations of the stochastic process are recorded, fixed the time horizon of the grid. 
Consequently, the speed of  convergence of the estimator to the true parameter  depends on the square root of the number of time steps.\\\\
The hypotheses on the threshold function are necessary for the consistency of the estimator, although, in practice, we can observe the realizations of   $Y_t$ only at discrete times and thus the time step $\delta t$ is fixed. Within this setting, the choice of the threshold function is not so easy. For example, the threshold 
function of the form $\theta\left(\delta t\right):= \left(\delta t\right)^{\beta}$ with $\beta \in \left]0,1\right[$  suggested in \cite{RePEc:bla:scjsta:v:36:y:2009:i:2:p:270-296} is  dominated by the $2\delta t \text{log}\frac{1}{\delta t}$ only in the limit for $\delta t \to 0$ and thus, if $\delta t$ is sufficiently high, the relation $\left(\Delta_i Y\right)^2>\theta \left(\delta t\right)>2\delta t \text{log}\frac{1}{\delta t}$ does not hold anymore and  some jumps could not be detected by the estimator. In addition, as $\Delta_i Y$ depends on the scale parameter $\sigma$, in order to assure that the relation holds, a suitable renormalization procedure must be considered in order to neutralize the effect of the scale parameter.\\  
In accordance with this reasoning, in our implementation of the estimator in Eq.~\ref{PAeq3}, we choose the threshold function exactly equal to
\begin{equation}\label{PAeq5}
\bar{\theta}\left(\delta t\right):=2 \text{log}\frac{1}{\delta t}
\end{equation}
rescaled by the sample variance of $\left\{\Delta_1 Y,\dots,\Delta_n Y\right\}$ (that is proportional to $\delta t$), obtaining a threshold proportional to $\delta t \text{log}\frac{1}{\delta t}$.\\
It is worth noting that the choice of this threshold function agrees with the results proved  in~\cite{RePEc:flo:wpaper:2017-01} concerning the optimal threshold  definition by means of  the mean square error technique.
\subsection{From Maximum to Order Statistics}
Since Eq.~\ref{PAeq4} holds in the limit, in case of fixed $\delta t$ the following relation must be true
\begin{equation}\label{PAeq6}
\begin{split}
\mathbb{P}\left(\underset{i=1,\dots,n}{\text{max}} \frac{|W_{\left(i+1\right)\delta t}-W_{i\delta t}|}{\sqrt{2\delta t \text{log}\frac{1}{\delta t}}}> 1\right)=\bar{p} &\iff  \mathbb{P}\left(\underset{i=1,\dots,n}{\text{max}} |W_{\left(i+1\right)\delta t}-W_{i\delta t}| > \sqrt{2\delta t \text{log}\frac{1}{\delta t}}\right)=\bar{p}
\end{split}
\end{equation}
where $\bar{p}\in\left[0,1\right]$ is a certain probability value which refers to the random variable $\underset{i=1,\dots,n}{\text{max}} |W_{\left(i+1\right)\delta t}-W_{i\delta t}|$. Eq.~\ref{PAeq6} provides a link between the threshold exploited in \cite{RePEc:bla:scjsta:v:36:y:2009:i:2:p:270-296} in order to define the estimator reported in Eq.~\ref{PAeq3} and a probability threshold $\bar{p}$. In particular, by this formulation, the choice of the threshold $\bar{\theta}\left(\delta t\right)$ results into the definition of the confidence level ($\bar{p}$) corresponding to the error in the jump detection that one is willing to accept.\\\\
More in general, it is possible to extend this approach for a generic tolerance level $p$ (without considering the absolute value)
\begin{equation}\label{PAeq7}
\mathbb{P}\left(\underset{i=1,\dots,n}{\text{max}} \left(W_{\left(i+1\right)\delta t}-W_{i\delta t}\right) > \theta^W\right)=p
\end{equation}
for a threshold $\theta^W$. Since in our approach, we want to invert Eq.~\ref{PAeq7} in order to obtain a threshold for a fixed value of the probability $p$, we make explicit this dependence using the notation $\theta^W\left(p;n,n\right)$, where the double $n$ refers to the fact that the maximum operator in Eq.~\ref{PAeq7} can be interpreted as the $n-$th minimum of a sample made by $n$ realizations. This notation will be exploited in the next sections, where we will generalize this setup. In addition, we apply our choice to consider a Gaussian cumulative distribution function (c.d.f.) as a reference distribution, obtaining the following results
\begin{equation}
\begin{array}{lll}
\theta^W\left(p;n,n\right)&=&\sqrt{\delta t}\,\theta\left(p;n,n\right)\\
\theta\left(p;n,n\right)&=&F^{-1}_{Z_{n:n}}\left(1-p\right)\\
\end{array}
\end{equation}
where $F_{Z_{n:n}}\left(x\right)=\left(\Phi\left(x\right)\right)^n$ is the c.d.f.~of the maximum of $n$ Standard Normal random variables and $\Phi\left(x\right)$ is the c.d.f.~of a Standard Normal random variable (refer to Appendix \ref{App1} for further details on such derivation). As mentioned in Section~\ref{Vol_Estimators_Introduction}, here we assume that our reference distribution is Normal but the approach can be easily extended to other reference distributions. The notation $Z_{k:j}$ stands for the $k-$th realization over an (increasing) ordered sample of $j$ elements.\\
It is worth noting that, since the Normal distribution is symmetric, the random variables $Z_{n:n}$ and $-Z_{1:n}$ (where $Z_{1:n}$ is the minimum of $n$ Standard Normal random variables) have the same distribution. In the following, we will consider Eq.~\ref{PAeq7} as starting point for the definition of our new volatility estimator.\\\\
In the light of the approximation stated in Appendix \ref{App2}, $\theta \left(p;n,n\right)$  multiplied by the square root of the sample variance of   $\left\{\Delta_1 Y,\dots,\Delta_n Y\right\}$ can be set as   threshold function to plug-in in Eq.~\ref{PAeq3} to estimate the IV of the Lévy process $Y_t$, once the tolerance level $p$ is fixed (note that here we are assuming that the volatility is constant in time, whereas in the next section we will relax this hypothesis).\\
This choice  implies that all the realizations in the set $\left\{\Delta_1 y,\dots,\Delta_n y\right\}$ larger, in absolute terms,  than the maximum value  acceptable for the discrete increments of $Y_t^c$  are excluded from the IV computation.  
\\
Nevertheless, this choice does not take into account the possibility of having realizations in the set $\left\{\Delta_1 y,\dots,\Delta_n y\right\}$ which, even being smaller than the maximum acceptable value, are too frequent to be Gaussian and, thus, to be realizations of the discrete increments of $Y_t^c$ only.
According to this reasoning, the threshold $\theta \left(p;n,n\right)$ can be replaced by a realization-varying threshold $\theta \left(p;k,n'\right)$, whose value depends on the progressive  classification of the ordered realizations of $\left\{\Delta_1 y,\dots,\Delta_n y\right\}$. \\
More precisely, starting from Eq.~\ref{PAeq7} and substituting the maximum over the $n$ observations with the $k$-th order statistic in the dataset of size $n'$, a more general threshold function $\theta\left(p;k,n'\right)$ can be defined. Formally,
\begin{equation}\label{PAeq8}
\begin{aligned}
\mathbb{P}\left(\text{$k$-th ordered statistic in the dataset} \right.& \left.\left\{\left(W_{\left(i+1\right)\delta t}-W_{i\delta t}\right)\right\}_{i=1}^{n'}  > \theta^W\left(p;k,n'\right)\right)=p\\
\ArrowBetweenLines
\theta^W\left(p;k,n'\right)=\sqrt{\delta t}\theta\left(p;k,n'\right)\quad \text{and}&\quad \theta\left(p;k,n'\right)=F^{-1}_{Z_{k:n'}}\left(1-p\right)
\end{aligned}
\end{equation}
where 
\begin{equation}\label{PAeq9}
F_{Z_{k:n'}}\left(x\right)=I_{\Phi\left(x\right)}\left(k,n'-k+1\right)=\frac{n'!}{\left(k-1\right)!\left(n'-k\right)!}\int_{0}^{\Phi\left(x\right)} u^{k-1}\left(1-u\right)^{n'-k} \mathrm{d}  u
\end{equation}
is the c.d.f.~of the $k$-th order statistic in the dataset of $n'$ Standard Normal random variables and $I_{x}\left(a,b\right)$ is an  Incomplete Beta function with parameters $a,b$. (The whole derivation of the threshold $\theta\left(p;k,n'\right)$ is reported in Appendix \ref{App1}). Moreover, thanks to the symmetry of the Normal distribution,  $Z_{k:n'}$ has the same distribution of $- Z_{\left(n'-k+1\right):n'}$. \\\\
Therefore, the procedure can be summarized as follows:
$\left\{\Delta_1 y,\dots,\Delta_n y\right\}$ are ordered and the largest observation in this set is compared with the threshold $\theta\left(p;n,n\right)$ rescaled by the square root of the sample variance of $\left\{\Delta_1 y,\dots,\Delta_n y\right\}$; depending on whether the largest observation has been classified as jump or not, the threshold itself can be updated to $\theta\left(p;n-1,n-1\right)$ or $\theta\left(p;n-1,n\right)$, always rescaling by the square root of the sample variance of $\left\{\Delta_1 y,\dots,\Delta_n y\right\}$. The new threshold so obtained can then be compared with the second largest element in the sample and so on and so forth.
The formal definition of such new estimator of the IV based on the order statistics theory is reported hereafter.

\subsection{Order Statistic Estimator}\label{OS}
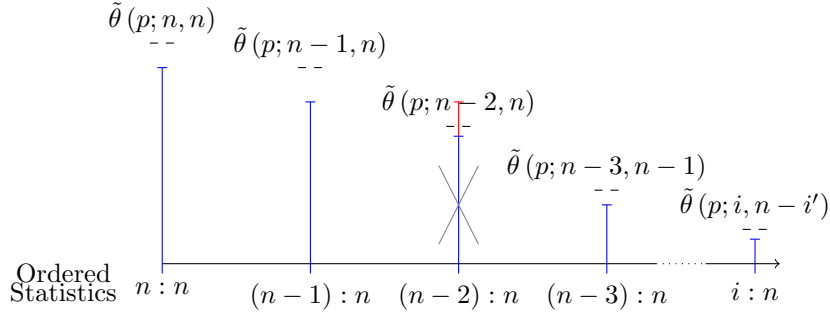
\begin{figure}[t]
\begin{center}
%
%
%
%
\begin{tikzpicture} [scale=0.65] 
\draw [-](0,0)--(10,0); 
\draw [dotted](10,0)--(11,0); 
 \draw [->](11,0)--(12.5,0); 
\draw [blue](0,-.2)--(0,4); 
 \draw [blue](3,-.2)--(3,3.3); 
 \draw[blue] (9,-.2)--(9,1.2); 
 \draw [blue](6,-.2)--(6,2.6); 
 \draw [blue](12,-.2)--(12,0.5); 
\draw[red] (6,2.6)--(6,3.3); 


\draw[dashed] (-.25,4.5)--(0.25,4.5);
\draw[dashed] (2.75,4)--(3.25,4);
\draw[dashed] (5.75,2.8)--(6.25,2.8); 
\draw[dashed] (8.75,1.5)--(9.25,1.5); 
\draw[dashed] (11.75,.7)--(12.25,.7); 

\draw [blue](-.1,4)--(0.1,4);
\draw[blue] (2.9,3.3)--(3.1,3.3);
\draw[red] (5.9,3.3)--(6.1,3.3);
\draw [blue](5.9,2.6)--(6.1,2.6); 
\draw [blue](8.9,1.2)--(9.1,1.2); 
\draw[gray] (5.6,.4)--(6.4,2);
\draw[gray] (5.6,2)--(6.4,.4);
\draw [blue](11.9,0.5)--(12.1,0.5);

 \node[align=left,below] at (-2,.2) {Ordered};
 \node[align=left,below] at (-2,-0.2) {Statistics}; 
 \node[align=left,above] at (0,4.5) {$\tilde{\theta}\left(p;n,n\right)$}; 
 \node[align=left,above] at (3,4) {$\tilde{\theta}\left(p;n-1,n\right)$};  
 \node[align=left,above] at (6,2.8) {$\tilde{\theta}\left(p;n-2,n\right)$}; 
  \node[align=left,above] at (9,1.5) {$\tilde{\theta}\left(p;n-3,n-1\right)$}; 
\node[align=left,above] at (12,0.7) {$\tilde{\theta}\left(p;i,n-i'\right)$}; 
\node[align=left,below] at (0,-0.2) {$n:n$}; 
\node[align=left,below] at (3,-0.2) {$\left(n-1\right):n$};  
 \node[align=left,below] at (6,-0.2) {$\left(n-2\right):n$};  
 \node[align=left,below] at (9,-0.2) {$\left(n-3\right):n$};  
\node[align=left,below] at (12,-0.2) {$i:n$};  
\end{tikzpicture} 
\end{center}
\caption{Some possible ordered realizations of Standard Normal random variables (in blue) with a jump component realization addition (in red)  and their relative OS threshold levels (dashed line), in a simplified framework.} \label{figE:1}
\end{figure}

First of all, we denote the ordered set of the discrete increments of the $Y_t$ process  as $\left\{\Delta_{n:n} Y,\dots,\Delta_{1:n} Y\right\} $ with $ \Delta_{i:n-1} Y\leqslant \Delta_{\left(i+1\right):n} Y\quad i=1,\dots,n$. Moreover, we introduce the increasing sequences of sets $\left\{\mathcal{G}_i \right\}_{i=0}^{m}$ and $\left\{\mathcal{G}^{'}_i \right\}_{i=0}^{m'}$, with $m= \left \lceil \frac{n}{2}  \right \rceil$ and  $m'= \left \lfloor \frac{n}{2}  \right \rfloor$. Assuming $n$ even for simplicity, we define
\begin{equation}\label{PAeq11}
\begin{split}
&\mathcal{G}_0 = \o \quad\quad \mathcal{G}_i =\left\{I_1,\dots,I_i\right\} \quad i=1,\dots,m\\
&\mathcal{G}^{'}_0 = \o \quad\quad \mathcal{G}^{'}_i =\left\{I^{'}_1,\dots,I^{'}_i\right\}\quad i=1,\dots,m'
\end{split}
\end{equation}
where $I_i$ and $I^{'}_i$  are random variables  progressively defined as
\begin{equation}\label{PAeq10}
\begin{split}
&I_i=\left\{\begin{array}{ll}
1& \text{if }\Delta_{\left(n-i+1\right):n} Y > \tilde{\theta}\left(i, \mathcal{G}_{i-1},\mathcal{G}^{'}_{i-1}\right)\\ 
0& \text{otherwise}
\end{array}\right.  \quad i=1,\dots,m\\ 
&I^{'}_i=\left\{\begin{array}{ll}
1& \text{if }-\Delta_{i:n} Y > \tilde{\theta}\left(i, \mathcal{G}_{i},\mathcal{G}^{'}_{i-1}\right)\\ 
0& \text{otherwise}
\end{array}\right.  \quad i=1,\dots,m'
\end{split}
\end{equation}
While the threshold function $\tilde{\theta }\left(i,\mathcal{G}_{a},\mathcal{G}_{b}^{'}\right)$ has the following definition
\begin{equation}\label{PAeq15}
\tilde{\theta }\left(i,\mathcal{G}_{a},\mathcal{G}_{b}^{'}\right) := \sqrt{\hat{\sigma}^2}\cdot\left\{\begin{array}{ll}
\theta\left(p;n,n\right)& \text{if}\; a=0,b=0\\ 
\theta\left(p;n-\sum_{j=1}^{a}I_j,n-\sum_{j=1}^{a}I_j\right)& \text{if}\; a\ne 0, b=0\\
\theta\left(p;n-i+1-\sum_{j=1}^{b}I^{'}_j,n-\sum_{j=1}^{a}I_j-\sum_{j=1}^{b}I^{'}_j\right)& \text{if}\; a \neq b \; \text{and}\;  b \neq 0\\
\theta\left(p;n-i+1-\sum_{j=1}^{a}I_j,n-\sum_{j=1}^{a}I_j-\sum_{j=1}^{b}I^{'}_j\right)& \text{otherwise}
\end{array}\right. 
\end{equation}
where $\hat{\sigma}^2$ indicates the sample variance estimator of $\left\{\Delta_1 Y,\dots,\Delta_n Y\right\}$ and $\theta\left(p;k,n'\right)$ is computed according to the formulas in Eq.~\ref{PAeq8} and Eq.~\ref{PAeq9} given $p$. \\\\
Fixed a tolerance level $p \in \left[0,1\right]$, the Order Statistic (OS) estimator is defined as follows
\begin{equation}\label{PAeq14}
\hat{IV}_{OS} = \sum_{i=1}^m \left(\Delta_{\left(n-i+1\right):n} Y\right)^2\mathbbm{1}_{\left\{\Delta_{\left(n-i+1\right):n} Y\leqslant \tilde{\theta }\left(i,\mathcal{G}_{i-1},\mathcal{G}_{i-1}^{'}\right) \right\}}+\sum_{i=1}^{m'} \left(\Delta_{i:n} Y\right)^2\mathbbm{1}_{\left\{-\Delta_{i:n} Y\leqslant \tilde{\theta }\left(i,\mathcal{G}_{i},\mathcal{G}^{'}_{i-1}\right) \right\}}
\end{equation}
It should be noticed that, in the OS estimator formulation,  each random variable corresponding to an increment of the process $Y_t$ is compared with a suitable threshold that is updated according to the new information flow arrival, while $\hat{\sigma}^2$ is used to appropriately rescale the realization-varying threshold.\\
Basically, the algorithm  which updates the information to define the thresholds starts from the maximum observation, then moves to  the  minimum observation, than removes towards the second largest realization and so on and so forth. The sequences $\left\{\mathcal{G}_i \right\}_{i=1}^{m}$ and $\left\{\mathcal{G}^{'}_i \right\}_{i=1}^{m'}$ allow to take into account the information history to update the thresholds and the introduction of the second sequence $\left\{\mathcal{G}^{'}_i \right\}_{i=1}^{m'}$ has the goal of exploiting the property of symmetry of the Normal distribution in the definition of our new estimator. In the practical use of the estimator, it should be noticed that the observations should be half positive and half negative in order to avoid possible distortion, this is clearly coherent with the symmetric choice of the distribution.\\\\
To give an intuition of the behaviour of this new estimator, we propose hereafter a simplified setting. In  Figure \ref{figE:1}, possible realizations of Standard Normal random variables (blue segments) with their threshold levels  (dashed lines) are shown in decreasing order. The third ordered observation is composed by a Gaussian realization   with the addition of a small   jump  realization (red interval). The comparison between the second and the third ordered observations  underlines the need for taking into account  the frequency of the realizations as well as their sizes, in the jump detection procedure. 
Indeed, if a fixed threshold, corresponding for example to the maximum admissible value for Normal random variables, having set the tolerance level $p$, was used to discriminate observations contaminated by jumps or not, the third realization of Figure \ref{figE:1} would not be detected as jump. On the contrary, the use of realization-varying thresholds allows to identify this kind of realizations interested by a jump component. 
Roughly speaking, an unusually high frequency of observations near the second and the third greatest values of the observations is interpreted by our estimator as a signal that these realizations cannot come from a Gaussian random variable.
 Obviously,  our estimator is not able to disentangle which realization is really affected by a jump.\\ Moreover, for the sake of completeness, pure Gaussian realizations could also be detected as jumps by the OS estimator with a certain probability (whose maximum level is $p\%$). 
\\\\This estimator is practically implemented by the simple  iterative algorithm reported in Algorithm \ref{alg1}. For simplicity, as already stated, we consider the case of an even number of $n$ observations.
\begin{algorithm}[t]
 \KwIn{$\left(\Delta_{n:n} y,\dots,\Delta_{1:n} y\right)$ vector of ordered observations, $p$ tolerance level, $s^2$ estimate of the sample variance of realizations }
 \KwOut{$s_{IV}$ IV estimate, $\left(j_n,\dots,j_1\right)$ vector of boolean jump indicators  associated to the vector $\left(\Delta_{n:n} y,\dots,\Delta_{1:n} y\right)$ }
 Initialize  $\left(j_n,\dots,j_1\right)=\left(0,\dots,0\right)$,  $n'=n$, $k=1$, $k'=1$\;
 \For{$i=n,n-1,\dots,\frac{n}{2}$}{
  Compute $\theta\left(p;n'-k+1,n'\right)$  according to Eq.~\ref{PAeq8} and Eq.~\ref{PAeq9} \;
 Evaluate $\tilde{\theta_1} =\sqrt{s^2}\theta\left(p;n'-k+1,n'\right)$\;
  \eIf{$\Delta_{i:n} y>\tilde{\theta_1}$}{
   $j_i \leftarrow1$\;
$n' \leftarrow n' -1$\;
   }
{
$k \leftarrow k+1$\;
}
Compute $\theta\left(p;n'-k'+1,n'\right)$ according to Eq.~\ref{PAeq8} and Eq.~\ref{PAeq9} \;
Evaluate $\tilde{\theta_2} =\sqrt{s^2}\theta\left(p;n'-k'+1,n'\right)$ \;
\eIf{-$\Delta_{\left(n-i+1\right):n} y>\tilde{\theta_2}$}{
   $j_{n-i+1} \leftarrow1$\;
$n' \leftarrow n' -1$\;
   }{$k' \leftarrow k'+1$\;}
 }
Evaluate $s_{IV}=\sum_{i=1}^{n}\left(\Delta_{i:n}y\right)^2\cdot j_i$
 \caption{Base algorithm to compute the IV by means of the OS estimator. \label{alg1}}
\end{algorithm}

\subsection{Iterative Algorithm for Local Volatility Estimation}\label{IA}
A further extension concerning  the OS estimator is its application to an iterative algorithm in order to estimate  the time-varying volatility of the $Y_t$ process. \\\\
The estimation of the local volatility $\sigma_t$ of the $Y_t$ process   is not completely straightforward and it involves  an iterative two-step estimation procedure. 
At first,  the process realizations must be renormalized by the corresponding local volatility estimates and, then, the OS estimator, including a kernel function,  must be applied to such renormalized observation set, obtaining new time-varying volatility estimates. 
The formal definition is the following. \\\\
First of all, a kernel function with bandwidth $h$ is defined as  
\begin{equation}\label{PAeq16}
 K_h\left(s\right):=\frac{1}{h}K\left(\frac{s}{h}\right) 
\end{equation}
where $K\left(s\right)$  is a non-negative weighting function, i.e.   such that $\int_{-\infty}^{+\infty}K\left(s\right)\mathrm{d}s = 1$ and $K\left(s\right) \geqslant  0 \quad\forall s$.\\\\
Considering the iteration $j>1$ 
and let $\left\{s_i^{\left(j\right)}\right\}_{i=1}^{n}$ be the $j$-th sequence of local volatility estimates associated to the record time of each random variable in the set $\left\{\Delta_1 Y,\dots,\Delta_n Y\right\}$, then we consider the $j$-th set of renormalized random variables 
\begin{equation}\label{PAeq160}
\left\{\Delta_i \bar{Y}^{\left(j\right)}\right\}_{i=1}^{n}:=\left\{\Delta_i \bar{Y}^{\left(j\right)} : \Delta_i \bar{Y}^{\left(j\right)}=\frac{\Delta_i Y}{s_i^{\left(j\right)}}\;\text{for}\; i=1,\dots,n\right\}
\end{equation}
Once named $\left\{\Delta_{n:n} \bar{Y}^{\left(j\right)},\dots,\Delta_{1:n} \bar{Y}^{\left(j\right)}\right\}$ the ordered set of these variables, we introduce the function $f^{\left(j\right)}$ that maps the index $i$ of each random variable in the ordered set $\left\{\Delta_{\left(n-i+1\right):n}\bar{Y}^{\left(j\right)}\right\}_{i=1}^{n}$ to the index $\tilde{i}=f^{\left(j\right)}\left(i\right)$ of the corresponding random variable in the set $\left\{\Delta_i \bar{Y}^{\left(j\right)}\right\}_{i=1}^{n}$. \\
The kernel OS estimator, for the $j$-th iteration,  is defined as
 \begin{equation}\label{PAeq35}
\hat{\sigma}^{2\quad\left(j\right)}_{OS}\left(t\right)=\frac{\splitfrac{\sum_{i=1}^m K_{h, \tilde{i}}^t\left(\Delta_{\left(n-i+1\right):n} \bar{Y}^{\left(j\right)}\right)^2 \mathrm{max} \left\{\mathbbm{1}_{\left\{\Delta_{\left(n-i+1\right):n} \bar{Y}^{\left(j\right)}\leqslant \bar{\theta }\left(i,\bar{\mathcal{G}}^{\left(j\right)}_{i-1},\bar{\mathcal{G}}^{'\left(j\right)}_{i-1}\right) \right\}},1-\bar{X}^{\left(j-1\right)}_{n-i+1}\right\}}{+\sum_{i=1}^{m'} K_{h,\tilde{i}}^t \left(\Delta_{i:n} \bar{Y}^{\left(j\right)}\right)^2 \mathrm{max} \left\{\mathbbm{1}_{\left\{-\Delta_{i:n} \tilde{Y}^{\left(j\right)}\leqslant \bar{\theta }\left(i,\bar{\mathcal{G}}^{\left(j\right)}_{i},\bar{\mathcal{G}}^{'\left(j\right)}_{i-1}\right) \right\}},1-\bar{X}^{\left(j-1\right)}_{i}\right\}}}{\sum_{i=1}^n K_{h,\tilde{i}}^t \delta t}
\end{equation}
where the sequences $\left\{\bar{\mathcal{G}}^{\left(j\right)}_i \right\}_{i=0}^{m}$ and  $\left\{\bar{\mathcal{G}}^{'\left(j\right)}_i \right\}_{i=0}^{m'}$  are such that
\begin{equation}\label{PAeq36}
\begin{split}
&\bar{\mathcal{G}}^{\left(j\right)}_0 = \o \quad\quad \bar{\mathcal{G}}^{\left(j\right)}_i =\left\{\bar{I}^{\left(j\right)}_1,\dots,\bar{I}_i^{\left(j\right)}\right\}\\
&\bar{\mathcal{G}}^{'\left(j\right)}_0 = \o \quad\quad \bar{\mathcal{G}}^{'\left(j\right)}_i =\left\{\bar{I}^{'\left(j\right)}_1,\dots,\bar{I}^{'\left(j\right)}_i\right\}
\end{split}
\end{equation}
and $\bar{I}^{\left(j\right)}_i$  and $\bar{I}^{'\left(j\right)}_i$   are  random variables  progressively defined as
\begin{equation}\label{PAeq37}
\begin{split}
&\bar{I}^{\left(j\right)}_i=\left\{\begin{array}{ll}
1& \text{if }\Delta_{\left(n-i+1\right):n} \bar{Y}^{\left(j\right)} > \bar{\theta}\left(i, \bar{\mathcal{G}}^{\left(j\right)}_{i-1}, \bar{\mathcal{G}}^{'\left(j\right)}_{i-1}\right) \quad \text{or}\quad \bar{X}^{\left(j-1\right)}_{n-i+1}=1\\ 
0& \text{otherwise}
\end{array}\right. \quad i=1,\dots,m\\
&\bar{I}^{'\left(j\right)}_i=\left\{\begin{array}{ll}
1& \text{if }-\Delta_{i:n} \bar{Y} > \bar{\theta}\left(i, \bar{\mathcal{G}}^{\left(j\right)}_{i}, \bar{\mathcal{G}}^{'\left(j\right)}_{i-1}\right) \quad\text{or}\quad \bar{X}^{\left(j-1\right)}_{i}=1\\ 
0& \text{otherwise}
\end{array}\right. \quad i=1,\dots,m'
\end{split}
\end{equation}
with $\bar{X}^{\left(j-1\right)}_{k}:=\left\{\begin{array}{ll}
\bar{I}^{\left(j-1\right)}_{n-f^{\left(j-1\right)^{-1}}\left(f^{\left(j\right)}\left(k\right)\right) +1}& \text{if }f^{\left(j-1\right)^{-1}}\left(f^{\left(j\right)}\left(k\right)\right) \in \left\{n,\dots,\left \lceil \frac{n}{2}  \right \rceil\right\}\\
\bar{I}^{'\left(j-1\right)}_{f^{\left(j-1\right)^{-1}}\left(f^{\left(j\right)}\left(k\right)\right)}& \text{if }f^{\left(j-1\right)^{-1}}\left(f^{\left(j\right)}\left(k\right)\right) \in \left\{0,\dots,\left \lfloor \frac{n}{2}  \right \rfloor\right\}
\end{array}\right. $ and $m,m'$ as specified in Subsection \ref{OS}. \\
The threshold function $\bar{\theta }\left(i,\bar{\mathcal{G}}_{a}^{\left(j\right)},\bar{\mathcal{G}}_{b}^{'\left(j\right)}\right)$ writes as
\begin{equation}\label{PAeq34}
\bar{\theta }\left(i,\bar{\mathcal{G}}_{a}^{\left(j\right)},\bar{\mathcal{G}}_{b}^{'\left(j\right)}\right) := \left\{\begin{array}{ll}
\theta\left(p;n,n\right)& \text{if}\; a=0,b=0\\ 
\theta\left(p;n-\sum_{j=1}^{a}\bar{I}^{\left(j\right)}_j,n-\sum_{j=1}^{a}\bar{I}^{\left(j\right)}_j\right)& \text{if}\; a\ne0, b=0\\
\theta\left(p;n-i+1-\sum_{j=1}^{b}\bar{I}^{'\left(j\right)}_j,n-\sum_{j=1}^{a}\bar{I}^{\left(j\right)}_j-\sum_{j=1}^{b}\bar{I}^{'\left(j\right)}_j\right)& \text{if}\; a \neq b \; \text{and}\;  b \neq 0\\
\theta\left(p;n-i+1-\sum_{j=1}^{a}\bar{I}^{\left(j\right)}_j,n-\sum_{j=1}^{a}\bar{I}^{\left(j\right)}_j-\sum_{j=1}^{b}\bar{I}^{'\left(j\right)}_j\right)& \text{otherwise}
\end{array}\right. 
\end{equation}
and the kernel weight  $K_{h,\tilde{i}}^t:=K_h\left(\tilde{i}\delta t-t \right)$ is the kernel value associated with the order statistic $\Delta_{i:n} \bar{Y}$.\\\\
Note that $f^{\left(j-1\right)^{-1}}\left(f^{\left(j\right)}\left(i\right)\right)$ stands for the index of the ordered random variable of $\left(j-1\right)$-th iteration  corresponding to the  ordered random variable with index $i$ of the $j$-th iteration. The variables $\bar{X}^{\left(j-1\right)}_{k}\;k=1,\dots,n$ allow to take into account the observations that have already been detected as jumps in the previous iterations of the algorithm  in order to exclude them from the volatility estimation of the current iteration.\\ Furthermore, the inclusion of kernel weights in the definition of this estimator allows to give different importance to the observations in the sample in accordance with their temporal proximity to the time instant at which the local volatility needs to be estimated. 
\\\\
In addition, once ${\hat{\sigma}^{2\quad\left(j\right)}}_{OS}\left(t\right)$ has been computed for all $t\in \left\{0,\delta t, \dots, n\delta t=T\right\}$, the set $\left\{s_i^{\left(j-1\right)}\right\}_{i=1}^{n}$ can be updated with these new estimates and the estimator can be applied again to the new renormalized random variables.\\\\
As starting volatility estimates, the sample variance of the whole set $\left\{\Delta_1 Y,\dots,\Delta_n Y\right\}$ can be  coupled to each random variable $\Delta_i Y$ for $i=1,\dots,n$. \\
It is worth mentioning that, with respect to the first iteration ($j=1$),  the indicators $\bar{I}_i^{\left(1\right)}$ and $\bar{I}_i^{'\left(1\right)}$ are defined as  
\begin{equation}\label{PAeq31}
\begin{split}
&\bar{I}_i^{\left(1\right)}=\left\{\begin{array}{ll}
1& \text{if }\Delta_{\left(n-i+1\right):n} \bar{Y}^{\left(1\right)} > \bar{\theta}\left(i, \bar{\mathcal{G}}_{i-1}^{\left(1\right)},\bar{\mathcal{G}}_{i-1}^{'\left(1\right)}\right)\\ 
0& \text{otherwise}
\end{array}\right. \quad i=1,\dots,m\\
&\bar{I}_i^{'\left(1\right)}=\left\{\begin{array}{ll}
1& \text{if }-\Delta_{i:n} \bar{Y}^{\left(1\right)} > \bar{\theta}\left(i, \bar{\mathcal{G}}_{i}^{\left(1\right)},\bar{\mathcal{G}}_{i-1}^{'\left(1\right)}\right)\\ 
0& \text{otherwise}
\end{array}\right.\quad i=1,\dots,m'
\end{split}
\end{equation}
and $ \bar{X}^{\left(0\right)}_{i}=0 \quad \forall i=1,\dots,n$. 
All the other variables and parameters coincide with the ones previously stated.\\\\
Finally, it is worth noting that, in general, the kernel function is an even function (i.e. $K\left(s\right)=K\left(-s\right)$), nevertheless, in order not to use information time preceding the volatility calculation time,  in our implementation of such estimator we use a  one-sided kernel function, that is $K\left(s\right)\neq0$ only for $ s\leqslant0$. Thus, without loss of generality, we assume that the kernel function can also be non-symmetrical.\\More precisely, the weighting function that we use in our practical implementation of such estimator is $K_h\left(t_i -t\right)=\left\{\begin{array}{ll}
\frac{1}{h}&\text{if}\quad|t_i -t| \leqslant \left(h+1\right) \delta t\quad \text{and} \quad t_i -t<0 \\ 
0& \text{otherwise}
\end{array}\right.$ for each $t_i \in \left\{0,\delta t, \dots, n\delta t=T\right\}$.\\\\
The iterative algorithm to compute the time-varying volatility of the $Y_t$  process is reported in Algorithm \ref{alg2}.\\\\
\begin{algorithm}[t]
 \KwIn{$\left(\Delta_1 y,\dots,\Delta_n y\right)$  whole sample vector, $p$ tolerance level }
 \KwOut{$\left(s_{1},\dots,s_n\right)$ vector of  volatility estimates, $\left(j_1,\dots,j_n\right)$ vector of boolean jump indicators associated to $\left(\Delta_1 y,\dots,\Delta_n y\right)$ }
Set  each component of the  vector $\left(s^2_{1},\dots,s^2_n\right)$ equal to the sample variance estimate of  $\left(\Delta_{1} y,\dots,\Delta_{n} y\right)$\;
Set the vector of boolean jump indicators $\left(j_1,\dots,j_n\right)=\left(0,\dots,0\right)$\;
Compute the normalized vector $\bar{y}:=\left(\Delta_{1} \bar{y},\dots,\Delta_{n} \bar{y}\right)$ with $\Delta_{i} \bar{y}=\frac{\Delta_{i}y}{s_i}$\;
Compute the kernel vector associated with $\bar{y}$\;
Order the vector $\bar{y}$  and  rearrange the kernel  and boolean vectors in the same way\;
\Repeat{the identified jumps before and after the application of the estimator
are the same}{
Update the vector of boolean jump indicators $\left(j_1,\dots,j_n\right)$ according to Eq.~\ref{PAeq37}\;
\For{t=1,\dots,n}{
Apply the kernel OS estimator defined in Eq.~\ref{PAeq35}  to the vector $\bar{y}$ using the rearranged kernel and boolean  vector\;
Update $s_t$ with the square root of estimate produced by the kernel OS estimator\;
} 
Update the normalized vector $\bar{y}$ using the new estimates $\left( s_{1},\dots,s_n\right)$\;
}
 \caption{Base iterative algorithm to compute the time-varying volatility of the $Y_t$ process  by  means of the OS estimator. \label{alg2}}
\end{algorithm}
An on-line version (in Python code) of such algorithm is also available on the website\footnote{\url{https://github.com/sigmaquadro/VolatilityEstimator}}.

\subsection{Further Observations}\label{FO}
By some approximations, we establish a correspondence between the threshold estimator proposed in Eq.~\ref{PAeq3} and the OS estimator defined in Eq.~\ref{PAeq14}. For practical purposes, this match can  be employed to perform a fair comparison between these two estimators.\\\\
In detail, starting from the OS estimator defined in Eq.~\ref{PAeq14} and fixing the threshold level  equal to $\tilde{\theta }\left(i,\mathcal{G}_{a},\mathcal{G}^{'}_{b}\right) \equiv\tilde{\theta }= \sqrt{\hat{\sigma}^2}\theta\left(\bar{p};n,n\right)$ for all $i=1,\dots,n$, we get that each observation in the sample $\left\{\Delta_1 y,\dots,\Delta_n y\right\}$ is compared with the maximum admissible value for the discrete increments of $Y_t^c$, having set the tolerance level $\bar{p}$.\\
This particular implementation of the OS estimator is comparable to the threshold estimator in Eq.~\ref{PAeq3} with threshold function corresponding  to $\theta\left(\delta t\right)=2 \text{log}\frac{1}{\delta t}\hat{\sigma}^2$ with $\delta t$  chosen such that 
\begin{equation}\label{PAeq18}
\begin{aligned}
\bar{p}=\mathbb{P}\left(\underset{i=1,\dots,n}{\text{max}} |W_{\left(i+1\right)\delta t}-W_{i\delta t}| > \sqrt{2\delta t \text{log}\frac{1}{\delta t}}\right)&\approx 2 \mathbb{P}\left(\underset{i=1,\dots,n}{\text{max}} \left(W_{\left(i+1\right)\delta t}-W_{i\delta t}\right) > \sqrt{2\delta t \text{log}\frac{1}{\delta t}}\right)\\
\ArrowBetweenLines
\bar{p}=\mathbb{P}\left(\underset{i=1,\dots,n}{\text{max}} |Z_i| > \sqrt{2 \text{log}\frac{1}{\delta t}}\right)&\approx 2 \mathbb{P}\left(\underset{i=1,\dots,n}{\text{max}} Z_i > \sqrt{2 \text{log}\frac{1}{\delta t}}\right)\\
\ArrowBetweenLines
\theta\left(\frac{\bar{p}}{2};n,n\right)&\approx  \sqrt{2\text{log}\frac{1}{\delta t}}
\end{aligned}
\end{equation}
where $Z_i=\frac{W_{\left(i+1\right)\delta t}-W_{i\delta t}}{\sqrt{\delta t}} \sim N\left(0,1\right) \quad \forall i=1,\dots,n$.
The approximation holds for large $\sqrt{2  \text{log}\frac{1}{\delta t}}$ and its derivation is reported in Appendix \ref{App2}.\\
We use this correspondence between the two estimators in order to be able to produce  their fair performance comparison in the simulation tests. \\\\
Moreover, employing a different c.d.f.~in the order statistic computation involved in the OS estimator represents a further extension of such estimator.\\
Indeed, as previously mentioned, recognizing that all the formulas proposed in Subsection \ref{OS} hold substituting the c.d.f.~of the Standard Normal random variable with a symmetric distribution, the OS estimator can be implemented also assuming that the  increments of the continuous part of $Y_t$  have a distribution different from the Normal one but still symmetric. For instance, assuming that their distribution is the t-Student one.\\
Obviously, this is just an observation and additional comments and details are not included in this paper.
\section{Numerical and Empirical Tests}\label{NETest}
\subsection{Simulation Tests}\label{St}
\begin{figure}[t]
\centering
\begin{subfigure}[b]{0.475\textwidth}
\includegraphics[width=\textwidth]{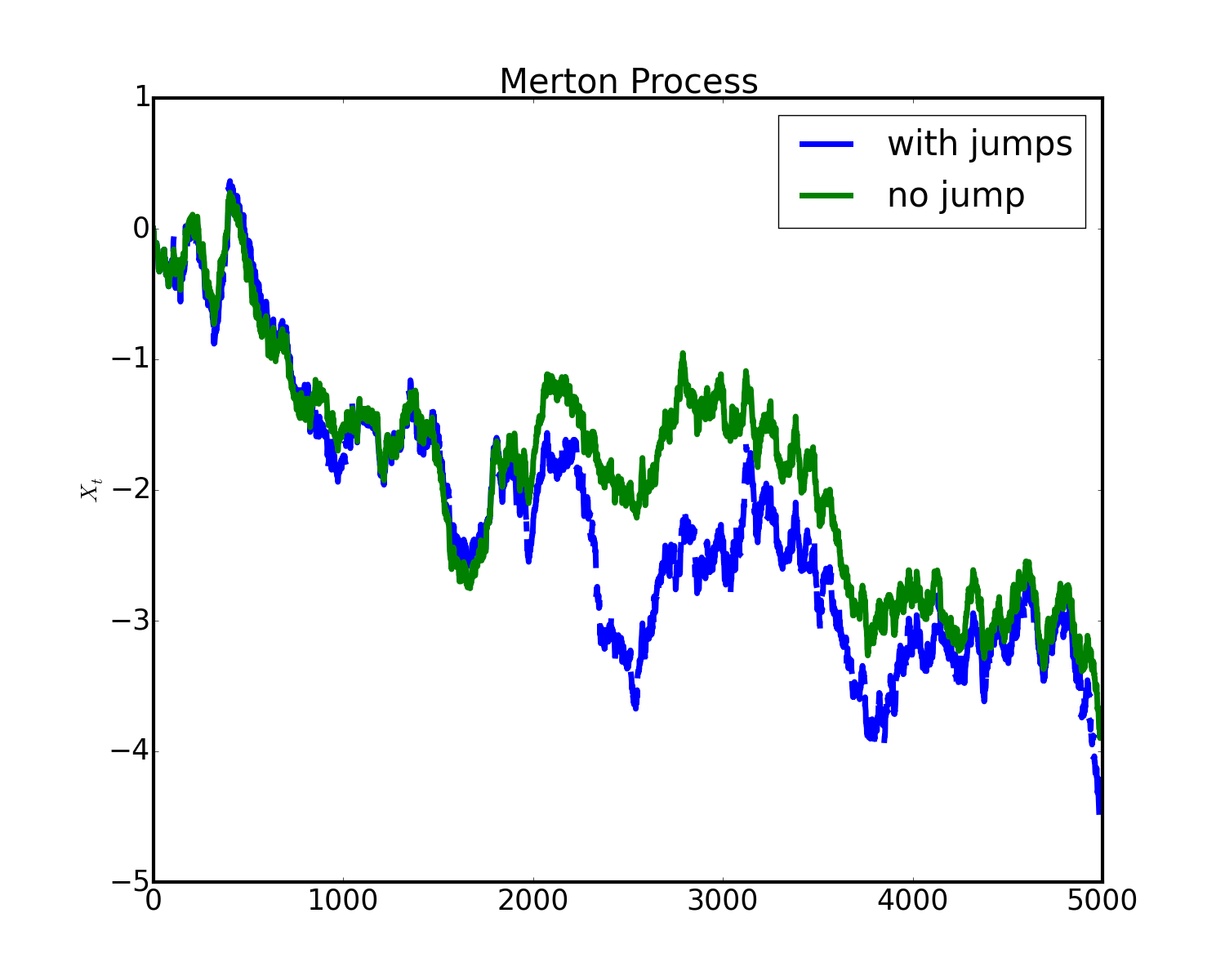}
\caption{}\label{fig:VE1a}
\end{subfigure}
\begin{subfigure}[b]{0.475\textwidth}
\includegraphics[width=\textwidth]{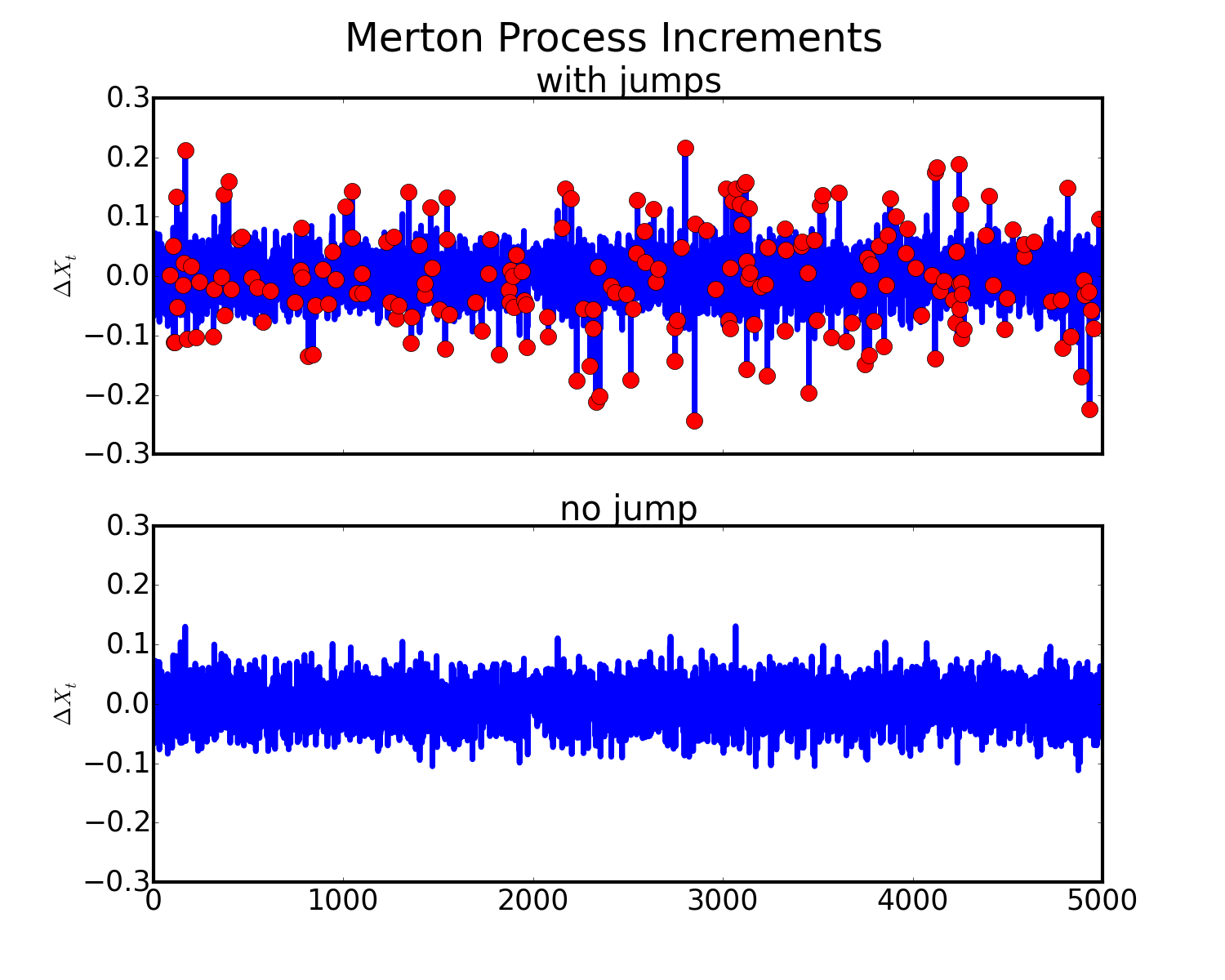}
\caption{}\label{fig:VE1b}
\end{subfigure}
\caption{In the left panel, a Merton process simulated trajectory (blue line) and the corresponding diffusion-only trajectory (green line) are shown, while, in the right panel, the increment time series computed using such simulated trajectories are reported, respectively. The red dots stand for the true jump realizations. Model parameters: $b=0$, $\sigma=0.5$, $\lambda=10$,  $\delta=1.5$ and $\mu=0$. 
Simulation parameters:  number of time-steps for each simulation $N_{TS}=5000$, simulation time horizon $T=20$. \label{figVE1}}
\end{figure}

\begin{figure}[t]
\centering
\begin{subfigure}[b]{0.475\textwidth}
\includegraphics[width=\textwidth]{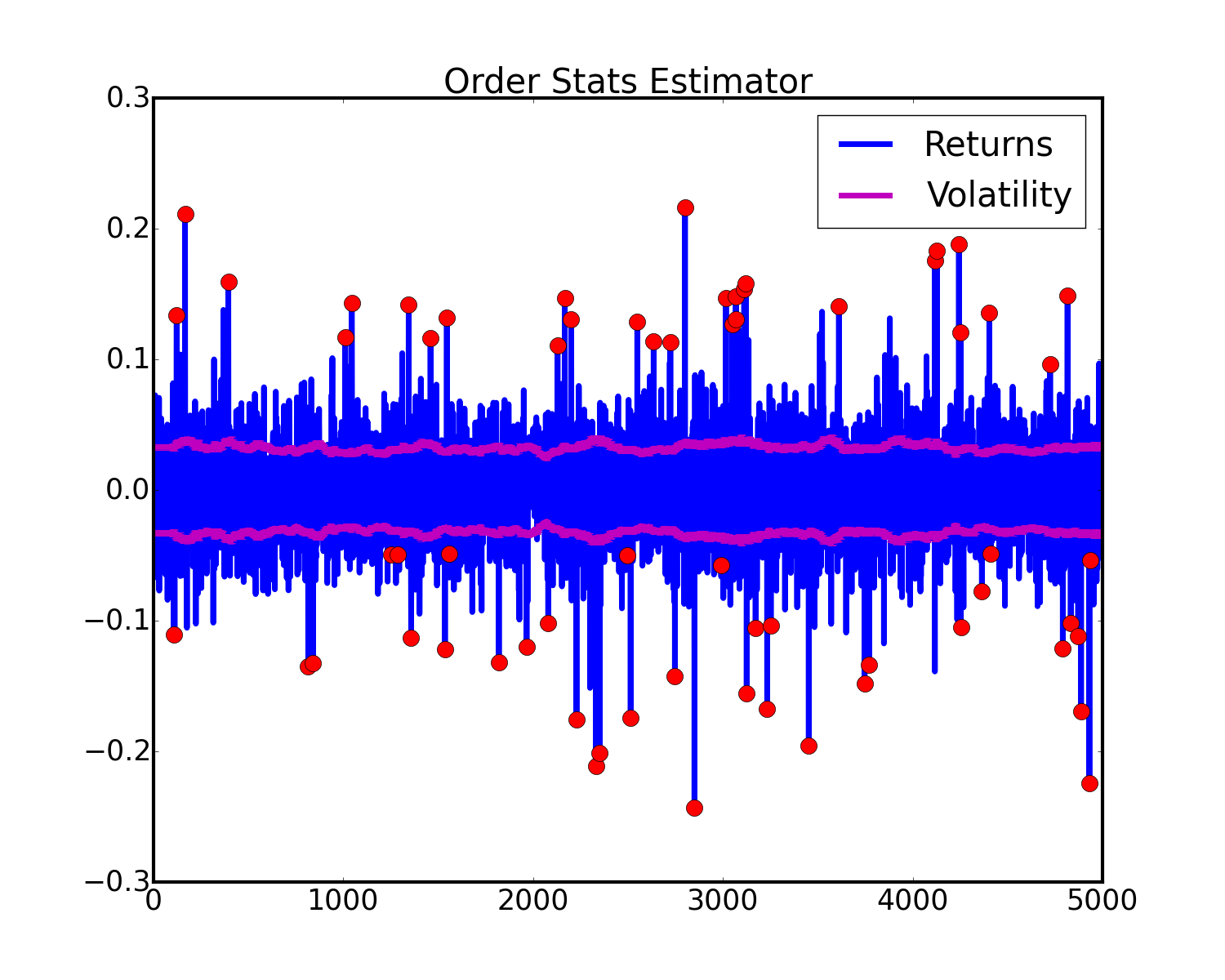}
\caption{}\label{fig:VE2a}
\end{subfigure}
\begin{subfigure}[b]{0.475\textwidth}
\includegraphics[width=\textwidth]{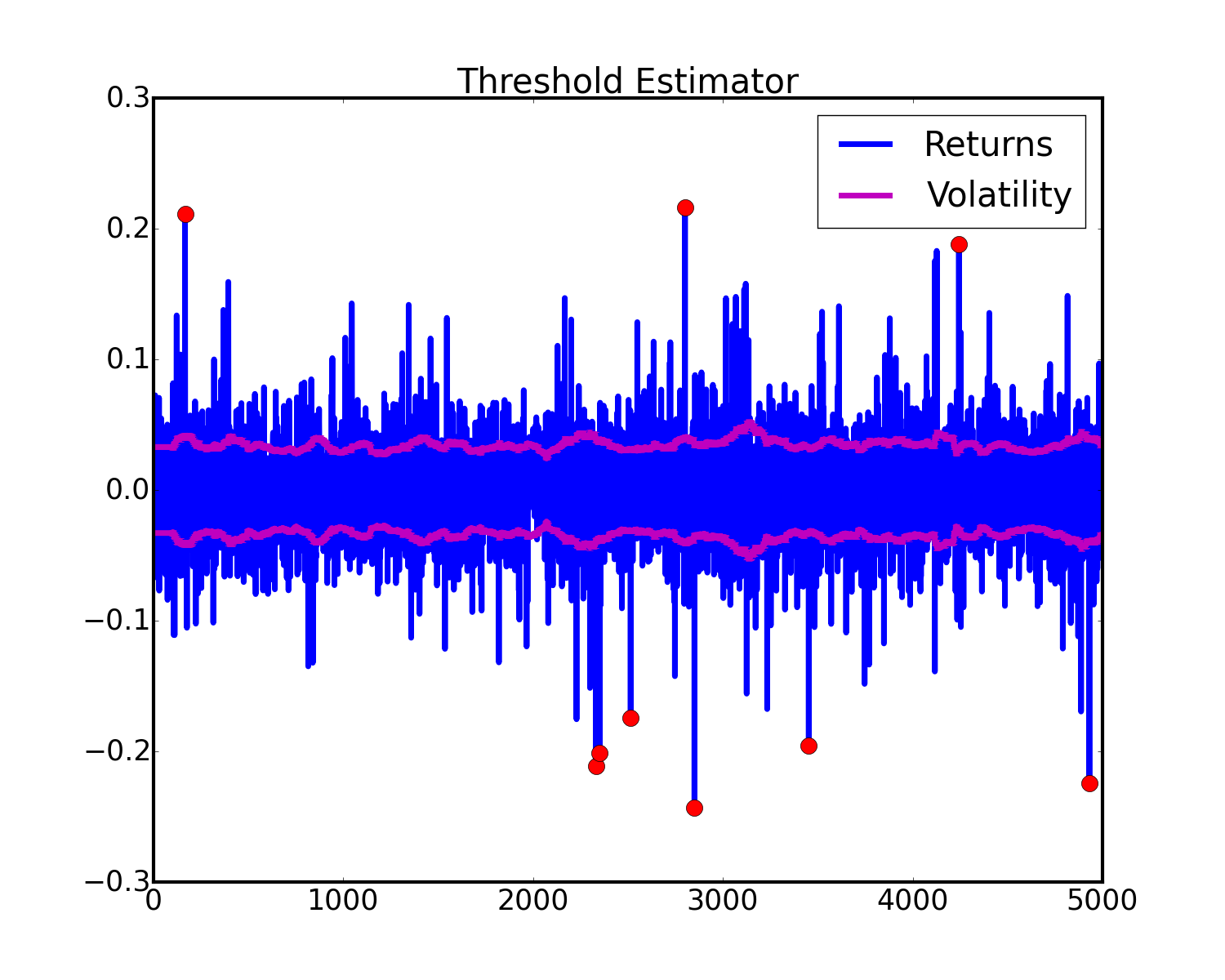}
\caption{}\label{fig:VE2b}
\end{subfigure}
\caption{In the left panel, the detected jumps  and the volatility estimates provided by the OS estimator on the Merton process simulated increment time series shown in Figure \ref{figVE1} are reported, while, in the right panel, the same outcomes provided by the threshold estimator are shown. Model parameters: $b=0$, $\sigma=0.5$, $\lambda=10$,  $\delta=1.5$ and $\mu=0$. 
Simulation parameters:  number of time-steps for each simulation $N_{TS}=5000$, simulation time horizon $T=20$. Estimator parameters: OS estimator tolerance level $p=5\%$, bandwidth $h=100$. \label{figVE2}}
\end{figure}

\begin{figure}[t]
\centering
\begin{subfigure}[b]{0.475\textwidth}
\includegraphics[width=\textwidth]{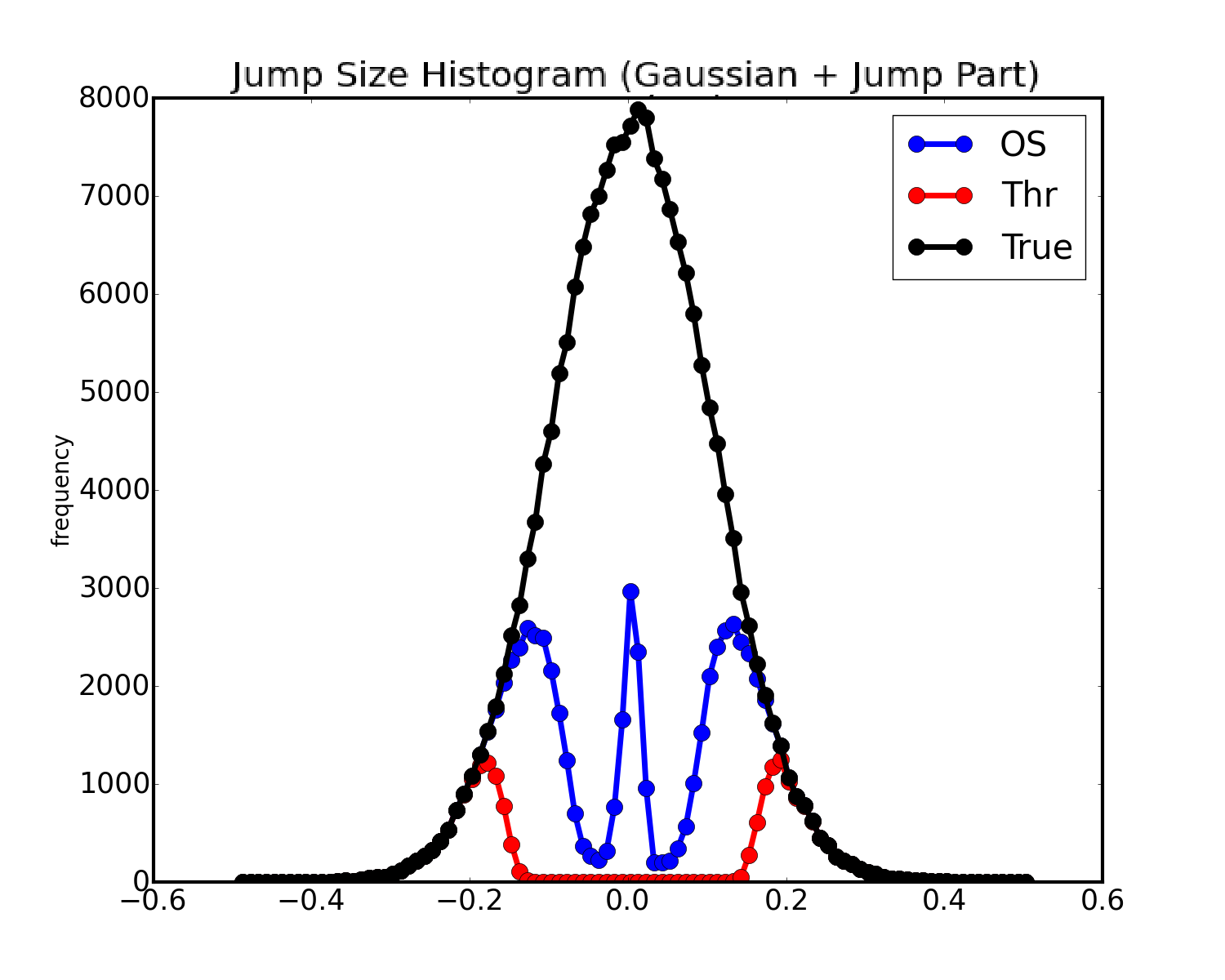}
\caption{}\label{fig:VE10a}
\end{subfigure}
\begin{subfigure}[b]{0.475\textwidth}
\includegraphics[width=\textwidth]{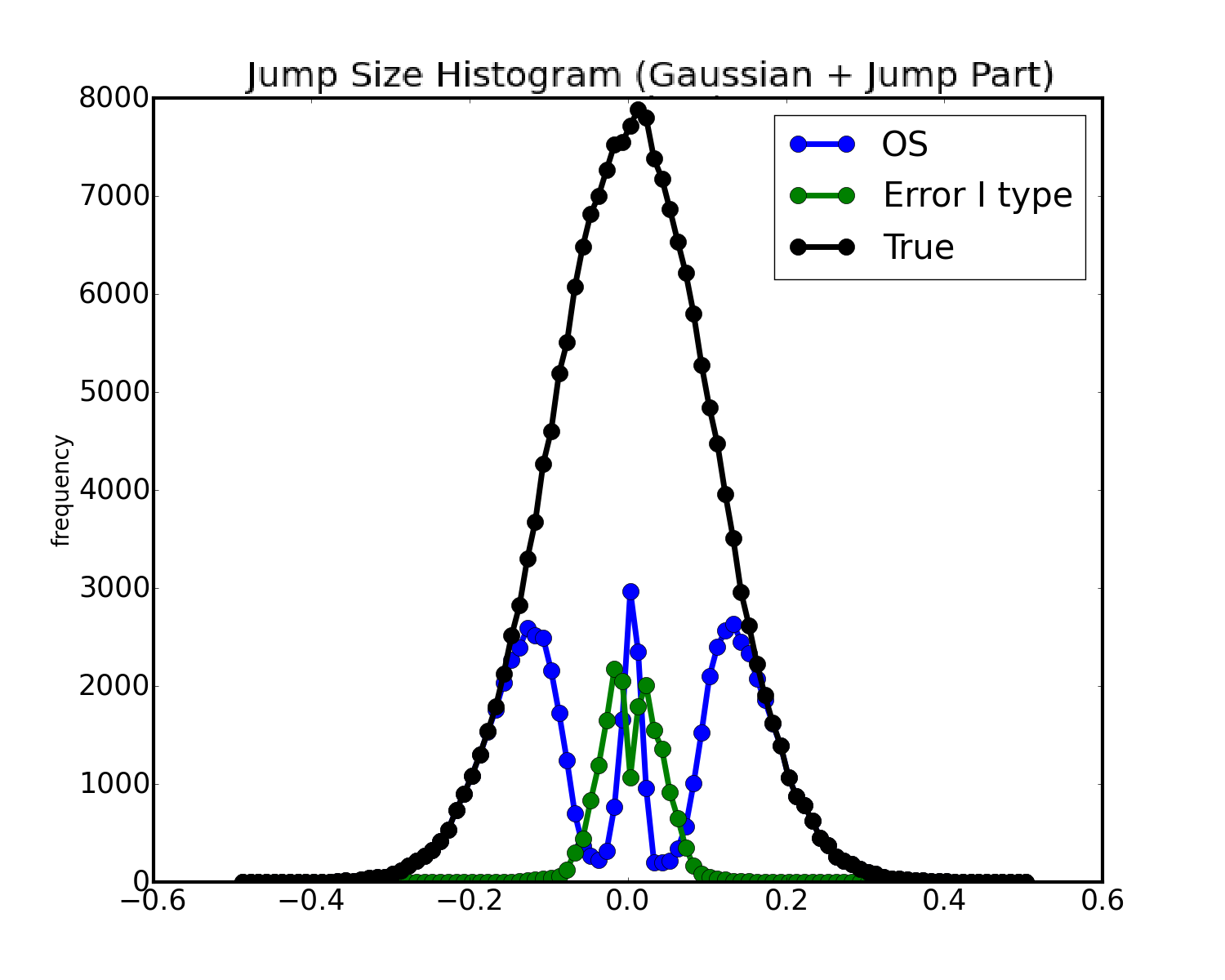}
\caption{}\label{fig:VE10b}
\end{subfigure}
\caption{In the left panel, the cumulative jump size (i.e.~including both the Brownian Motion and the pure jump realizations) histograms created using the  realizations affected by true jumps (black line), the jump realizations identified by the OS estimator (blue line) and the ones identified by the threshold estimator (red line)  are shown, while, in the right panel, the cumulative jump size histograms created using the jump realizations identified through the OS estimator application on the process increment time series cleaned out from the Compounded Poisson contributions is presented (green line), in addiction to the previously described data. Model parameters: $b=0$, $\sigma=0.5$, $\lambda=10$,  $\delta=1.5$ and $\mu=0$. 
Simulation parameters:  number of time-steps for each simulation $N_{TS}=5000$, simulation time horizon $T=20$, number of simulations $M=1000$. Estimator parameters: OS estimator tolerance level $p=5\%$, bandwidth $h=100$. \label{figVE10}}
\end{figure}

\begin{figure}[!h]
\centering
\includegraphics[width=0.6\textwidth]{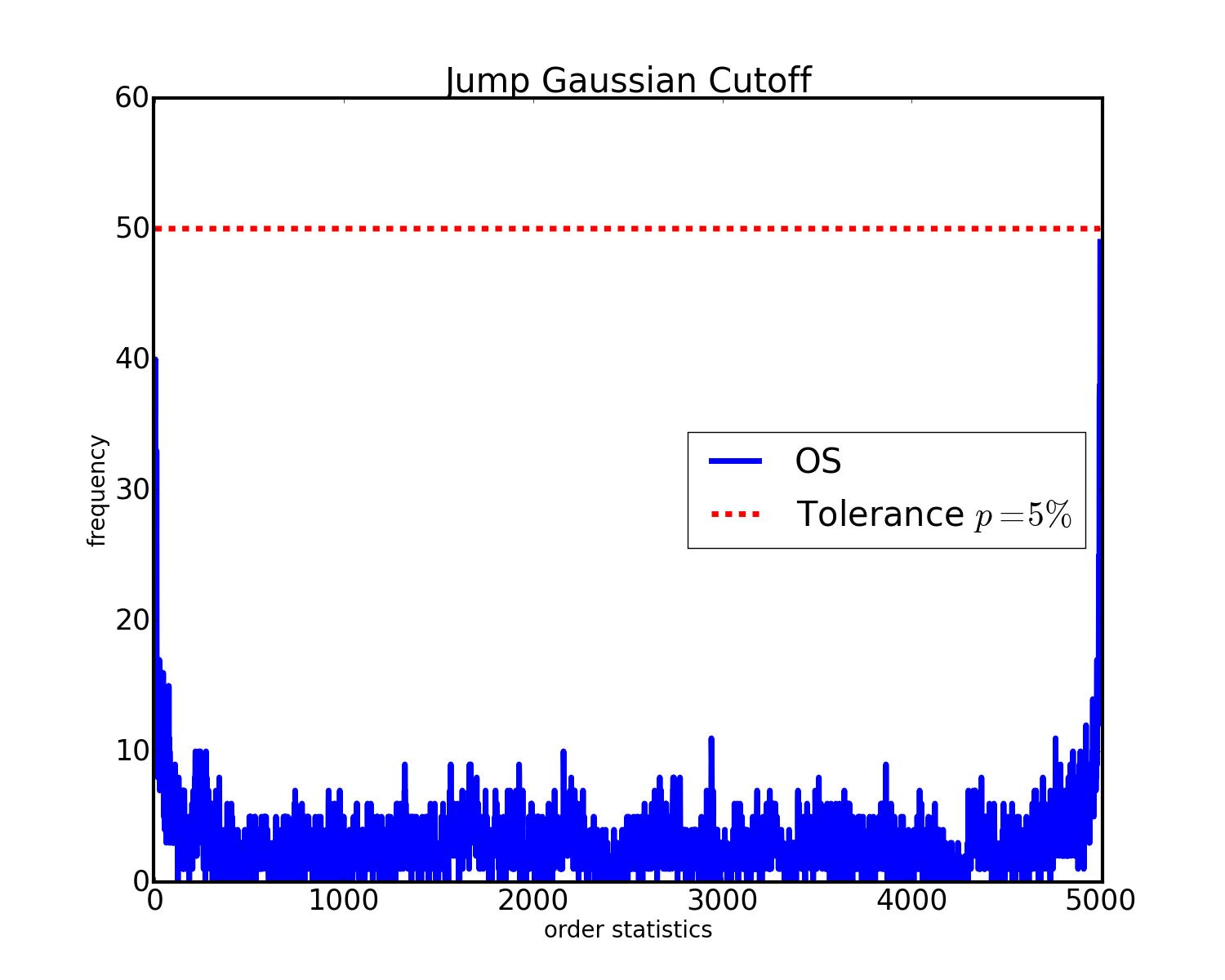}
\caption{Frequency of the Gaussian realizations that are wrongly classified as jumps by the OS estimator (blue line) applied on the Merton process increment time series. The red dashed line stand for the maximum frequency of error of type I that the OS estimator can commit, i.e. $Mp=50$. Model parameters: $b=0$, $\sigma=0.5$, $\lambda=10$,  $\delta=1.5$ and $\mu=0$. Simulation parameters:  number of time-steps for each simulation $N_{TS}=5000$, simulation time horizon $T=20$, number of simulations $M=1000$.  Estimator parameters: OS estimator tolerance level $p=5\%$, bandwidth $h=100$. \label{figVE11}}
\end{figure}

\begin{figure}[!h]
\centering
\includegraphics[width=0.6\textwidth]{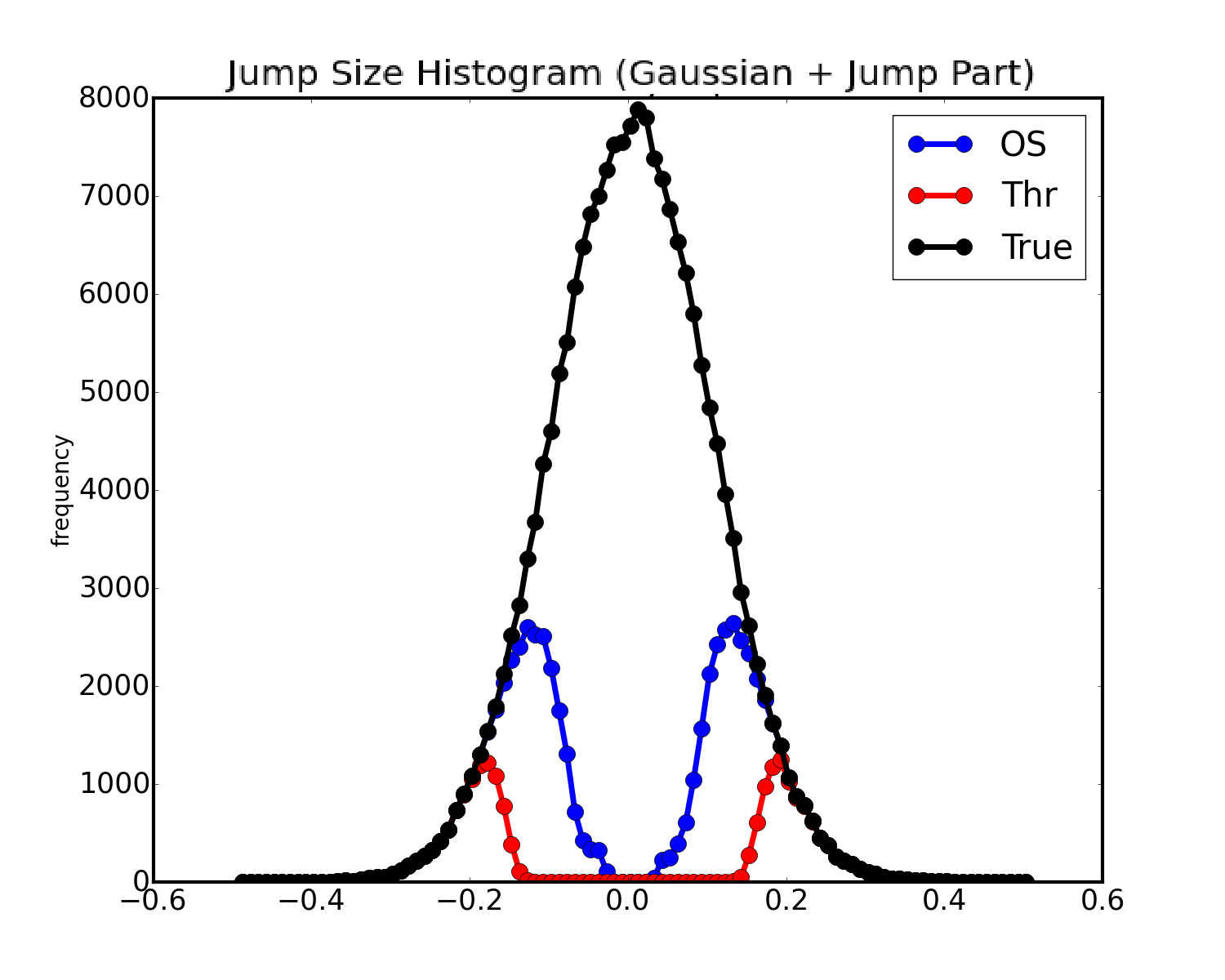}
\caption{Cumulative jump size (i.e.~including both the Brownian Motion and the pure jump realizations) histograms created using the  realizations affected by true jumps (black line), the jump realizations identified by the OS estimator including the error of I type control (blue line) and the ones identified by the threshold estimator (red line). Model parameters: $b=0$, $\sigma=0.5$, $\lambda=10$,  $\delta=1.5$ and $\mu=0$. Simulation parameters:  number of time-steps for each simulation $N_{TS}=5000$, simulation time horizon $T=20$, number of simulations $M=1000$.  Estimator parameters: OS estimator tolerance level $p=5\%$, bandwidth $h=100$. Note that the  control on the error of I type corresponds to reclassifying all the realizations identified as jumps but with size in absolute value smaller than the estimated volatility as ordinary observations. \label{figVE12}}
\end{figure}

\begin{figure}[!h]
\centering
\includegraphics[width=0.6\textwidth]{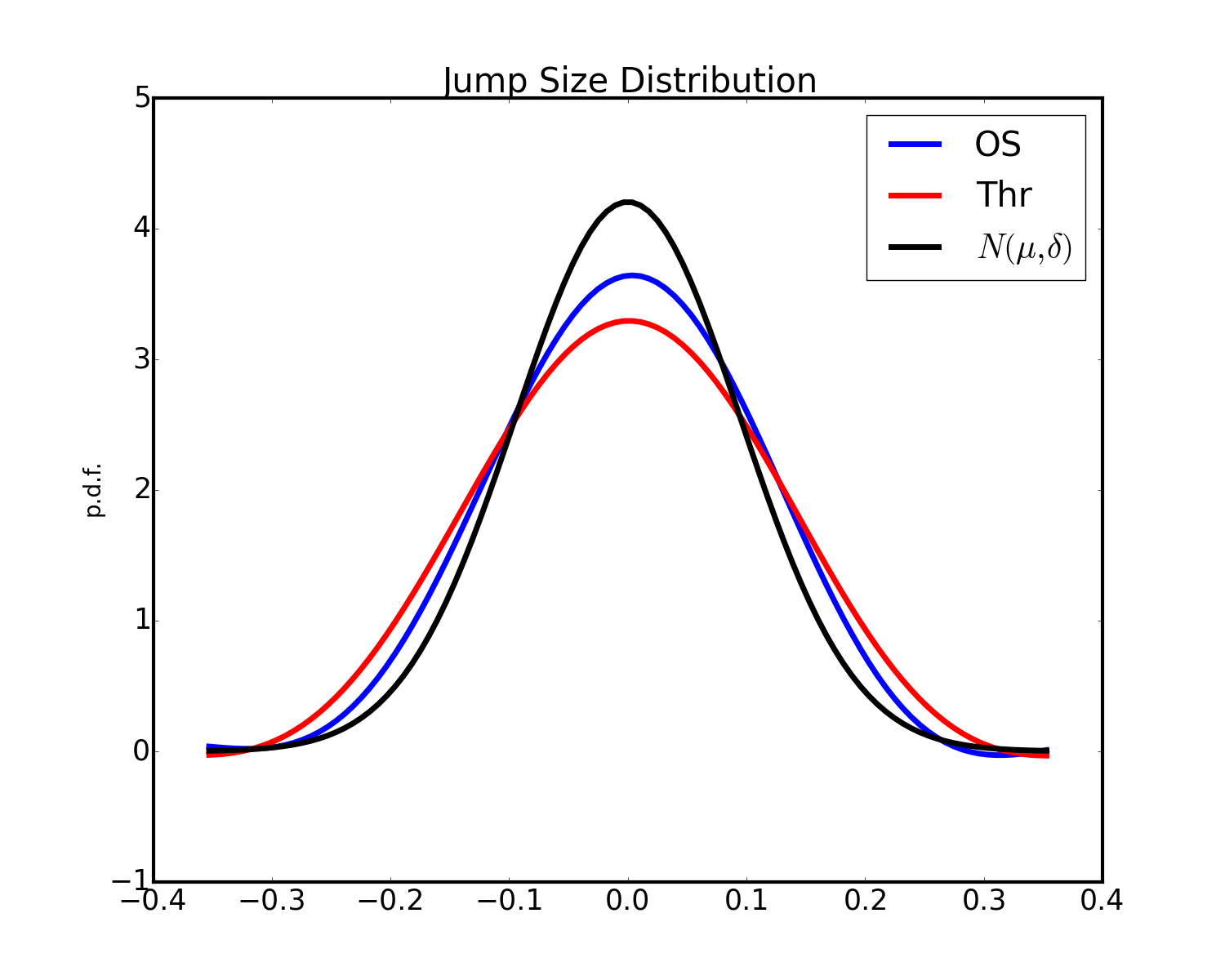}
\caption{True jump size distribution (black line) and jump size distributions isolated from the cumulative ones obtained through  the OS estimator (blue line) and the threshold estimator (red line) application on the simulated increment time series. Model parameters: $b=0$, $\sigma=0.5$, $\lambda=10$,  $\delta=1.5$ and $\mu=0$. Simulation parameters:  number of time-steps for each simulation $N_{TS}=5000$, simulation time horizon $T=20$, number of simulations $M=1000$.  Estimator parameters: OS estimator tolerance level $p=5\%$, bandwidth $h=N_{TS}$. \label{figVE14}}
\end{figure}

\begin{figure}[!h]
\centering
\includegraphics[width=0.6\textwidth]{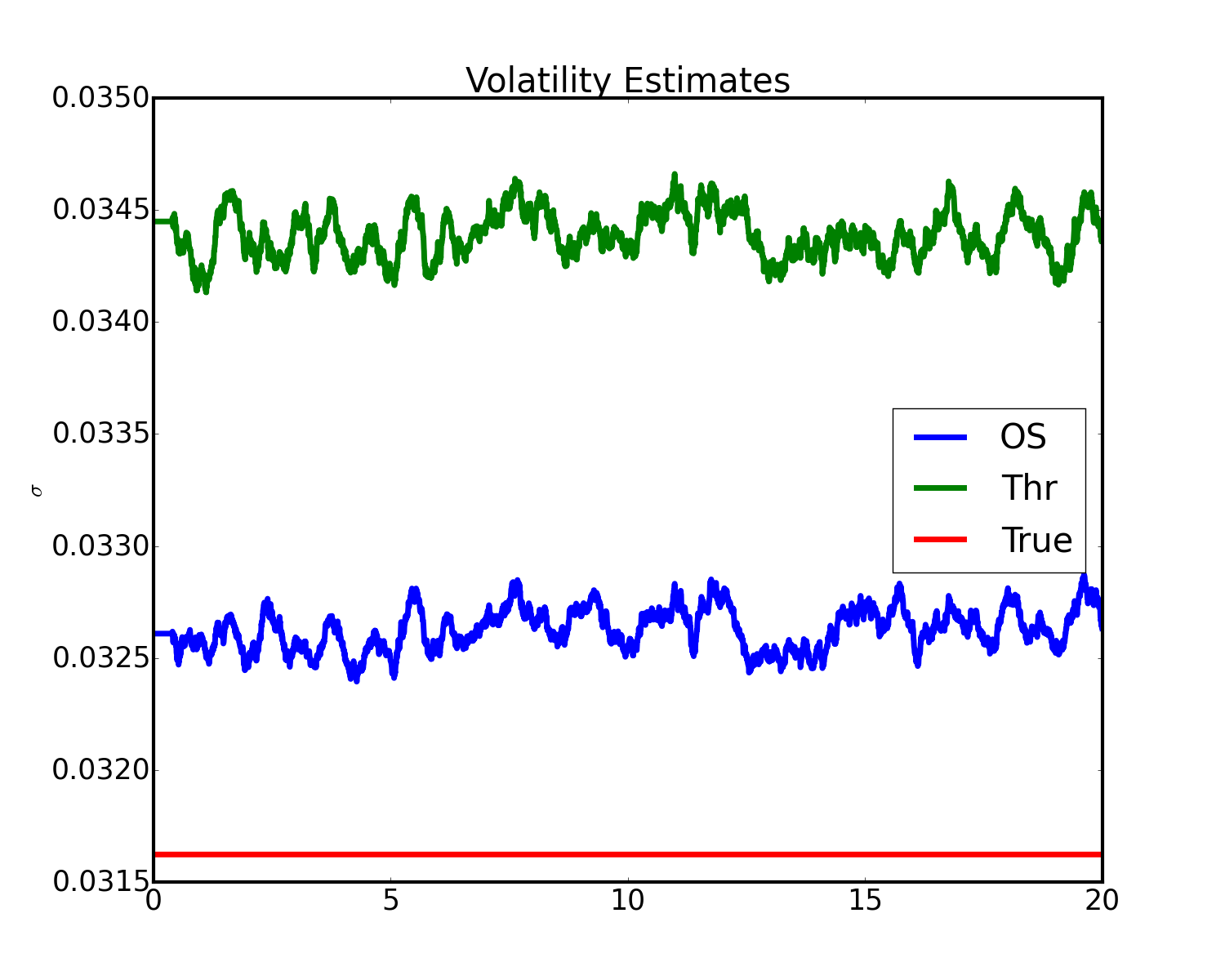}
\caption{Comparison among the average  time-varying  volatility estimates provided by the OS estimator (blue line) and the threshold estimator (green line) application on the simulated Merton process increment time series  and the true values (red line). Model parameters: $b=0$, $\sigma=0.5$, $\lambda=10$,  $\delta=1.5$ and $\mu=0$. Simulation parameters:  number of time-steps for each simulation $N_{TS}=5000$, simulation time horizon $T=20$, number of simulations $M=1000$.  Estimator parameters: OS estimator tolerance level $p=5\%$, bandwidth $h=100$. Note that the true volatility value which must be compared with the estimates  is  $\sigma \sqrt{\Delta t}$ where, in our implementation, $\Delta t =\frac{T}{N_{TS}}=0.004$. \label{figVE13}}
\end{figure}

\begin{figure}[t]
\centering
\begin{subfigure}[b]{0.475\textwidth}
\includegraphics[width=\textwidth]{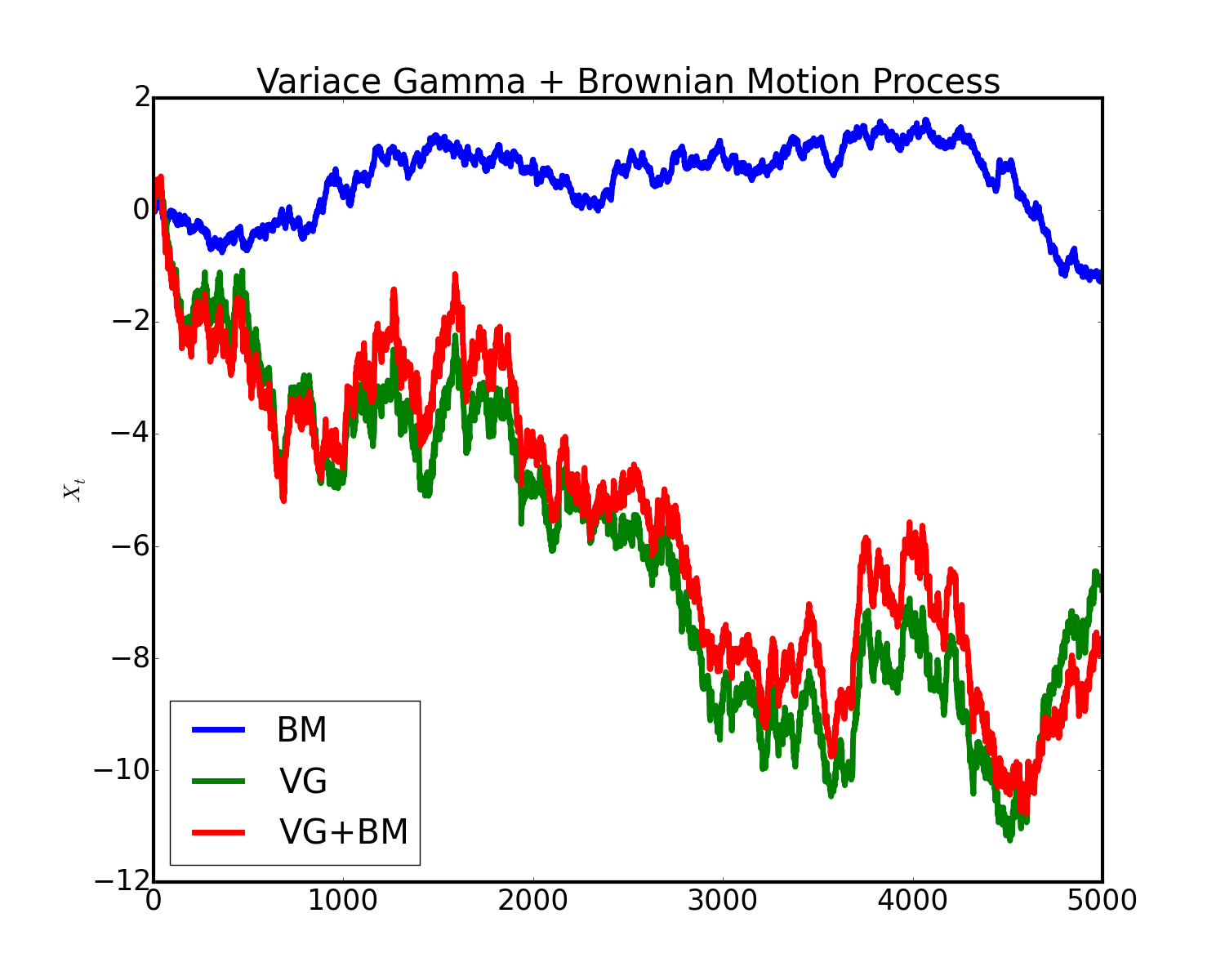}
\caption{}\label{fig:VE4a}
\end{subfigure}
\begin{subfigure}[b]{0.475\textwidth}
\includegraphics[width=\textwidth]{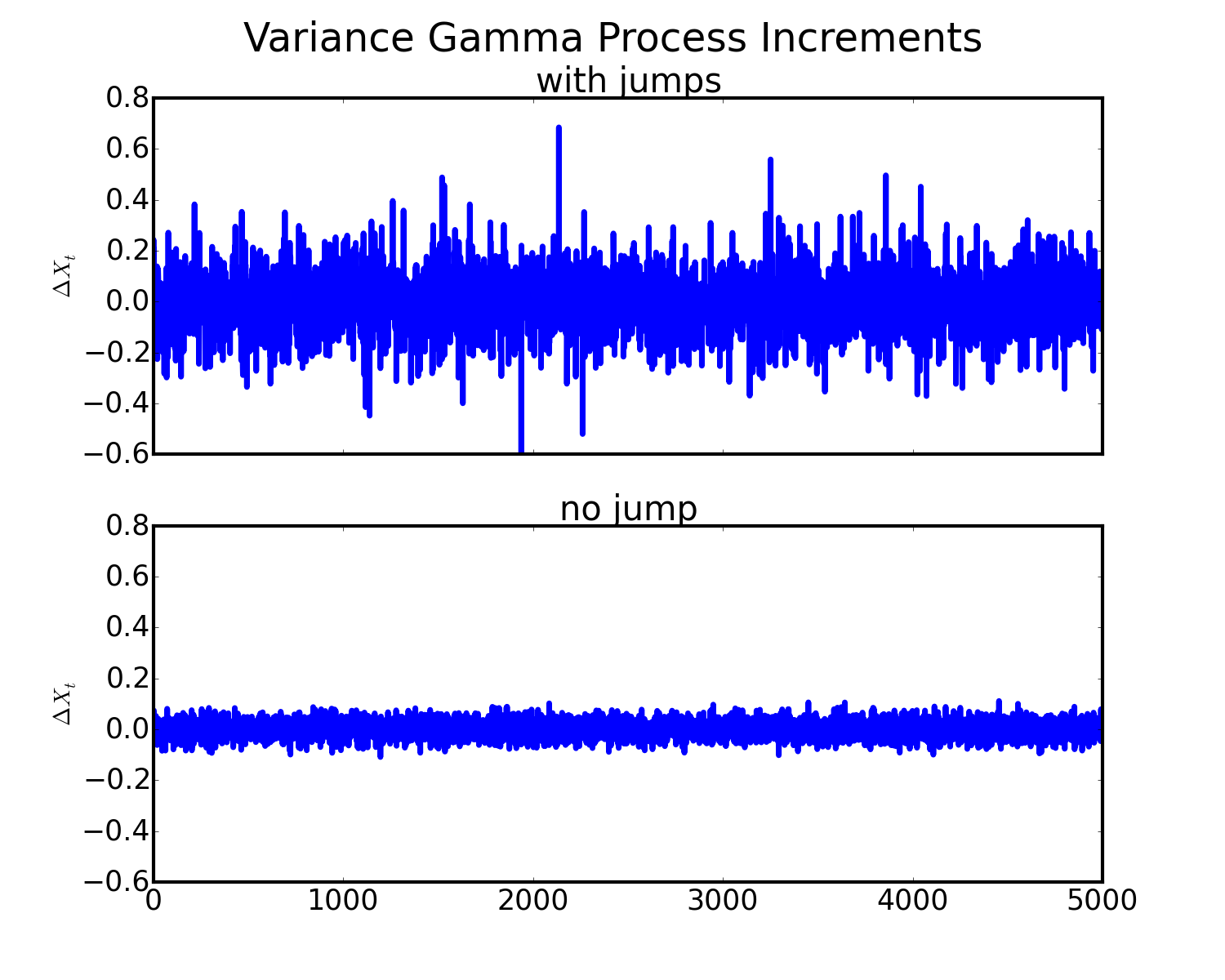}
\caption{}\label{fig:VE4b}
\end{subfigure}
\caption{In the left panel, a Variance-Gamma (VG) process simulated trajectory (green line), a Brownian motion (BM) simulated trajectory (blue line) and the trajectory resulting by their sum (VG+BM) (red line) are shown. In the right panel, the increment time series computed using the VG+BM and the BM  simulated  trajectories are reported, respectively. Model parameters: $b=0$, $\sigma=0.5$, $\alpha=0$,  $\beta=1.5$ and $k=0.004$. 
Simulation parameters:  number of time-steps for each simulation $N_{TS}=5000$, simulation time horizon $T=20$. \label{figVE4}}
\end{figure}

\begin{figure}[!h]
\centering
\begin{subfigure}[b]{0.475\textwidth}
\includegraphics[width=\textwidth]{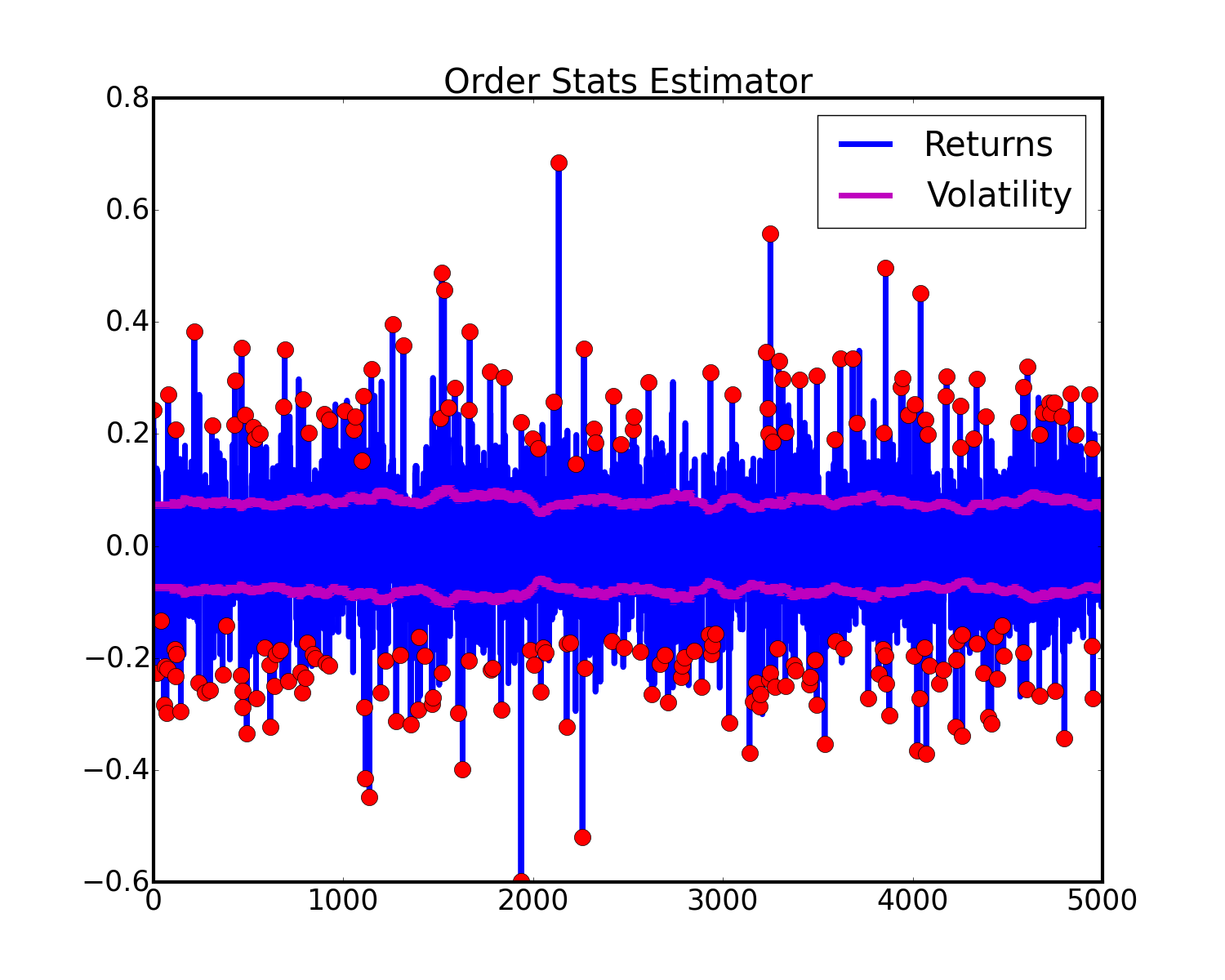}
\caption{}\label{fig:VE5a}
\end{subfigure}
\begin{subfigure}[b]{0.475\textwidth}
\includegraphics[width=\textwidth]{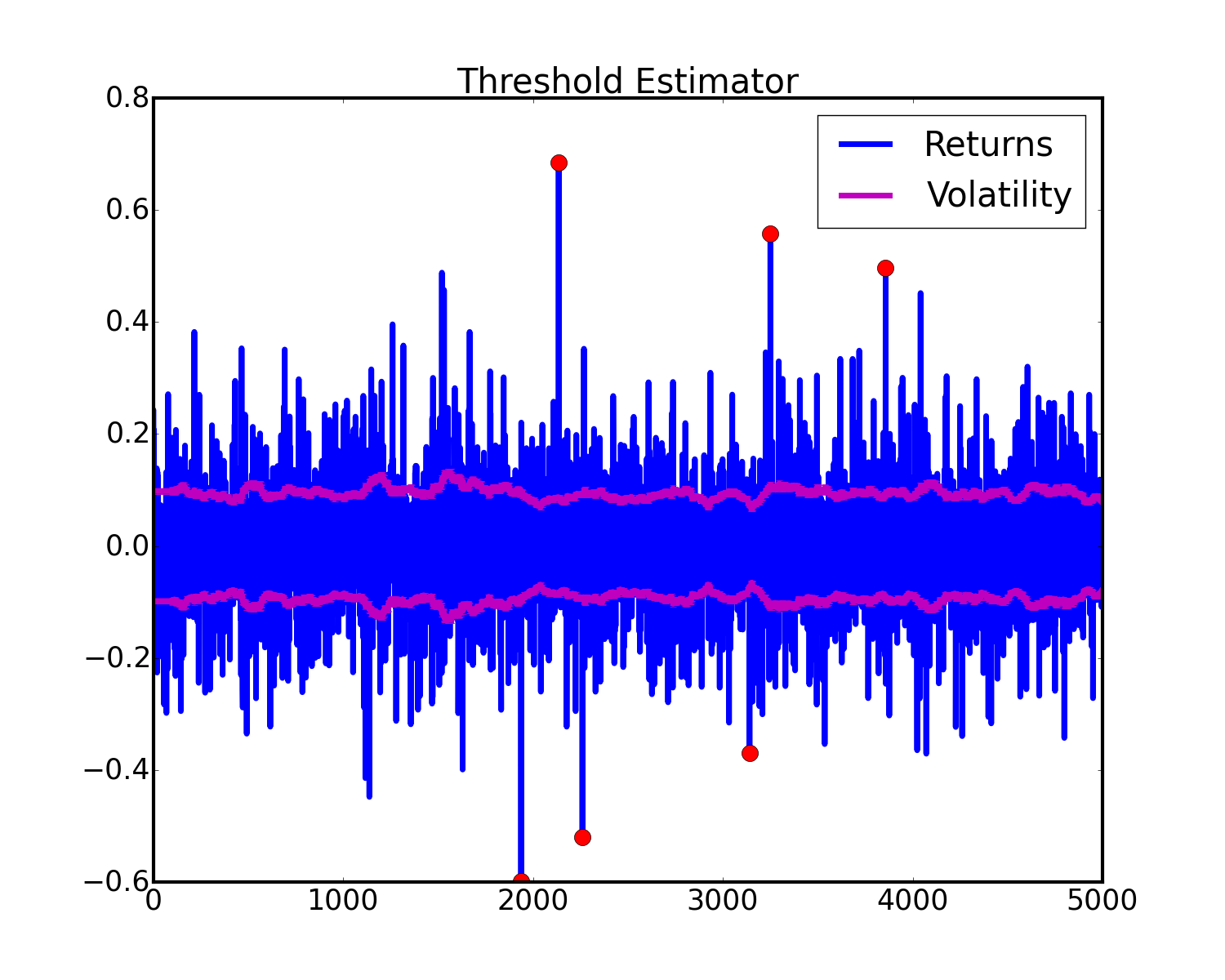}
\caption{}\label{fig:VE5b}
\end{subfigure}
\caption{In the left panel, the detected jumps  and the volatility estimates provided by the OS estimator on the VG+BM process simulated increment time series shown in Figure \ref{figVE4} are reported, while, in the right panel, the same outcomes provided by the threshold estimator are shown. Model parameters: $b=0$, $\sigma=0.5$, $\alpha=0$,  $\beta=1.5$ and $k=0.004$. 
Simulation parameters:  number of time-steps for each simulation $N_{TS}=5000$, simulation time horizon $T=20$. Estimator parameters: OS estimator tolerance level $p=5\%$, bandwidth $h=100$. \label{figVE5}}
\end{figure}

\begin{figure}[!h]
\centering
\begin{subfigure}[b]{0.475\textwidth}
\includegraphics[width=\textwidth]{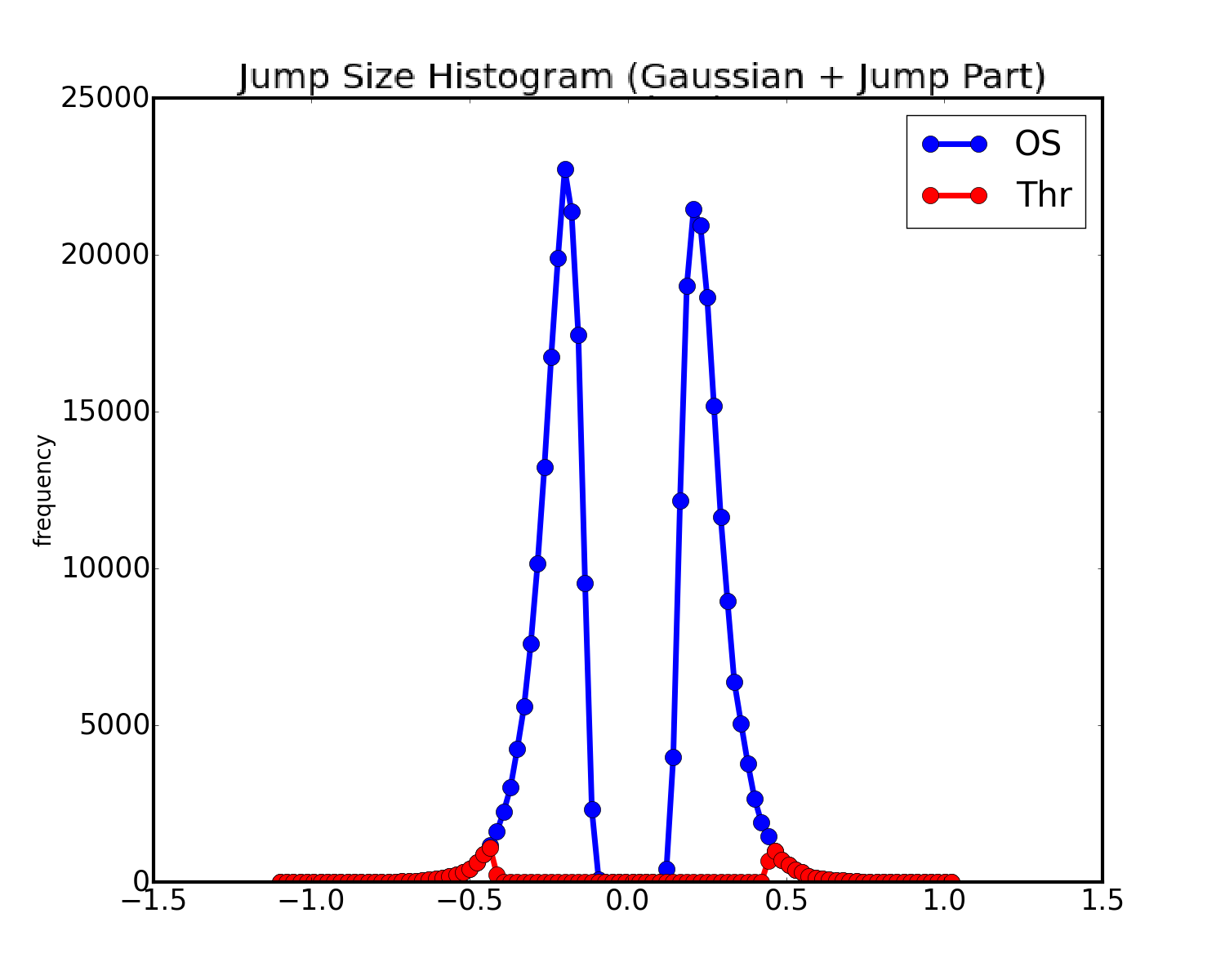}
\caption{}\label{fig:VE15a}
\end{subfigure}
\begin{subfigure}[b]{0.475\textwidth}
\includegraphics[width=\textwidth]{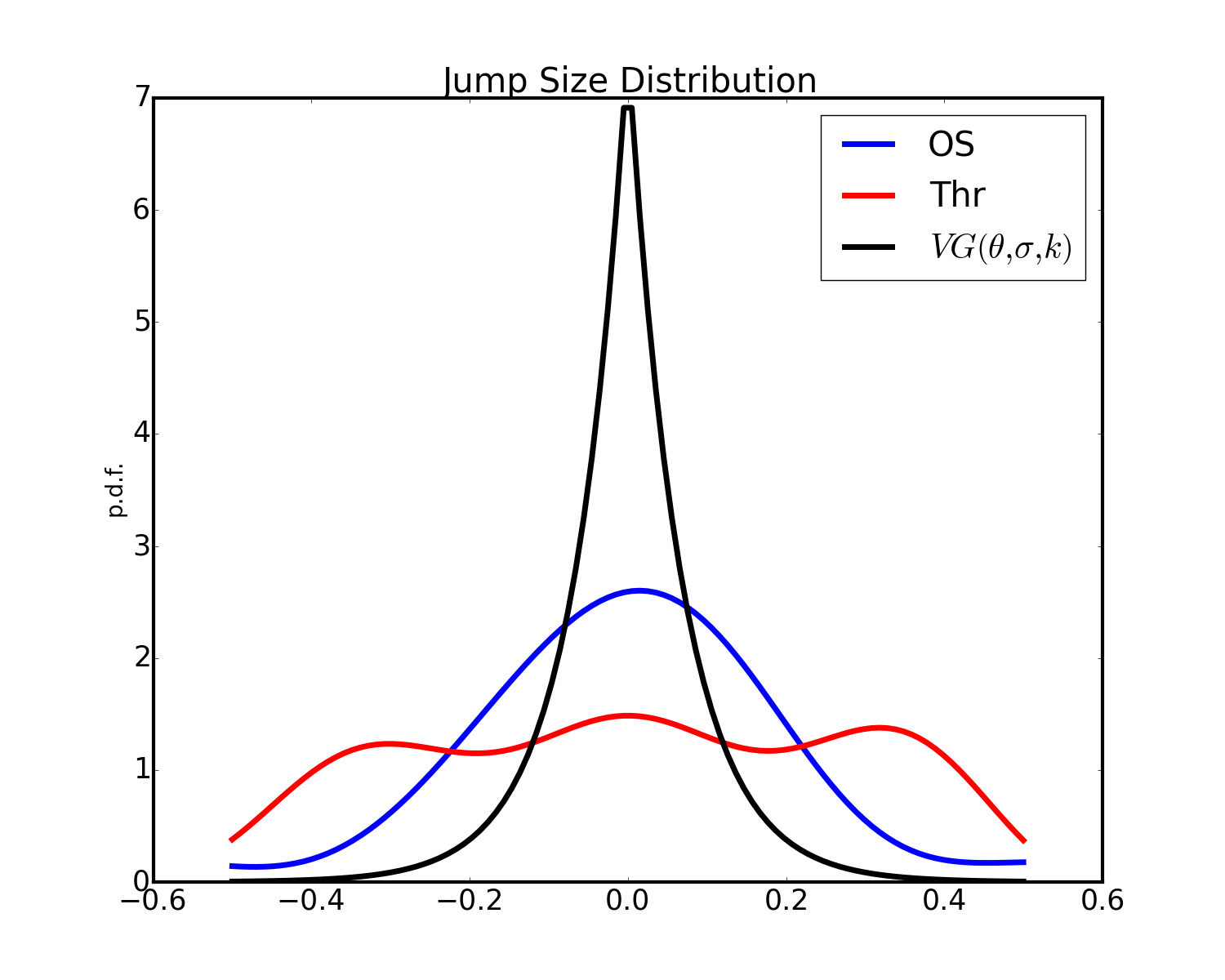}
\caption{}\label{fig:VE15b}
\end{subfigure}
\caption{In the left panel, the detected jumps  and the volatility estimates provided by the OS estimator on the VG+BM process simulated increment time series shown in Figure \ref{figVE4} are reported, while, in the right panel, the same outcomes provided by the threshold estimator are shown. Model parameters: $b=0$, $\sigma=0.5$, $\alpha=0$,  $\beta=1.5$ and $k=0.004$. 
Simulation parameters:  number of time-steps for each simulation $N_{TS}=5000$, simulation time horizon $T=20$, number of simulations $M=1000$. Estimator parameters: OS estimator tolerance level $p=5\%$, bandwidth $h=N_{TS}$. \label{figVE15}}
\end{figure}

\begin{figure}[!h]
\centering
\includegraphics[width=0.6\textwidth]{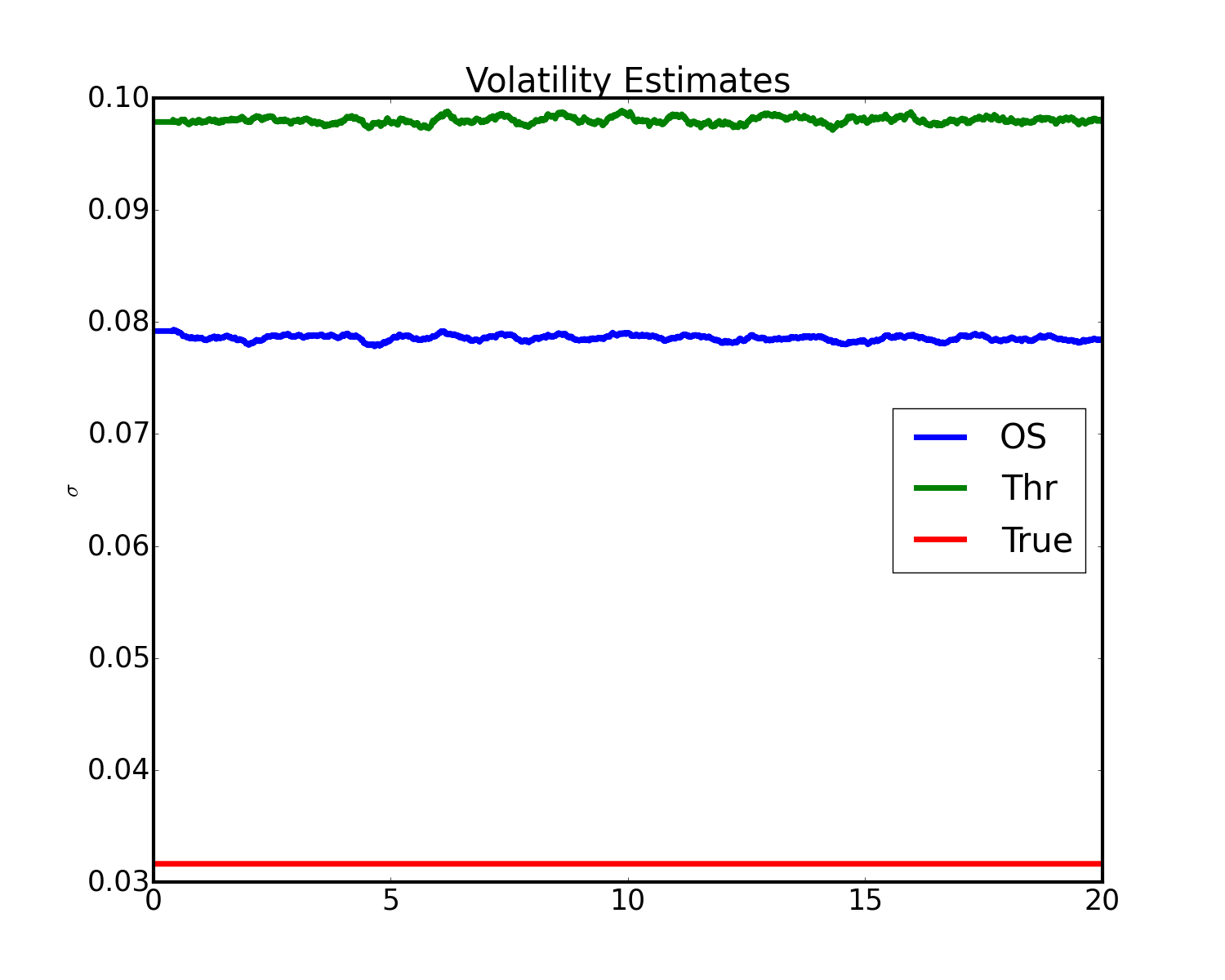}
\caption{Comparison among the average time-varying  volatility estimates provided by the OS estimator (blue line) and the threshold estimator (green line) application on the simulated Merton process increment time series  and the true values (red line). Model parameters: $b=0$, $\sigma=0.5$, $\alpha=0$,  $\beta=1.5$ and $k=0.004$. Simulation parameters:  number of time-steps for each simulation $N_{TS}=5000$, simulation time horizon $T=20$, number of simulations $M=1000$. Estimator parameters: OS estimator tolerance level $p=5\%$, bandwidth $h=100$. Note that the true volatility value which must be compared with the estimates  is  $\sigma \sqrt{\Delta t}$ where, in our implementation, $\Delta t =\frac{T}{N_{TS}}=0.004$. \label{figVE6}}
\end{figure}

We evaluated the performances of the OS estimator by applying it on simulated samples coming from stochastic processes with either finite activity or  infinite activity jump part. \\
More precisely, we simulated two different processes: the Merton process, having a jump part corresponding to a compounded Poisson Process (finite activity), and a process composed by the superposition of a Brownian motion and an independent Variance-Gamma (VG) process, having infinite activity jump part. \\\\
For all tests, we also compared the results concerning the jump detection obtained by the OS estimator with the ones produced by the threshold estimator. As previously explained, in order to make a fair comparison of these two estimators, we  implemented the threshold estimator with threshold function stated in Subsection \ref{FO} and with $\delta t$ calibrated in accordance with Eq.~\ref{PAeq18}.\\ In practice, as far as the OS estimator is concerned,  we used the iterative algorithm proposed in Subsection \ref{IA}, while, with respect to the threshold estimator,  we properly adapted the considered  algorithm to the threshold estimator and we applied it.\\\\ 
We considered the following  Merton model dynamics 
\begin{equation}\label{PAeq19}
\left\{\begin{matrix}\begin{split}
&\mathrm{d}X_t=b\mathrm{d}t + \sigma\mathrm{d}W_t+ \gamma\mathrm{d}N_t \\ 
&X_0=0  
\end{split}\end{matrix}\right.
\end{equation}
where $W_t$ is a standard Brownian motion, $\sigma$ is the volatility of the continuous part, $N_t$ is a Poisson process with constant intensity $\lambda$ and, conditionally on a jump occurring, $\gamma$ is the jump size $\sim N\left(\mu,\delta^2\right)$ with mean $\mu$ and variance $\delta^2$. $W_t$ and $N_t$ are independent. \\
Chosen a parameter set (reported in the caption of the Figure \ref{figVE1} and \ref{figVE2}), we simulated $M=1000$ trajectories of such process and computed the relative increments, $\Delta X_t$, in order to estimate the volatility $\sigma$ for each path.  We  applied both the OS estimator and the threshold estimator to the increment time series and recorded the size of the  realizations detected as jumps. 
As an example, Figures \ref{fig:VE1a} and \ref{fig:VE1b} show a  Merton model trajectory and its increments, respectively. The corresponding jump detections performed by the OS estimator and the threshold estimator are reported in Figures \ref{fig:VE2a} and \ref{fig:VE2b}, respectively. The red dots stand for the realizations identified as jumps by the applied estimator. 
\\Figure \ref{fig:VE10a} shows the convoluted ( i.e.~including both the Brownian Motion and the pure jump realizations) jump size frequency distribution recreated from the simulated data provided by the two estimators considering  all the samples. The peak in the OS  cumulative jump size histogram is due to the error (of  type I) in the jump detection that one is willing to accept fixing the tolerance level $p$. This behaviour is confirmed by the comparison between Figure \ref{fig:VE10a} and Figure \ref{fig:VE10b} where we show the convoluted jump size histogram obtained by applying the OS estimator to the same simulations cleaned-out from the Compounded Poisson process contributions. More precisely, in case of the application of the OS estimator on a rescaled Brownian motion increment time series, the error in the jump detection occurs in the $p\%$ of cases with respect to the maximum (or the minimum, depending on the starting point, i.e.~the maximum or the minimum of the ordered realizations, of the recursive algorithm in proposed in Subsection \ref{IA}). In Figure~\ref{figVE11} we report the observed probability when applying the OS estimator to a Gaussian noise. Contrarily to our expectations, the probability is not stuck and equal to $p$, but it decreases moving into the central region of the order statistics. This effect is essentially due to the fact that we are neglecting the correlation among ordered realizations when we compute Eq.~\ref{PAeq9}.
Despite this decrease of the probability, a peak around 0 is evident in Figure \ref{figVE10} and it is due to the higher probability of observing a realization around zero. As a consequence, even if the probability of committing an error of type I in the jump detection is smaller around zero, the number of misclassified realizations is quite high and this explains the presence of the peak.\\ To control this error, all realizations identified as jumps by the estimator that are smaller, in absolute value, than the estimated volatility were reclassified as ordinary observations. The new convoluted jump size histogram is presented in Figure \ref{figVE12}. 
Such figure  shows the threshold estimator limit in recognizing as non-Brownian realizations many jumps whose size is comparable with the one of the continuous part process increments (hidden jumps), while it shows as the OS estimator, based on order statistics, tries to reduce this bias.
 The area between the red and blue line  represents the gain in the hidden jump identification produced by the OS estimator. 
\\
Moreover, starting from the empirical characteristic function (e.c.f.)~of the cumulative jump size, we isolated the distribution of the pure jump size to provide a deeper comparison between the two estimator results. Figure \ref{figVE14} reports the obtained distributions and confirms the improvement produced by the OS estimator in the jump detection. The technique used to filter and  isolate the jump size distribution from its convolution with the Gaussian distribution  is described in Appendix \ref{App3}.  
\\Finally, mean time-varying volatility estimates produced by the two estimators, obtained by averaging across the $M$ simulations, are reported in Figure \ref{figVE13}. As expected, the OS estimator produces volatility estimates nearer to the real volatility values, since it is capable of  detecting jumps even when their size is comparable to the one of the realizations of the continuous part of the process.
\\\\For what concerns the infinite activity, the process composed by the sum of a Brownian motion and a VG process (VG+BM process) has the following dynamics
\begin{equation}\label{PAeq20}
\left\{\begin{matrix}\begin{split}
&\mathrm{d}X_t= \mathrm{d}X^c_t + \mathrm{d}X^{VG}_t = \underbrace{b\mathrm{d}t+\sigma\mathrm{d}Z_t}_{ \mathrm{d}X^c_t}+\underbrace{\alpha\mathrm{d}V_t + \beta\mathrm{d}W_{V_t}}_{\mathrm{d}X^{VG}_t} \\ 
&X_0=0 
\end{split}\end{matrix}\right.
\end{equation}
where $V_t \sim \Gamma_t \left( \frac{1}{k},\frac{1}{k}\right)$ is the Gamma subordinator process  with variance $kt$ and, thus,  $X^{VG}_t$ is a Variance-Gamma process with parameters $\alpha$ and $\beta$, while  $\sigma$ is the volatility of the continuous part process ($X_t^c$) and $Z_t$ and $W_t$ are two independent Brownian motions.\\
As in the Merton model case, we set the model parameters (as described in the caption of the Figure \ref{figVE4} and \ref{figVE5}), we simulated $M=1000$ trajectories of such process and computed the relative increments $\Delta X_t$ in order to estimate the volatility $\sigma$. To the resulting time series we applied the two estimators obtaining the local volatility estimates. \\As an example, Figures \ref{fig:VE4a} and \ref{fig:VE4b} show a trajectory of this model and its increments, respectively. While, the corresponding jump detections performed by the OS estimator and the threshold estimator are reported in Figures \ref{fig:VE5a} and \ref{fig:VE5b}, respectively. 
\\
Moreover, the histograms of the realizations identified as jumps by the two different estimators across the simulations are reported in Figure \ref{fig:VE15a}, while the corresponding jump size distributions isolated from the cumulative ones are shown in Figure \ref{fig:VE15b}.  
\\The volatility estimates obtained from the two estimators averaging across the $M$ simulations are shown in Figure \ref{figVE6}. As in the finite activity jump case, the OS estimator is able to identify a larger number of non-Brownian realizations and to produce better volatility estimates than the threshold one. Nevertheless, with this kind of process, the error in the volatility estimate is very high. 
\subsection{Real Data Examples}
\begin{figure}[!h]
\centering
\begin{subfigure}[b]{0.6\textwidth}
\includegraphics[width=\textwidth]{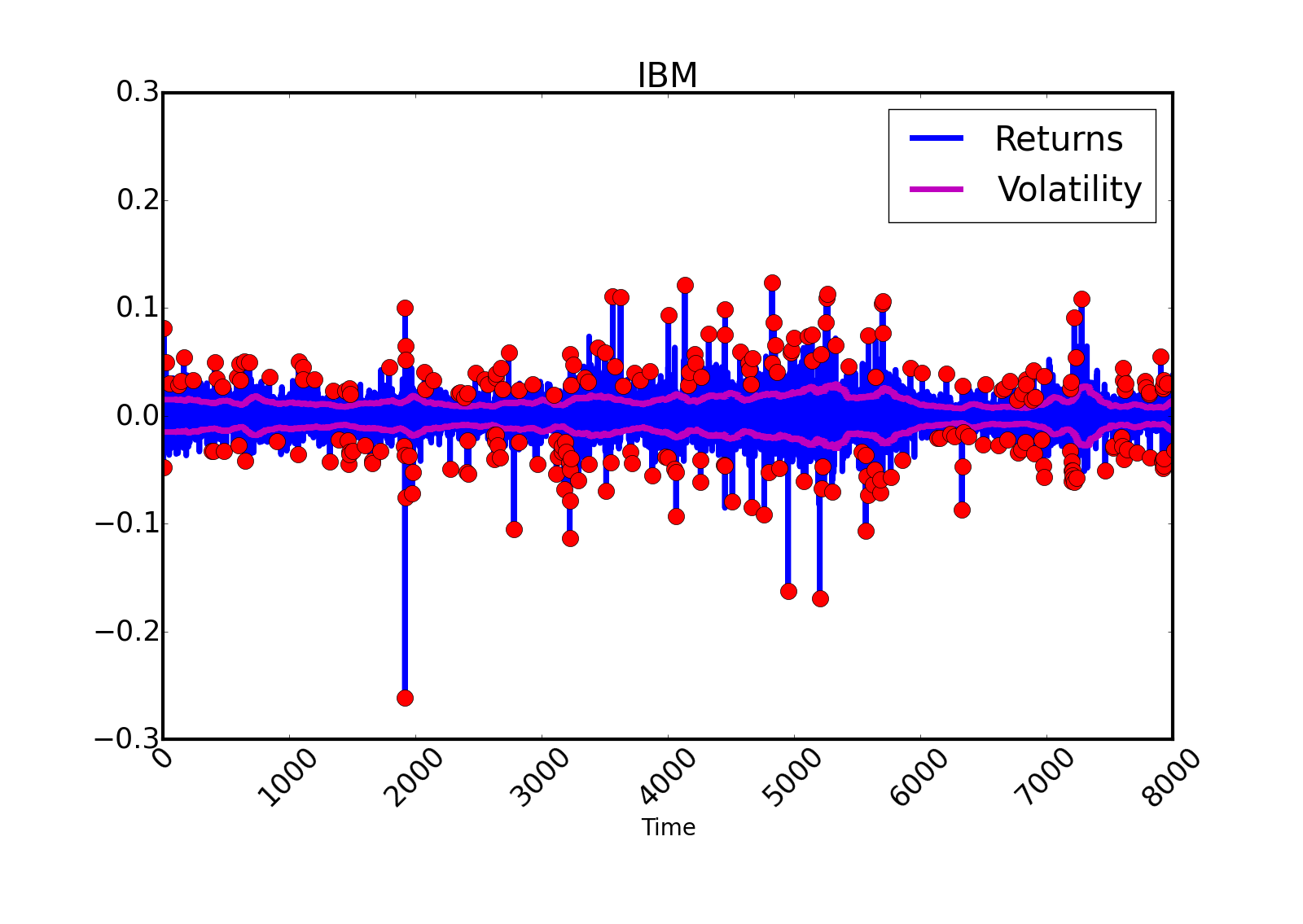}
\caption{}\label{fig:VE20a}
\end{subfigure}
\begin{subfigure}[b]{0.475\textwidth}
\includegraphics[width=\textwidth]{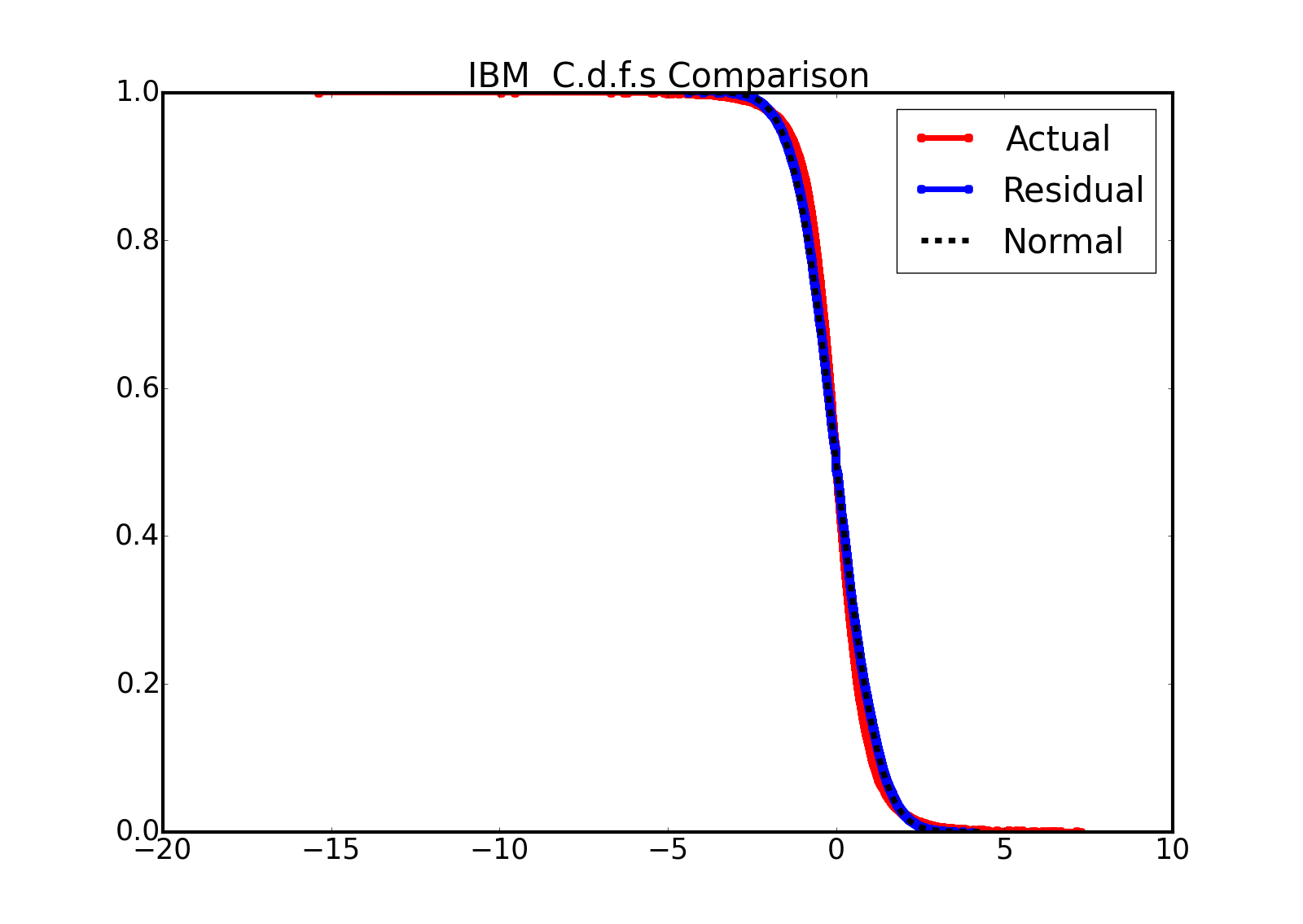}
\caption{}\label{fig:VE20b}
\end{subfigure}
\begin{subfigure}[b]{0.475\textwidth}
\includegraphics[width=\textwidth]{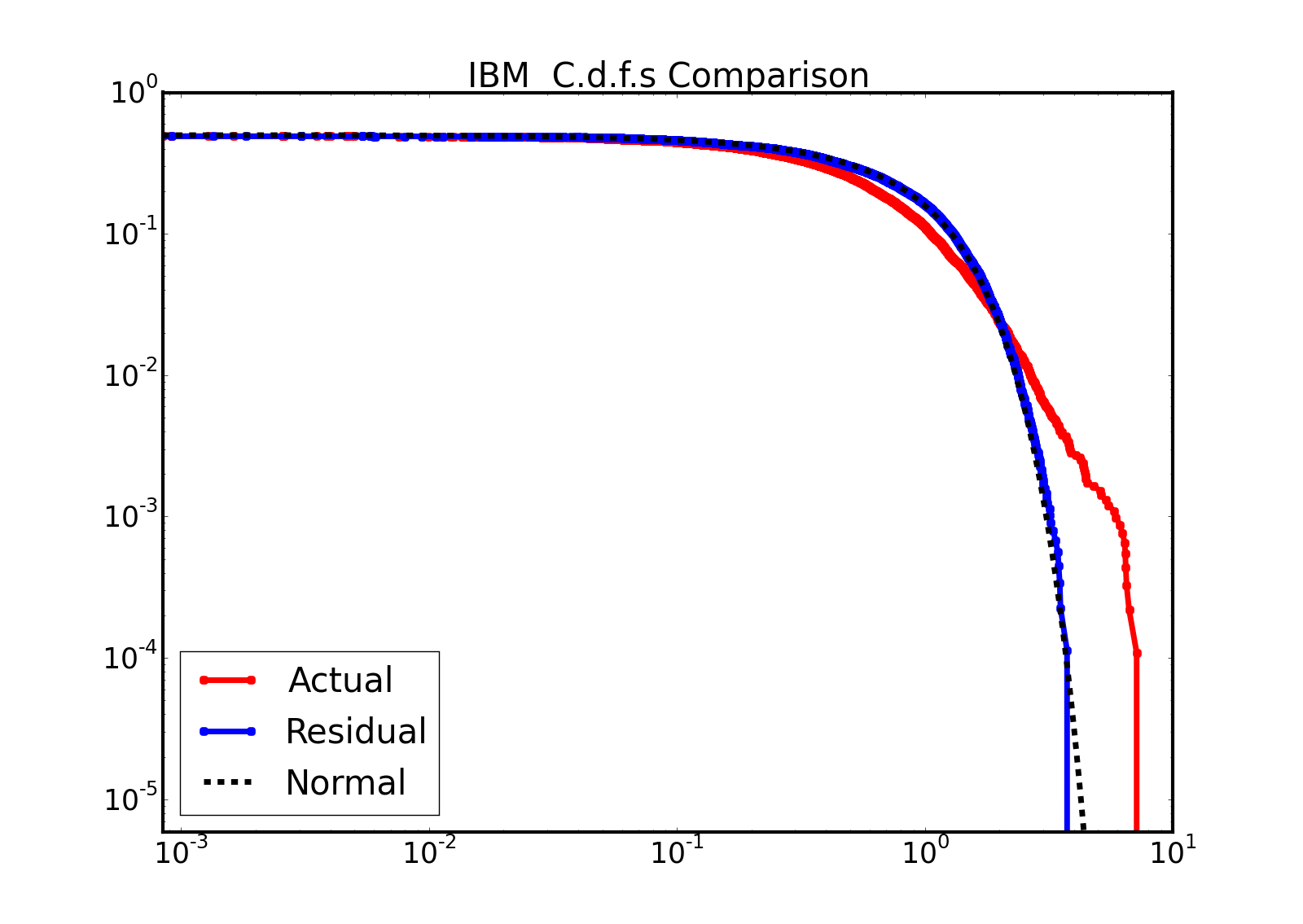}
\caption{}\label{fig:VE20c}
\end{subfigure}
\caption{In the first panel (\ref{fig:VE20a}), the IBM stock daily log-return time series and the volatility estimate time series obtained by applying the OS estimator  are reported. The red dots stand for the realizations identified as jumps by the OS estimator. While, in the other panels (\ref{fig:VE20b} and \ref{fig:VE20c}), the empirical c.d.f.~of the whole renormalized sample (red line), the empirical c.d.f.~of the renormalized observations identified as non-jumps (blue line) and the c.d.f.~of a Standard Normal random variable (black dashed line) are shown, in linear scale and logarithmic scale respectively. 
Time series  parameters:  number of observations $N=9180$ (from 17 March 1980 to 9 August 2016), daily recording frequency. Estimator parameters: OS estimator tolerance level $p=5\%$, bandwidth $h=100$. \label{figVE20}}
\end{figure}

\begin{figure}[!h]
\centering
\begin{subfigure}[b]{0.6\textwidth}
\includegraphics[width=\textwidth]{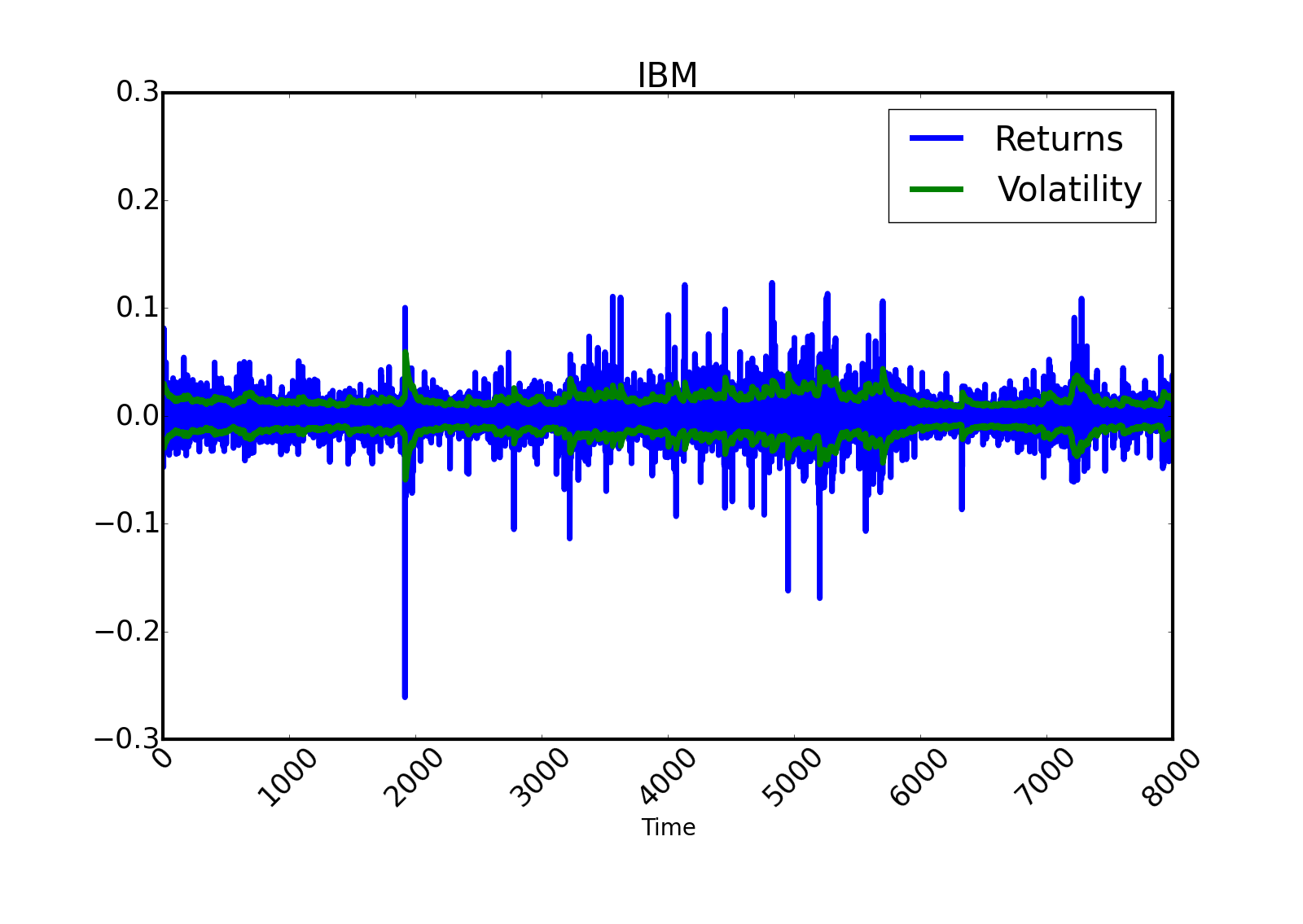}
\caption{}\label{fig:VE21a}
\end{subfigure}
\begin{subfigure}[b]{0.475\textwidth}
\includegraphics[width=\textwidth]{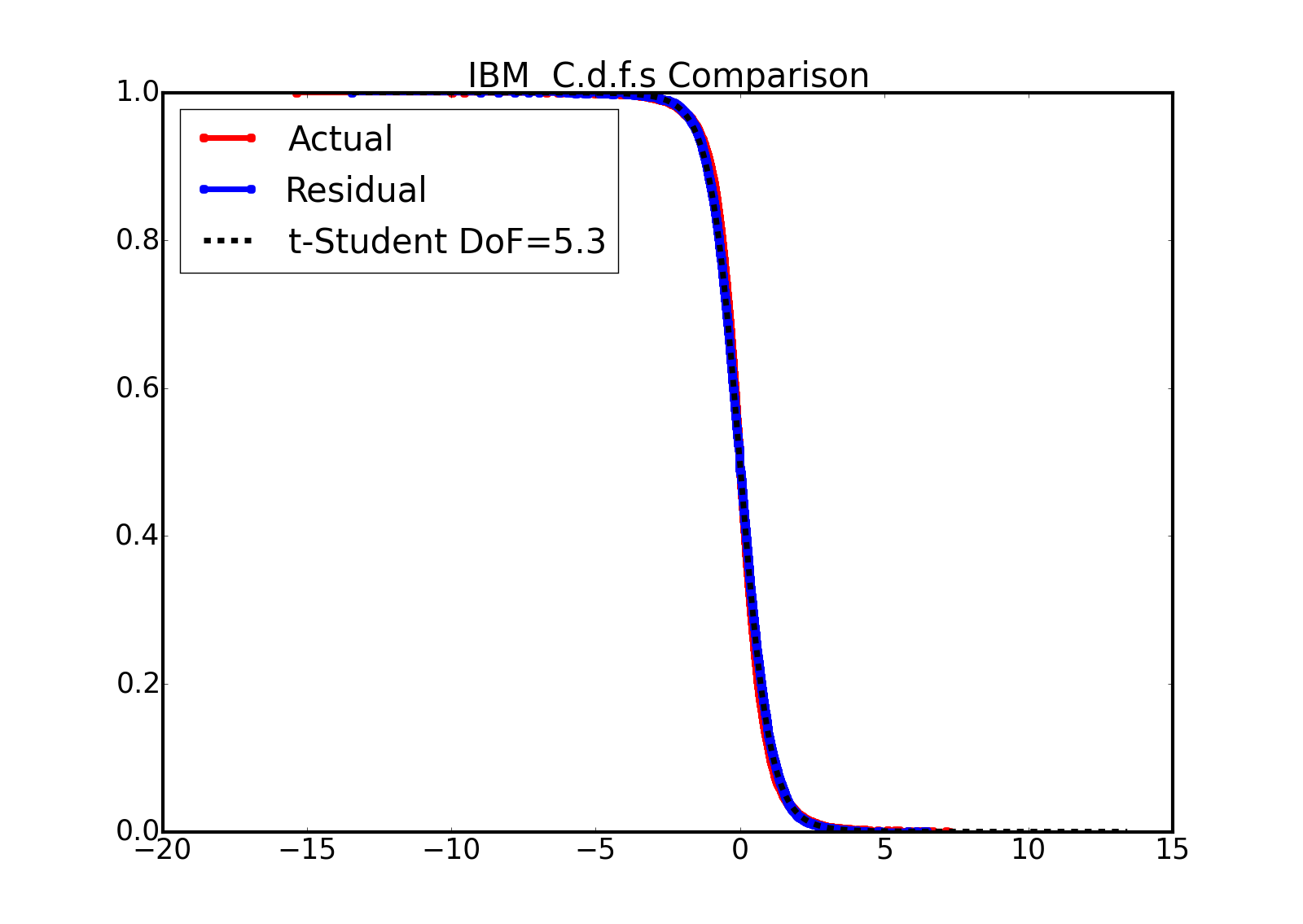}
\caption{}\label{fig:VE21b}
\end{subfigure}
\begin{subfigure}[b]{0.475\textwidth}
\includegraphics[width=\textwidth]{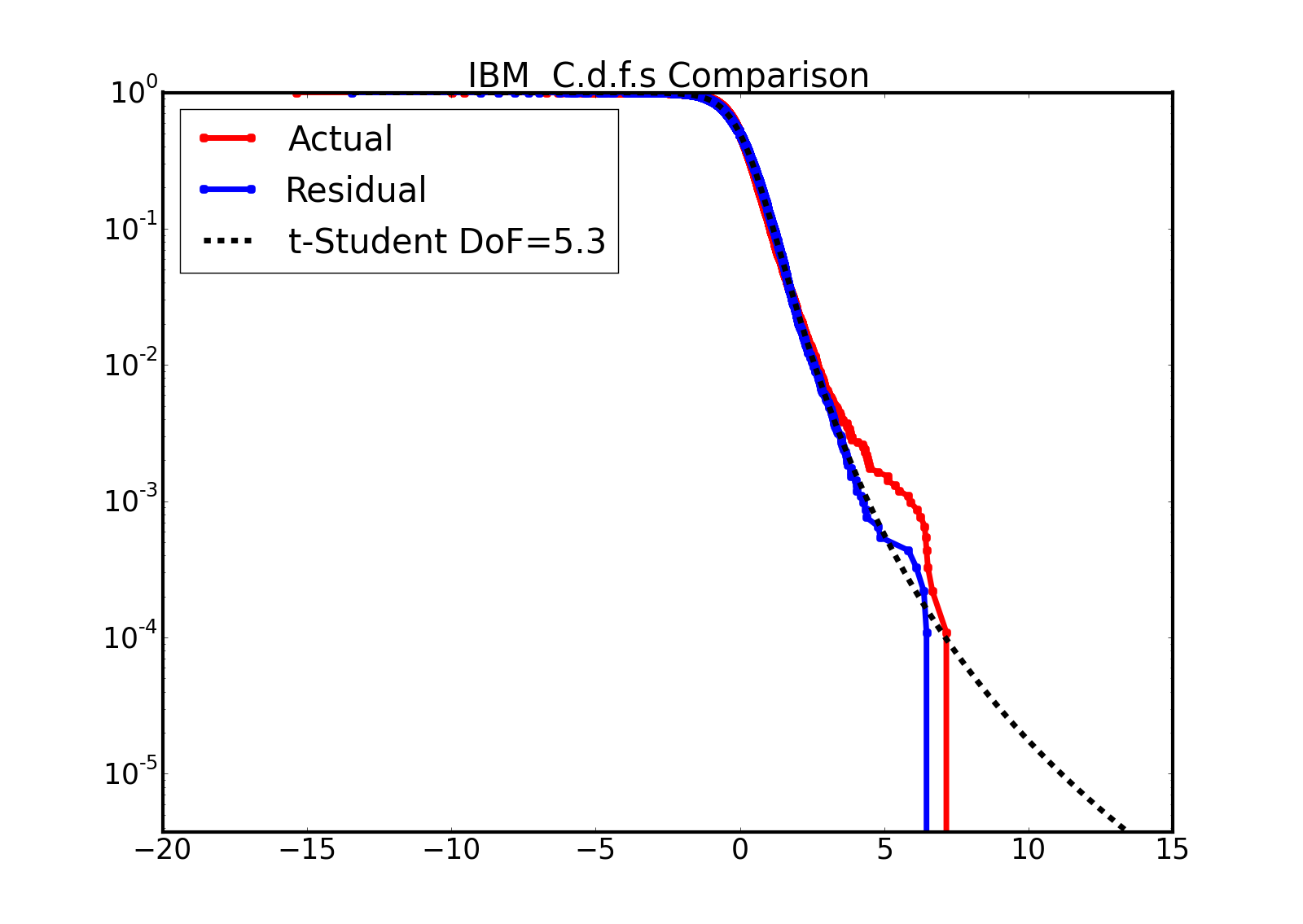}
\caption{}\label{fig:VE21c}
\end{subfigure}
\caption{In the first panel (\ref{fig:VE21a}), the IBM stock daily log-return time series and the volatility estimate time series obtained by considering a GARCH model with T-Student innovations  are reported. While, in the other panels (\ref{fig:VE21b} and \ref{fig:VE21c}),  the empirical  c.d.f.~of the whole sample (red line), the empirical  c.d.f.~of the GARCH residuals (blue line) and the   c.d.f.~of the T-Student random variable  with degrees of freedom, location and scale parameters  assumed for the GARCH model innovations (black dashed line)   are shown, in linear scale and logarithmic scale respectively. 
Time series  parameters: number of observations $N=9180$ (from 17 March 1980 to 9 August 2016), daily recording frequency. Estimator parameters: GARCH model with symmetric T-Student innovations. \label{figVE21}}
\end{figure}

We also tested the performances of our estimator on real data samples. In detail, we applied the OS algorithm to estimate the local volatility on the daily log-return time series of the IBM stock. 
We compared the obtained results with the ones provided by the time-varying volatility estimation via GARCH model with t-Student innovations. For practical purposes, we used the  demeaned time series.\\
Figure \ref{figVE20} shows the volatility estimates provided by the OS estimator, while Figure \ref{figVE21} reports the GARCH volatility estimates. The red dots stand for the realizations identified as jumps by the OS estimator. Moreover, in Figures \ref{fig:VE20b}, \ref{fig:VE20c} the empirical c.d.f.~of the  whole sample renormalized realizations   (red line), the empirical c.d.f.~of the renormalized realizations identified as non-jumps by the OS estimator (blue line) and the Standard Normal random variable c.d.f.~(black dashed line) are reported, both with linear and logarithmic scale. Instead, Figures \ref{fig:VE21b}, \ref{fig:VE21c} show the empirical c.d.f.~of the whole sample (red line), the empirical c.d.f.~of the GARCH residuals  (blue line) and the c.d.f.~of  a t-Student random variable with degrees of freedom, location and scale parameters  assumed for the GARCH model innovations (black dashed line), both with linear and logarithmic scale.
Comparing Figure \ref{fig:VE20a} with Figure \ref{fig:VE21a}, 
the volatilities estimated  by the OS estimator result smoother than the ones obtained by considering the GARCH model.\\
Moreover, it is worth noting that, in general, the OS estimated volatilities are smaller than the GARCH ones, since the former evaluates the Gaussian process volatility, filtering the detected jumps, while the latter provides an estimate of the whole process volatility.
\section{Sanity Check: Empirical Results}\label{Data_Description}
In order to provide a sanity check for the OS estimator, we  massively applied it on real data log-return times series and we
tested whether the renormalized realizations classified as non-jumps 
 really came from a standard Normal distribution as supposed by the OS estimator model. 
\subsection{Dataset Description and Diagnostic Procedure}
The dataset used for the residual diagnostics consists of  the S\&P 500 stock daily log-return time series with more than $2000$ observations in the period of interest from 1 January 2001 to December 2016, with a total of $307$ time series. The stock log-returns were computed by using the close-prices.\\\\
For each time series, we checked whether the renormalized realizations not classified as jumps by the OS estimator were distributed as a standard Normal random variable by applying the Anderson-Darling (AD) test. 
Roughly speaking, the AD test compares a theoretical distribution with an empirical one measuring their distance in an appropriate metric. 
If such distance is too large, then the null hypothesis that the analyzed sample comes from the theoretical distribution must be rejected. For further details about the AD test, refer to \cite{ad1954}. We stress that in this case, as we are interested in proving the null hypothesis that the no-jump classified returns are Gaussian, the higher is the confidence level, the more severe  the test  is.\\ 
To provide strong evidences of the consistency check success, we set the confidence level of the AD test equal to $15\%$, allowing for a $15\%$ of probability in the type I error   occurrence (i.e.~in the 15\% of the cases we could reject the null hypothesis even if it was true). 
\subsection{AD Test Results}
\begin{figure}[t]
\centering
\begin{subfigure}[b]{0.475\textwidth}
\includegraphics[width=\textwidth]{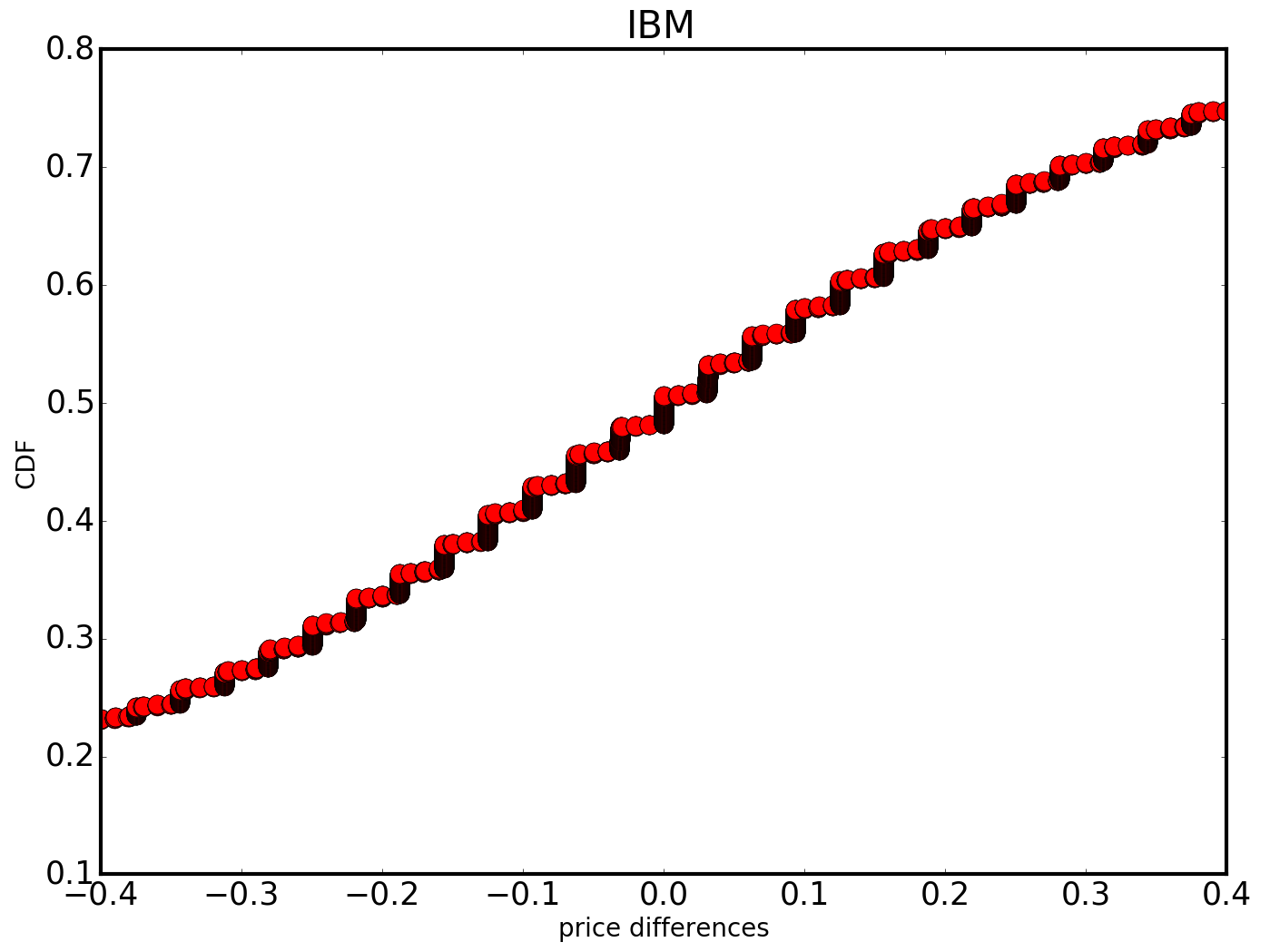}
\caption{}\label{fig:VE25a}
\end{subfigure}
\begin{subfigure}[b]{0.475\textwidth}
\includegraphics[width=\textwidth]{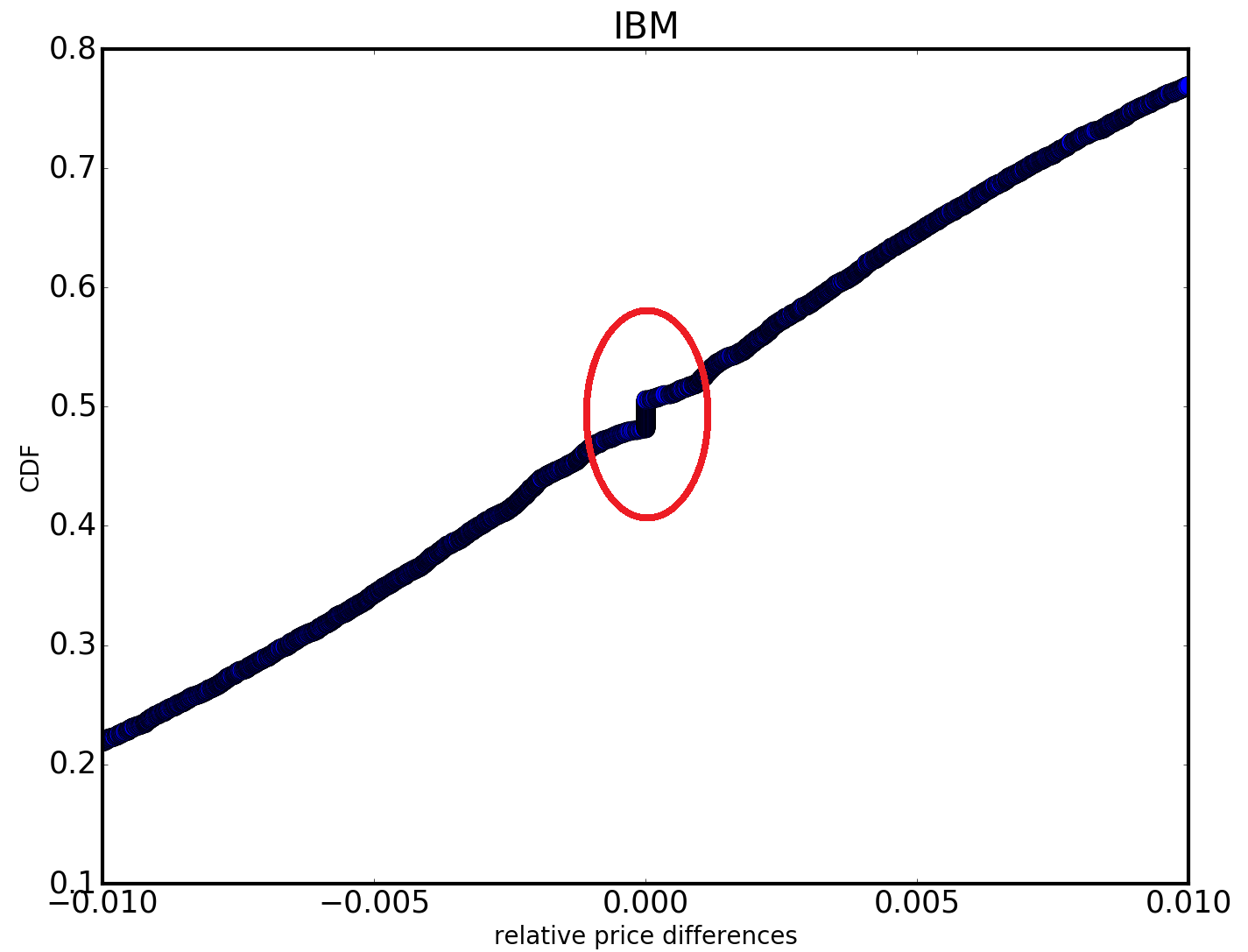}
\caption{}\label{fig:VE25b}
\end{subfigure}
\caption{In the first panel (\ref{fig:VE25a}), the empirical c.d.f.~of the IBM stock daily price difference time series is reported. While, in the second  panel  (\ref{fig:VE25b}), the empirical c.d.f.~of the IBM stock daily relative return time series is shown.  
Time series  parameters:  observations  from 3 September 1980 to 12 September  2012, daily recording frequency. \label{figVE25}}
\end{figure}

The results of the application of the AD test with null hypothesis
\[
H_0: \text{the sample data comes from the Normal distribution}
\]
on the $307$  samples of the residuals classified as non-jumps are summarized in Table \ref{tab:1} (``Standard Prices'' row).\\
\begin{table}[htbp]
  \centering
    \begin{tabular}{l|cc}
          & \textbf{\# TS } & \textbf{Rejection } \\
          & \textbf{rejected Ho} & \textbf{Rate} \\
\midrule
    \textbf{Standard Prices} & 60    & 20\% \\
    \textbf{Adjusted Prices} & 50    & 16\% \\
    \end{tabular}%
  \caption{Number of time series that did not pass the AD test on the Normal distribution of the residuals classified as non-jumps and corresponding rejection rates. Time series  features:  daily residuals  corresponding to the log-returns of the S\&P 500 stocks  classified as non-jumps by the OS estimator, number of observations $N>2000$ (from  1 January 2001 to December 2016),  307 analyzed samples.  Estimator parameters: OS estimator tolerance level $p=5\%$, bandwidth $h=50$.}\label{tab:1}%
\end{table}%
The quite high percentage of time series for whom the null hypothesis of the AD test could not be rejected confirms the success of the performed consistency check. Nevertheless, in order to investigate the reasons leading to the rejection of the AD test null hypothesis for   some times series, we carried out  a deeper analysis on the log-return time series used as input in the consistency check. \\
 As a result, we found out that, due to the quotation conventions, for some time series the frequency of exactly zero log-returns is much higher than expected, influencing the result of the AD test and implying the rejection of the null hypothesis.\\
More in detail, since 2001 the tick size (i.e.~the minimum price increment in which the prices are quoted) in the U.S. stock market has been 0.01\$. Consequently, the stock price differences suffer from a discretization issue and 
can only take values in the set of the multiples of 0.01 (i.e.~$M=\left\{m : m=0.01 z , z \in \mathbb{Z}\right\}$), without spanning the interval $\left]m ,m-0.01\right[$ for each possible multiple  $m \in M$. For instance, Figure \ref{fig:VE25a} shows the empirical c.d.f.~of the IBM stock price differences and the resulting step function is a clear signal of the discretization effect induced by the quotation convention. \\ With  the relative return computation,   the discrete price differences are divided by the corresponding previous stock prices. However,  while the price differences exactly equal to zero stay to zero even if divided by the previous stock price (whatever its value is), all the other discrete price differences lead to  real numbers whose values also depend on the (random) previous stock price amounts. As a result, the frequency of the relative returns exactly equal to 0 is higher than expected. As an example, Figure \ref{fig:VE25b} shows the empirical c.d.f.~of the relative returns of the IBM stock and the  huge vertical step centered in zero is the outcome of   the too high frequency of zero returns that can be probably caused by the price  discretization effect. Except for zero, the empirical c.d.f.~of the other returns results smooth. \\ The log-returns, being approximations of the relative returns, suffer from the same issue.
\\Adjusting the stock prices by completing them with random digit sequences leads to a slightly  improvement of  the AD test performances as reported in Table  \ref{tab:1} (``Adjusted Prices'' row).\\
\\For completeness, it should be mentioned that before 2001 the tick size typical of the U.S. stock market was $\frac{1}{8}$\$$=0.125\$$ (or other fractions) and thus the procedure to adjust  these prices with random digits result slightly   more tricky than adjusting the prices based on the decimal system quotation convention.\\
To summarize, the analyzes confirm the overall agreement between the estimator reference distribution hypothesis and the empirical results.

\section{Jumping VaR Model}\label{JumpingVaR_Models}
The Order Statistic (OS) volatility estimator has a  practical application  in the field of Market Risk modeling. Indeed, it can be used as tool to improve the historical simulation method for VaR forecasting.\\\\ First of all, the volatility estimates can be employed to filter (re-normalize)  the loss process   random variables in order to make them identically distributed from the second moment point of view. Moreover, the distinction between the jump component and the ordinary (Gaussian) component realizations of the loss process identified through the estimator can be used to study jump probabilities  and, therefore, to make a more accurate VaR prediction.\\\\
Given the sequence $\left\{l_i\right\}_{i=1}^{N}$ of the $N$ realizations of the loss process from time $t-1$ to time $t-N-1$,  the OS estimator allows to associate  a boolean indicator stating the occurrence  of a jump or not with each $l_i$. Formally, we define $\left\{j_i\right\}_{i=1}^{N}=\left\{j_1,j_2,\dots,j_N\right\}$ to be the sequence of  the $N$  indicators concerning the realizations of the jump component of the loss process from time $t-1$ to time $t-N-1$, with $j_i=\begin{cases}1&\text{if a jump occurs} \\ 0 & \text{otherwise}
\end{cases}\quad\forall i=1,\dots ,N$. \\
This additional information  on the loss process up to time $t-1$ is thus exploited to produce a  better estimate of the Value at Risk at time $t$.\\\\ 
The new model descriptions and the outcomes of their application are presented as follows.

\subsection{Normalized Risk Model}
The Normalized Risk model is the easiest model among the proposed ones. It is based on the assumption that the loss process random variables are independent but not identically distributed. Within this framework, in order to consider the loss process realizations drawn from i.i.d.~random variables, they are renormalized by their standard deviations.\\\\
Given the sequence of the $N$ realizations $\left\{l_i\right\}_{i=1}^{N}$ defined before and used to approximate the c.d.f.~of the loss process  at time $t$, let $\left\{\sigma_i\right\}_{i=1}^{N}$  be the volatility estimates associated with each realization of the loss process from  time $t-1$ to time $t-N-1$. Under the hypothesis of zero-mean  loss process random variables, the filter
\begin{equation}\label{eq1}
\hat{l_i}=\frac{l_i}{\sigma_i}\quad \forall i=1,\dots,N
\end{equation}
makes these realizations drawn from i.i.d.~random variables $\sim d\left(0,1\right)$, where $d$ is a generic distribution with zero mean and standard deviation equal to one.\\
The Value at Risk at time $t$ with confidence level $\lambda$, $VaR_{\lambda,t}$, is computed as the sample $\lambda$-quantile of the empirical c.d.f.~of the renormalized loss process at time $t$, built associating an uniform occurrence probability with each value $\hat{l_i}$, multiplied by the last disposable volatility $\sigma_N$, as we assumed that the last available volatility is the best forecast of the volatility over the VaR time horizon. 

\subsection{Jumping VaR Model}\label{Jumping_VaR_model_section}
Supposing the loss process can be decomposed into an ordinary component and a jump one, the Jumping VaR model assumes that 
the occurrence or not of a jump at time $t$ can be described by a Bernoulli random variable whose probability value is   estimated by using the past realizations of the loss process (from time $t-1$ to time $t-T$, for a fixed time interval $T$). \\
The jump size is assumed independent on the past realizations of the jump component.\\\\
More precisely, let  $X_t$ be a Bernoulli random variable that  assumes value 1 if a jump occurs at time $t$ with probability $ p_J\left(t\right)$ and 0 otherwise, associated with the corresponding loss process random variable. Formally,
\begin{equation}\label{eq11}
X_t=\begin{cases}1& p_J\left(t\right) \\ 0 &  1-p_J\left(t\right) 
\end{cases}
\quad  \forall t
\end{equation}
with $X_t, X_s$ independent $\forall t\neq s$.\\
The probability $p_J\left(t\right) $ is defined as a deterministic function of time only.\\\\ 
Within this framework, the occurrence probabilities $p_i$ $i=1,\dots N$ associated with the $N$ renormalized realizations of the loss process, $\left\{\hat{l_i}\right\}_{i=1}^{N}$, 
are computed as follows.
\subsubsection{Occurrence Probabilities \label{sec1}}
Let $J=\left\{j_i \in \left\{j_i\right\}_{i=1}^{N} : j_i=1 \right\}$ be the set of the boolean indicators corresponding to a  jump realization of the loss process from time $t-1$ to time $t-N-1$ and let $NOJ=\left\{j_i \in \left\{j_i\right\}_{i=1}^{N} : j_i=0 \right\}$ be the set of the boolean indicators corresponding to  no jump realization of the loss process from time $t-1$ to time $t-N-1$. Under the model assumptions,   the occurrence probabilities $p_i=\frac{\alpha}{N}$ assigned to the indicators in the set $J$ and thus to the corresponding loss process realizations, are determined imposing that their sum is equal to the jump probability at time $t$, $p_J\left(t\right)$. 
 Formally,
\begin{equation}\label{eq12}
\begin{split}
 p_J\left(t\right)=\sum_{i=1}^{\#J}\frac{\alpha}{N} &  \iff   p_J\left(t\right)=\alpha\sum_{i=1}^{\#J}\frac{1}{N}\\
&\iff \alpha =\frac{p_J\left(t\right)}{p_J}
\end{split}
\end{equation}
The last equality follows from the definition of the quantity $p_J:=\sum_{i=1}^{\#J}\frac{1}{N}=\frac{1}{N}\sum_{i=1}^{N}j_i=\frac{\#J}{N}$.\\
Under the hypothesis of jump sizes independent on the  past, the same occurrence probability is associated to all the jump indicators in the set $J$.
The realizations of the loss process due to a jump are indeed considered equiprobable.\\\\
The occurrence probabilities  $p_i=\frac{\beta}{N}$ assigned to the  jump indicators in the set $NOJ$ are computed requiring that all the occurrence probabilities $p_i$ sum to 1 
\begin{equation}\label{eq13}
\begin{split}
\sum_{i=1}^{N}p_i=1&\iff \sum_{i=1}^{\#J}\frac{\alpha}{N}+\sum_{i=1}^{\#NOJ}\frac{\beta}{N}=1\\ 
&\iff \alpha p_J+\beta\left(1-p_J\right)=1\\
&\iff \beta =\frac{1-\alpha p_J}{1- p_J}=\frac{1-p_J\left(t\right)}{ 1- p_J }
\end{split}
\end{equation}
\subsubsection{Jump Probability Forecasting \label{sec_jump_probability_forecasting}}
The previously defined sequence $\left\{j_i\right\}_{i=1}^{N}$ represents the realizations of the  Bernoulli random variables $X_s$ $s=t-1,\dots,t-N-1$. Using them and fixed a time interval of size $T\leq N$, the time-average jump probability within the interval from $s-T$ to $s$ can be estimated through the estimator
\begin{equation}\label{eq14}
\widehat{p_J\left(s\right)}= \frac{1}{T}\sum_{v=s-T}^{s}X_v 
\end{equation}
for each time $s \in \left\{t-1,\dots,t-\left(N+T+1\right)\right\}$ .
\\It is worth noting that $\widehat{p_J\left(s\right)}$ produces an unbiased estimate of the time-average jump probability within the interval from $s-T$ to $s$ (i.e.~$\mathbb{E}\left [\widehat{p_J\left(s\right)}  \right ]= \frac{1}{T}\sum_{v=s-T}^{s}p_J\left(v\right) $).
We also assume that the best forecast of the jump probability at time $t$ given all the information up to time $t-1$ is the estimate of the  time-average  jump probability at time $t-1$.
\subsubsection{$VaR_{\lambda,t}$ Computation\label{ss1}}
Fixed  time $t$ and knowing the realizations of the jump component from time $t-1$ to time $t-N-1$, the occurrence probabilities $p_i$ $ i=1,\dots,N$ associated with the jump boolean indicators, $\left\{j_i\right\}_{i=1}^{N}$,  and thus with the corresponding renormalized loss process realizations, $\left\{\hat{l_i}\right\}_{i=1}^{N}$, are computed as explained in the Subsections~\ref{sec1},~\ref{sec_jump_probability_forecasting}. The c.d.f.~of the filtered loss process at time $t$ is approximated assigning at each  realization $\hat{l_i}$ $i=1,\dots,N$  the relative occurrence probability $p_i$ $i=1,\dots,N$. The $VaR_{\lambda,t}$ is estimated as the $\lambda$-quantile of this distribution multiplied by the last disposable volatility $\sigma_N$.

\section{VaR Back-testing Performances}\label{Results}
\begin{figure}[!h]
\centering
\includegraphics[width=0.6\textwidth]{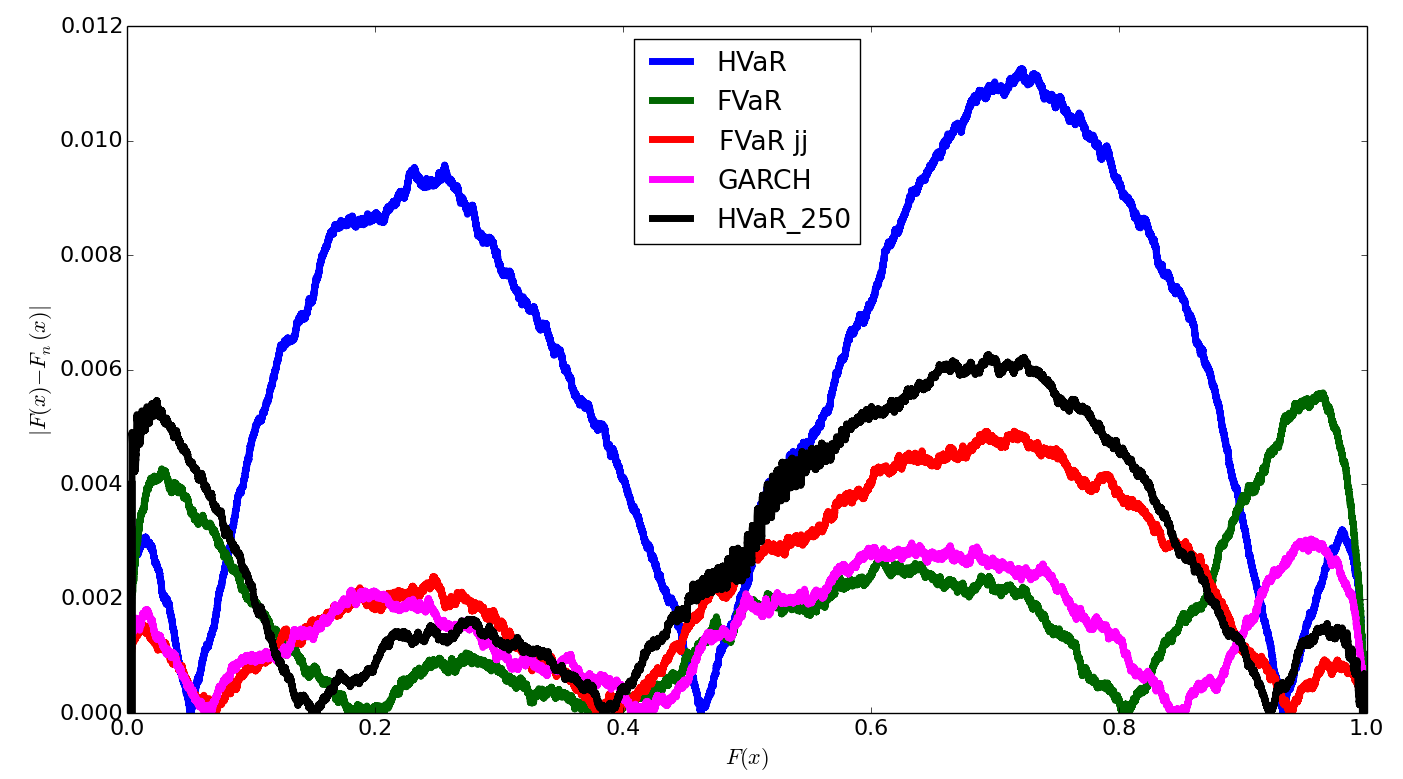}
\caption{Absolute differences between the empirical c.d.f., resulting from the  percentile ranks obtained comparing the daily back-testing log-returns and the predicted  log-return distributions, and the theoretical Uniform c.d.f. for the  S\&P500 stock log-return time series. VaR models:  standard historical simulation   method (HVaR with time window of 1000 observations and HVaR\_250  with time window of 250 observations), Normalized Risk model (FVaR),
Jumping VaR model (FVaR jj), filtered historical simulation method with T-GARCH (GARCH model with T-Student innovations) volatility estimates (GARCH).\label{figB2}}
\end{figure}

To practically test the goodness of the models  proposed in  Section \ref{JumpingVaR_Models}, we forecast the Value at Risk for the 307 S\&P500 stock log-return time series, assessing   the back-testing outcomes.\\\\
More in detail, we compared the performance  of $5$ models: 
\begin{itemize}
\item the filtered Jumping VaR model (FVaRjj), as presented in Section~\ref{Jumping_VaR_model_section}
\item the (filtered) Normalized Risk model (FVaR)
\item the standard historical simulation VaR using a sample made by $1000$ realizations (HVaR)
\item the standard historical simulation VaR using a sample made by $250$ realizations (HVaR\_250) and
\item the performance of the filtered historical simulation method with GARCH model (t-Student innovations) volatility estimates (GARCH).
\end{itemize}
\begin{figure}[!h]
\centering
\includegraphics[width=0.6\textwidth]{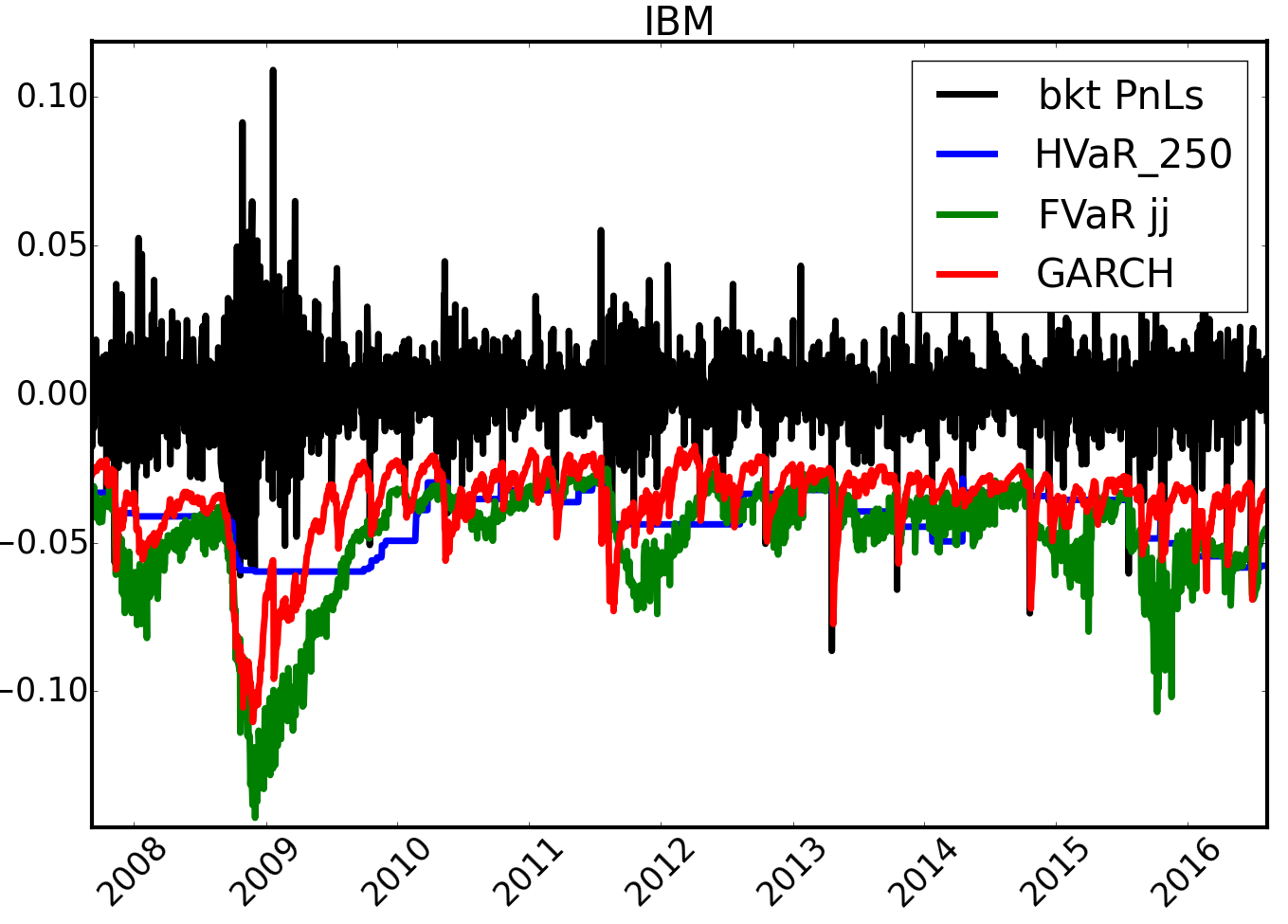}
\caption{IBM stock daily log-return time series and corresponding VaR predictions implied by the  Jumping VaR model (green line), the  standard historical simulation   method with 250-day time window (blue line), the  filtered historical simulation method with T-GARCH volatility estimates (red line). Jumping VaR model parameters: $T=60$, $N=250$. Estimator parameters: OS estimator tolerance level $p=5\%$, bandwidth $h=100$. \label{figB1}}
\end{figure}
Figure \ref{figB1} shows an example of the different VaR predictions for the IBM stock log-return time series (for sake of clarity, in this case we excluded the HVaR model from the plot). \\\\
In order to evaluate the goodness of   the  VaR models, we carried out a test involving the comparison between  the daily back-testing log-returns and the predicted  log-return distributions. For sake of clarity, the back-testing log-return is the log-return that effectively occurred at a fixed time $t$, while the predicted  log-return distribution is the distribution  predicted at time $t-1$ for time $t$ and used for the VaR forecast at time $t$. \\
 For each time series and for each disposable date, the daily back-testing log-return was compared to the predicted  log-return distribution and, assuming (null hypothesis) that the back-testing log-return is a realization of the  predicted  log-return distribution, the percentile rank corresponding to the considered   quantile  was computed.\\
If the null hypothesis held, the value of each obtained percentile rank would be a realization of an Uniform $[0,1]$ random variable. This backtesting approach is similar to the fundamental framework exploited in order to define the Anderson-Darling test, where two distributions are compared.
To test this hypothesis, we compared the empirical c.d.f., resulting from the occurred percentile ranks, and the theoretical Uniform c.d.f.. \\
Figure \ref{figB2} shows the average absolute differences between the resulting  empirical c.d.f.~points and the Uniform c.d.f.~for all the samples concerning the 307 S\&P500 stock log-return time series.
\\\\As far as the equity time series are concerned, the historical simulation   methods, both with the time window of 1000 observations (HVaR) and with the time window of  250 observations (HVaR\_250), have the worst performances, even if the reduction of the time window length improves the outcomes of this  method for the VaR estimation. The performances of the Normalized Risk model (FVaR) are slightly better than  the other models in the central part of the distribution that should correspond to the Gaussian regime of the true loss distribution.   Nevertheless, the introduction of the jump adjustment involved by the Jumping VaR model (FVaR jj) allows to improve the performance of this method in the tails of the   true loss distribution. Finally, the filtered historical simulation method with T-GARCH volatility estimates shows a performance comparable to the Jumping VaR model one.\\
 
\section{Summary and Conclusion}\label{Summary_Conclusion}
We propose a new integrated variance estimator that relies on a well-defined theoretical framework of integrated variance threshold estimators and that allows to define a new time-varying volatility estimator for return time series. Our approach is based on the order statistics theory which enables to discriminate the jump part realizations from the continuous part ones of a stochastic process not only according to the size of the realizations,  but also according to their  frequency  of occurrence.\\ Numerical and empirical examples 
compare  the OS volatility estimator performances with the ones of the threshold and T-GARCH estimators, showing the capability of this estimator of disentangling the process volatility due to its continuous part from the further ``volatility" due to the jump component. \\ An on-line version of the Python code implementing our estimator is also available on the website\footnote{\url{https://github.com/sigmaquadro/VolatilityEstimator}}.\\
The main advantages of the OS volatility estimator are the following.
\begin{itemize}
\item A formal definition of \textit{jump} is provided and a jump classification algorithm is described.
\item The jump classification method is based on the definition of a confidence interval instead of a threshold. This allows the identification of hidden jumps, i.e. the jumps below the threshold.
\item Only the Gaussian part of the volatility is estimated. This point can be helpful in order to calibrate a model where a stochastic differential equation, composed by a Wiener process and a jumping component, is specified (e.g. Local Stochastic Volatility Pricing Models).\\\\
\end{itemize}
Finally, we  introduce a new and parsimonious Jumping VaR model for the Value at Risk predictions to include jump
effects in the standard historical (filtered) VaR estimation framework. Empirical back-testing results on equity time series show that the Jumping VaR model outperforms the standard hystorical simulation method and gives comparable results to further advanced historical simulation methods.\\
The main advantages related to the Jumping VaR methodology are:
\begin{itemize}
\item specific assumptions on the tails of the returns distribution are not required (it is only required an assumption on the central part of the distribution), differently to GARCH-like models where an explicit assumption on innovators  distribution is compulsory;
\item thanks to the jump classification method, an ad-hoc methodology for the tail risk can be developed
\end{itemize}
\newpage

\appendix
\section{Volatility Estimator\label{Appendix_B}}

\subsection{Threshold $\theta\left(p;n,n\right)$ and $\theta\left(p;k,n'\right)$ Derivation \label{App1}}
Considering the framework introduced in Subsection \ref{VE_frame}, the threshold value $\theta\left(p;n,n\right)$ in Eq.~\ref{PAeq7} is computed as follows.
\begin{equation}\label{B1}
\begin{split}
\mathbb{P}\left(\underset{i=1,\dots,n}{\text{max}} \left(W_{\left(i+1\right)\delta t}-W_{i\delta t}\right) > \theta^W\left(p;n,n\right)\right)=p &\iff \mathbb{P}\left(\underset{i=1,\dots,n}{\text{max}} \frac{\left(W_{\left(i+1\right)\delta t}-W_{i\delta t}\right)}{\sqrt{\delta t}} > \frac{\theta^W\left(p;n,n\right)}{\sqrt{\delta t}}\right)=p\\
&\iff \mathbb{P}\left(\underset{i=1,\dots,n}{\text{max}} Z_i > \frac{\theta^W\left(p;n,n\right)}{\sqrt{\delta t}}\right)=p\\
&\iff \mathbb{P}\left(Z_{n:n} > \frac{\theta^W\left(p;n,n\right)}{\sqrt{\delta t}}\right)=p\\
&\iff \mathbb{P}\left(Z_{n:n} \leqslant \frac{\theta^W\left(p;n,n\right)}{\sqrt{\delta t}}\right)=1-p\\
&\iff F_{Z_{n:n}} \left( \frac{\theta^W\left(p;n,n\right)}{\sqrt{\delta t}}\right)=1-p\\
&\iff \theta^W\left(p;n,n\right)=\sqrt{\delta t}\underbrace{F^{-1}_{Z_{n:n}}\left(1-p\right)}_{=:\theta\left(p;n,n\right)}
\end{split}
\end{equation}
where $Z_i$ are i.i.d. $\sim N\left(0,1\right)$ for $i=1,\dots,n$ and $Z_{n:n}:= \underset{i=1,\dots,n}{\text{max}} Z_i$, $F_{Z_{n:n}}\left(x\right)=\left(\Phi\left(x\right)\right)^n$ is the cumulative distribution function (c.d.f.) of the maximum of $n$ Standard Normal random variables and $\Phi\left(x\right)$ is the c.d.f.~of a Standard Normal.
\\\\
Repeating the same reasoning, the threshold value $\theta\left(p;k,n'\right)$ in Eq.~\ref{PAeq8} is the following.
\begin{equation}\label{B2}
\begin{aligned}
\mathbb{P}\left(\text{$k$-th ordered statistic in the dataset} \right.& \left.\left\{\left(W_{\left(i+1\right)\delta t}-W_{i\delta t}\right)\right\}_{i=1}^{n'}  > \theta^W\left(p;k,n'\right)\right)=p\\
\ArrowBetweenLines
\mathbb{P}\left(\text{$k$-th ordered statistic in the dataset} \right.& \left.\left\{\frac{\left(W_{\left(i+1\right)\delta t}-W_{i\delta t}\right)}{\sqrt{\delta t}}\right\}_{i=1}^{n'}  > \frac{\theta^W\left(p;k,n'\right)}{\sqrt{\delta t}}\right.\left.\right)=p\\
\ArrowBetweenLines
\mathbb{P}\left(\text{$k$-th ordered statistic in the dataset} \right.& \left.\left\{Z_i\right\}_{i=1}^{n'}  > \frac{\theta^W\left(p;k,n'\right)}{\sqrt{\delta t}}\right.\left.\right)=p\\
\ArrowBetweenLines
\mathbb{P}\left(Z_{k:n'}  >\right.& \left. \frac{\theta^W\left(p;k,n'\right)}{\sqrt{\delta t}}\right.\left.\right)=p\\
\ArrowBetweenLines
\mathbb{P}\left(Z_{k:n'}  \leqslant\right.& \left. \frac{\theta^W\left(p;k,n'\right)}{\sqrt{\delta t}}\right.\left.\right)=1-p\\
\ArrowBetweenLines
F_{Z_{k:n'}}\left(\right.& \left.\frac{\theta^W\left(p;k,n'\right)}{\sqrt{\delta t}}\right.\left.\right)=1-p\\
\ArrowBetweenLines
\theta^W\left(p;k,n'\right)&=\sqrt{\delta t}\underbrace{F^{-1}_{Z_{k:n'}}\left(1-p\right)}_{=:\theta\left(p;k,n'\right)}
\end{aligned}
\end{equation}
where $F_{Z_{k:n'}}\left(x\right)$  is the c.d.f.~of the $k$-th order statistic in dataset of $n'$ Standard Normal random variables. According to the order statistic theory (\cite{book:12094})
\begin{equation}\label{B3}
\begin{split}
F_{Z_{k:n'}}\left(x\right)=&I_{\Phi\left(x\right)}\left(k,n'-k+1\right)\\
=& \frac{n'!}{\left(k-1\right)!\left(n'-k\right)!}\int_{0}^{\Phi\left(x\right)} u^{k-1}\left(1-u\right)^{n'-k} \mathrm{d}  u
\end{split}
\end{equation}
with $\Phi\left(x\right)$ c.d.f.~of a Standard Normal and $I_{x}\left(a,b\right)$  Incomplete Beta function with parameters $a,b$.
\subsection{Approximation $\mathbb{P}\left(\underset{i=1,\dots,n}{\text{max}} |Z_i| > \Lambda\right)\approx 2 \mathbb{P}\left(\underset{i=1,\dots,n}{\text{max}} Z_i > \Lambda\right)$ Derivation \label{App2}}
Let $Z_i$  $\overset{\text{i.i.d.}}{\sim}N\left(0,1\right)$ for $i=1,\dots,n$, let $F_Z\left(z\right)=\Phi\left(z\right)$ be the c.d.f.~of a Standard Normal random variable and let $\Lambda \in \mathbb{R^{+}}$.\\ Moreover, by simple calculations, the c.d.f.~of the folded Normal random variable $|Z_i|$ is $F_{|Z|}\left(z\right)=2\Phi\left(z\right)-1$.\\\\ 
 First of all, the following approximations hold for $\Lambda$ large  (\cite{bouchaud2000theory}) 
\begin{equation}\label{B4}
\begin{split}
\mathbb{P}\left(\underset{i=1,\dots,n}{\text{max}} |Z_i| \le \Lambda\right)&=\left[F_{|Z|}\left(\Lambda\right)\right]^n\\
&=\left[2\Phi\left(\Lambda\right)-1\right]^n\\
&=\left[1-2\left(1-\Phi\left(\Lambda\right)\right)\right]^n\\
&\approx 1-2n\left(1-\Phi\left(\Lambda\right)\right)\\
\end{split}
\end{equation}
where the first equality is consequence of the order statistic theory and, in particular, of the c.d.f.~of the maximum of $n$ identical distributed random variables.\\
Analogously,
\begin{equation}\label{B5}
\begin{split}
\mathbb{P}\left(\underset{i=1,\dots,n}{\text{max}} Z_i \le \Lambda\right)&=\left[F_{Z}\left(\Lambda\right)\right]^n\\
&=\left[\Phi\left(\Lambda\right)\right]^n\\
&=\left[1-\left(1-\Phi\left(\Lambda\right)\right)\right]^n\\
&\approx 1-n\left(1-\Phi\left(\Lambda\right)\right)\\
\end{split}
\end{equation}
Therefore, applying the approximations in Eq.~\ref{B4} and \ref{B5},
\begin{equation}\label{B6}
\begin{split}
\mathbb{P}\left(\underset{i=1,\dots,n}{\text{max}} |Z_i| > \Lambda\right)&=1-\mathbb{P}\left(\underset{i=1,\dots,n}{\text{max}} |Z_i| \le \Lambda\right)\\
&\approx n2\left(1-\Phi\left(\Lambda\right)\right)\\
&=2\left[n\left(1-\Phi\left(\Lambda\right)\right)\right]\\
&\approx 2 \mathbb{P}\left(\underset{i=1,\dots,n}{\text{max}} Z_i > \Lambda\right)
\end{split}
\end{equation}
where the last approximation holds since $ \mathbb{P}\left(\underset{i=1,\dots,n}{\text{max}} Z_i > \Lambda\right)= 1-\mathbb{P}\left(\underset{i=1,\dots,n}{\text{max}} Z_i  \le \Lambda\right) \approx n\left(1-\Phi\left(\Lambda\right)\right)$.
\section{Numerical and Empirical Tests}
\subsection{Jump Size Distribution Filtering and Isolation\label{App3}}
Starting from the characteristic function estimate of the convolution of two random variables and  knowing  the distribution function of the first random variable analytically, the probability density function of the second one can be estimated via Fourier Transform technique. 
\subsubsection{Characteristic Functions and Convolution Property}
Let $X_1$ and $X_2$ be two random variables with real-valued p.d.f.s $f_{X_1}\left(x\right)$ and $f_{X_2}\left(x\right)$, respectively. Moreover, let $\Psi_{X_1}\left(u\right)$ and $\Psi_{X_2}\left(u\right)$ be the corresponding characteristic functions and let $Y:=X_1+X_2$ be the random variable resulting from the convolution of $X_1$ and $X_2$.\\ 
The characteristic function of $Y$ equals 
\begin{equation}\label{B9}
\Psi_{Y}\left(u\right)=\Psi_{X_1}\left(u\right)\cdot\Psi_{X_2}\left(u\right)
\end{equation}
\subsubsection{From Fourier Transforms to p.d.f.s}
Let $Z$ be a random variable with real-valued p.d.f.~$f_{Z}\left(z\right)$ and  let $\Psi_{Z}\left(u\right)$  be the corresponding characteristic function. \\
Recalling that  the characteristic function of a generic real-valued random variable  corresponds to the Fourier Transform of the underlying p.d.f., i.e. $\Psi_{Z}\left(u\right):=\mathbb{E}\left[e^{iuZ}\right]=\int_{\mathbb{R}}e^{iuz}f_{Z}\left(z\right)dz=\hat{f}_{Z}\left(u\right)$, the following property
\begin{equation}\label{B10}
f_{Z}\left(z\right)=\frac{1}{\pi}\text{Re}\left(\int_0^{\infty}\hat{f}_{Z}\left(u\right)e^{iuz}du\right)
\end{equation}
allows to identify the p.d.f  of $Z$ starting from its characteristic function.
\subsubsection{Jump Size Distribution Isolation}
Assuming to know the characteristic function $\Psi_{Y}\left(u\right)$ of the convolution of the Gaussian random variable $X_1$ corresponding to the Brownian motion part and the jump size random variable $X_2$,  the characteristic function of $X_2$ can be computed inverting Eq. \ref{B9}, i.e.
\begin{equation}\label{B11}
\Psi_{X_2}\left(u\right)=\frac{\Psi_{Y}\left(u\right)}{\Psi_{X_1}\left(u\right)}
\end{equation}
with $\Psi_{X_1}\left(u\right)=e^{iu\mu -\frac{1}{2}\sigma^2 u^2}$ and $\mu$ and $\sigma^2$ the mean and variance of the Gaussian random variable, respectively. \\
Finally, the p.d.f. of the jump size random variable can be estimated by numerically evaluating formula \ref{B10}, where the Fourier Transform of $f_{X_2}\left(x\right)$ coincides with the characteristic function $\Psi_{X_2}\left(u\right)$.\\
Note that, in practice,  the characteristic function $\Psi_{Y}\left(u\right)$ can be empirically estimated by using the realizations of the variable $Y$.
\subsubsection{Filtering Technique}
\begin{figure}[!h]
\centering
\begin{subfigure}[b]{0.475\textwidth}
\includegraphics[width=\textwidth]{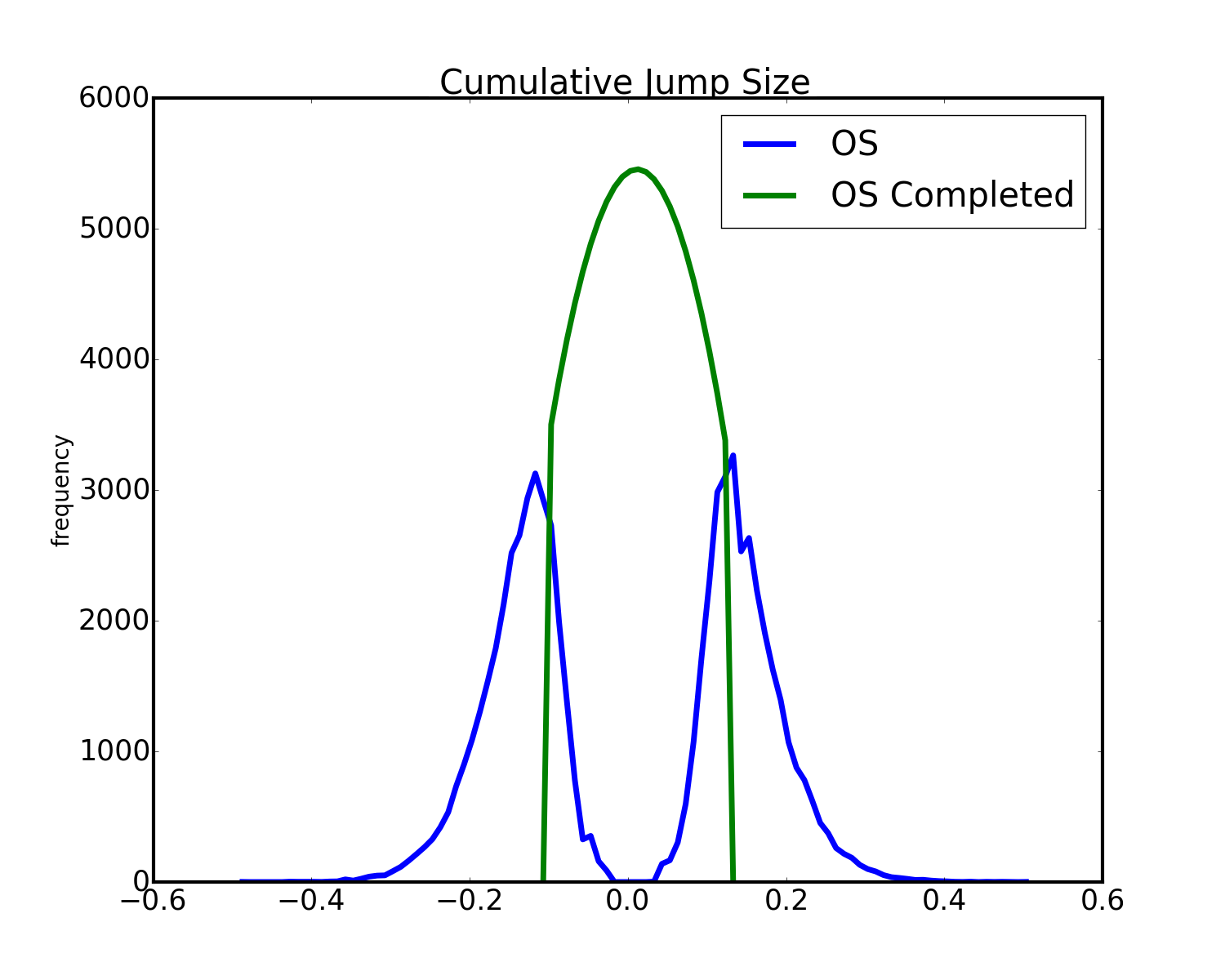}
\caption{}\label{figApp1a}
\end{subfigure}
\begin{subfigure}[b]{0.475\textwidth}
\includegraphics[width=\textwidth]{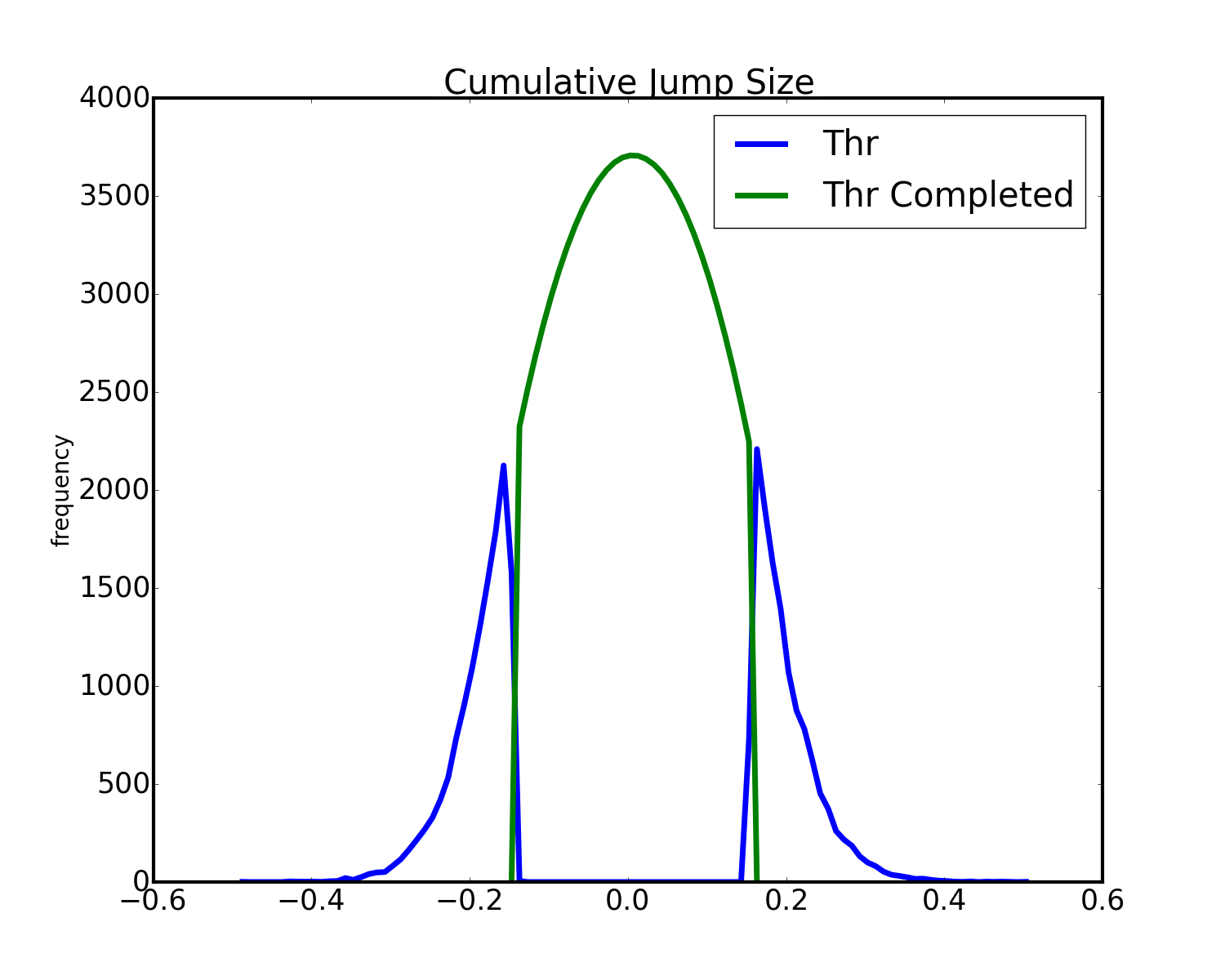}
\caption{}\label{figApp1b}
\end{subfigure}
\caption{Cumulative jump size (i.e.~including both the Brownian Motion and the pure jump realizations) histograms filtering with a cubic interpolating function. In the left panel, the cumulative jump size  histogram created using the jump realizations identified by the OS estimator (blue line) and its central part filling (green line)  are shown, while, in the right panel, the same data obtained using the threshold estimator are reported. Model parameters: $b=0$, $\sigma=0.5$, $\lambda=10$,  $\delta=1.5$ and $\mu=0$. 
Simulation parameters:  number of time-steps for each simulation $N_{TS}=5000$, simulation time horizon $T=20$, number of simulations $M=1000$. Estimator parameters: OS estimator tolerance level $p=5\%$, bandwidth $h=N_{TS}$. \label{figApp1}}
\end{figure}

By means of the OS or threshold estimator the realizations classified as sum of the Brownian component and the jump component are identified and thus their empirical characteristic functions can be estimated. Nevertheless, since the central part of the convoluted distributions is missed by definition of the threshold  estimator and by implementation choices of the OS estimator, a filter technique to complete such distributions must be applied.\\
In practice, we  interpolated the last central points of the histogram of the cumulative jump size with a cubic function and we added to the sample $y_1:=f\left(x_1\right),\dots,y_n:=f\left(x_n\right)$ realizations with sizes $x_1,\dots,x_n$, respectively. $f$ is the interpolating function used.\\ As an example, Figure \ref{figApp1a} and Figure \ref{figApp1b} show the filtering procedure applied to the distributions reported in Figure \ref{figVE14} of Subsection \ref{St}.

\newpage


\end{document}